\newcommand{\DRbarprime}{$\overline{\rm DR}'$}
\newcommand{\lnbar}{{\overline{\rm ln}}}
\newcommand{\xx}{X}

\newcommand{\tbar}{\overline{t}}
\newcommand{\bbar}{\overline{b}}
\newcommand{\taubar}{\overline{\tau}}
\newcommand{\phiOi}{\phi^0_i}
\newcommand{\phiOj}{\phi^0_j}
\newcommand{\phiOk}{\phi^0_k}
\newcommand{\phiOm}{\phi^0_m}

\newcommand{\phipi}{\phi^+_i}
\newcommand{\phimi}{\phi^-_i}
\newcommand{\phipj}{\phi^+_j}
\newcommand{\phimj}{\phi^-_j}
\newcommand{\phipk}{\phi^+_k}
\newcommand{\phimk}{\phi^-_k}
\newcommand{\phipm}{\phi^+_m}
\newcommand{\phimm}{\phi^-_m}
\newcommand{\suL}{\tilde u_L}
\newcommand{\suR}{\tilde u_R}
\newcommand{\sdL}{\tilde d_L}
\newcommand{\sdR}{\tilde d_R}
\newcommand{\seL}{\tilde e_L}
\newcommand{\seR}{\tilde e_R}
\newcommand{\scL}{\tilde c_L}

\newcommand{\ssL}{\tilde s_L}

\newcommand{\snutau}{\tilde \nu_\tau}
\newcommand{\nutau}{\nu_\tau}
\newcommand{\snu}{\tilde \nu}
\newcommand{\stopi}{\tilde t_i}
\newcommand{\stopj}{\tilde t_j}
\newcommand{\stopk}{\tilde t_k}
\newcommand{\stopm}{\tilde t_m}
\newcommand{\stopn}{\tilde t_n}
\newcommand{\stopp}{\tilde t_p}
\newcommand{\stopq}{\tilde t_q}
\newcommand{\sfermion}{\tilde f}
\newcommand{\sfermioni}{\tilde f_i}
\newcommand{\sfermionj}{\tilde f_j}
\newcommand{\sfermionk}{\tilde f_k}
\newcommand{\sfermionm}{\tilde f_m}
\newcommand{\sfermionn}{\tilde f_n}
\newcommand{\sfermionp}{\tilde f_p}

\newcommand{\sboti}{\tilde b_i}
\newcommand{\sbotj}{\tilde b_j}
\newcommand{\sbotk}{\tilde b_k}
\newcommand{\sbotm}{\tilde b_m}
\newcommand{\sbotn}{\tilde b_n}
\newcommand{\sbotp}{\tilde b_p}
\newcommand{\sbotq}{\tilde b_q}
\newcommand{\staui}{\tilde \tau_i}
\newcommand{\stauj}{\tilde \tau_j}
\newcommand{\stauk}{\tilde \tau_k}
\newcommand{\staum}{\tilde \tau_m}
\newcommand{\staun}{\tilde \tau_n}
\newcommand{\staup}{\tilde \tau_p}

\newcommand{\gluino}{\tilde g}
\newcommand{\squark}{\tilde q}
\newcommand{\neutk}{\tilde N_k}
\newcommand{\neutn}{\tilde N_n}
\newcommand{\chark}{\tilde C_k}
\newcommand{\gp}{g'}

\newcommand{\propA}{{\rm A}}
\newcommand{\propB}{{\rm B}}
\newcommand{\propS}{{\rm S}}
\newcommand{\propU}{{\rm U}}
\newcommand{\propM}{{\rm M}}
\newcommand{\propV}{{\rm V}}
\newcommand{\propW}{{\rm W}}
\newcommand{\propX}{{\rm X}}
\newcommand{\propY}{{\rm Y}}
\newcommand{\propZ}{{\rm Z}}

\newcommand{\propgaugeSS}{G_{SS}}
\newcommand{\propgaugeFF}{G_{FF}}
\newcommand{\propgaugeFbarFbar}{G_{\Fbar\Fbar}}
\newcommand{\Fbar}{\overline{F}}

\newcommand\beq{\begin{eqnarray}}
\newcommand\eeq{\end{eqnarray}}

\documentclass[twocolumn,
amsmath,
prd,
nofootinbib,
floatfix
]{revtex4}

\allowdisplaybreaks

\usepackage{axodraw}
\usepackage{graphicx}
\usepackage{bm}

\begin{document}
\renewcommand{\theequation}{\arabic{section}.\arabic{equation}}

\title{Strong and Yukawa two-loop contributions to Higgs scalar boson 
self-energies and pole masses in supersymmetry}

\author{Stephen P. Martin}
\affiliation{
Physics Department, Northern Illinois University, DeKalb IL 60115 USA\\
{\rm and}
Fermi National Accelerator Laboratory, PO Box 500, Batavia IL 60510}


\phantom{.}

\begin{abstract} I present results for the two-loop self-energy functions
for neutral and charged Higgs scalar bosons in minimal supersymmetry. The
contributions given here include all terms involving the QCD coupling, 
and those following from Feynman diagrams involving Yukawa
couplings and scalar interactions that do not vanish as the electroweak
gauge couplings are turned off. The impact of these contributions on the
computation of pole masses of the neutral and charged Higgs scalar bosons
is studied in a few examples.

\end{abstract}


\maketitle

\tableofcontents

\section{Introduction}\label{sec:introduction}
\setcounter{equation}{0}

The small ratio of the electroweak symmetry breaking scale to other
possible energy scales, including the Planck scale, is one of the most
important puzzles in high-energy physics today. This hierarchy is
stabilized by low-energy 
supersymmetry\cite{Haber:1984rc,Gunion:1984yn,Martin:1997ns}, 
but only if it is within discovery reach of
the Large Hadron Collider (LHC) and subject to detailed study at a future
TeV-scale $e^+e^-$ linear collider (LC). It is a pleasant feature of
supersymmetry that the Higgs sector is both perturbatively calculable, and
highly sensitive to radiative corrections at least at two-loop order.  
The precision of measurements at the next generation of high-energy
physics experiments will therefore allow precision tests of theoretical
model frameworks for low-energy supersymmetry. For example, the mass of
the lightest neutral Higgs scalar boson, $h^0$, may be obtained at the LHC
with an uncertainty of perhaps 100-200 MeV \cite{HiggsLHC}, and 
about 50 MeV at a LC \cite{HiggsLCa}-\cite{HiggsLCj}.

Much work (see for example 
\cite{Haber:1990aw}-\cite{Heinemeyer:2004by}
and references therein) has
already been done on radiative corrections to electroweak symmetry
breaking in supersymmetry and on the related problem of evaluating the
physical masses of the Higgs scalar bosons. The effective potential for
the minimal supersymmetric standard model (MSSM) has now been evaluated at
two-loop order \cite{effpot,effpotMSSM}. In addition to allowing an 
accurate
implementation of electroweak symmetry breaking in the MSSM, this allows a
calculation \cite{Martin:2002wn} of the physical mass of the lightest
neutral Higgs scalar, incorporating the complete one-loop results and all
two-loop results in the effective potential approximation. This means that 
the two-loop self-energy contributions to the pole mass are
estimated by setting the external momentum invariant equal to
zero, instead of evaluating them at the pole in the renormalized
propagator. This can be a good approximation for the lightest Higgs
scalar, $h^0$, since it is much lighter than most of the virtual particles
propagating in loops, in particular the top quark and the squarks.  
However, the error made in doing so is still significant compared to the
eventual experimental uncertainty in the mass of a light Higgs scalar 
boson, as we will see below. Also, it is generally not a valid 
approximation for the
other Higgs scalar bosons $H^0, A^0, H^\pm$, especially in the limit
that they are heavy.  In order to adequately compete with the accuracy
that can be obtained at the LHC and an LC, it will probably be necessary
to have the complete momentum-dependent set of corrections to the two-loop
self energy, and the leading three-loop contributions in the effective
potential approximation, at least for $h^0$.

In this paper, I extend previous work by presenting analytical
expressions for some of the leading contributions to the two-loop
self-energy functions for the Higgs scalars in the minimal supersymmetric
standard model. My calculations use the mass-independent \DRbarprime
scheme \cite{DRbarprime} based on regularization by dimensional reduction
\cite{DRED}. Of course, these results should eventually agree with
calculations done in the on-shell type schemes for all questions posed in
terms of physical observables, up to corrections of higher order. As a
matter of opinion, I find the calculations in the \DRbarprime scheme
to be simpler than in the on-shell schemes, and more flexible in the sense
that they can be performed once for generic field theories and then
applied to all kinds of special cases. (In on-shell schemes, the
organization of the calculations depends on a special choice of
observable input parameters; this choice 
will be different for different
particles and for different theories.)  Indeed, the results presented
below rely on calculations already performed in a generic renormalizable
quantum field theory in ref.~\cite{Martin:2003it}. In that paper, formulas 
for the two-loop scalar self-energy diagrams involving up to
two gauge couplings were presented in terms of a minimal basis of two-loop
integrals. Explicit definitions and procedures for the efficient numerical
evaluation of these basis integrals\footnote{The basis integrals are
renormalized versions of the ones whose recursion relations were worked
out in \cite{Tarasov:1997kx} and implemented in \cite{Mertig:1998vk}. The
strategy for their evaluation in \cite{evaluation} (soon to be implemented
in a computer program package \cite{program}) is similar to the one put
forward earlier in \cite{CCLR}. Some other useful two-loop 
self-energy basis integral strategies are found in
\cite{Weiglein:hd}-\cite{Ghinculov:1997pd}.} are described in
ref.~\cite{evaluation,program}. Comparisons with the
predictions of specific models for very high-energy physics and
supersymmetry breaking will require the evaluation of \DRbarprime scheme
parameters, by global fits of many observables to data.

The objects of interest in this paper are the one-loop and two-loop
contributions to the
self-energy function matrices for Higgs scalar fields $\phi_i$:
\beq
\Pi_{ij}(s) = 
\frac{1}{16 \pi^2}
\Pi^{(1)}_{ij}(s) + 
\frac{1}{(16 \pi^2)^2}
\Pi^{(2)}_{ij}(s)  + \ldots ,
\eeq
as functions of the squared-momentum invariant
\beq
s = -p^2.
\eeq
using a metric of signature ($-$$+$$+$$+$). Here $s$ is always given an  
infinitesimal positive imaginary part to resolve branch cuts above 
thresholds. Then the gauge-invariant 
\cite{Willenbrock:1991hu}-\cite{Gambino:1999ai} complex pole masses 
of the Higgs scalar bosons,
\beq
s_k = M_k^2 - i \Gamma_k M_k,
\eeq
can be found by iteratively solving the equation 
\beq
{\rm Det}\left [ (m_i^2 - s_k)\delta_{ij} + \Pi_{ij}(s_k) \right ] = 0,
\label{eq:iteratepole}
\eeq
where the $m_i^2$ are the tree-level renormalized running squared masses.
Here, the self-energy function must be evaluated in the sense of a Taylor
series around a nearby point on the real $s$ axis; in other words, the
self-energy and its derivatives are first evaluated for $s$ with an
infinitesimal positive imaginary part, and this data is then used
to construct a Taylor series expansion for complex $s$. This is necessary
because the imaginary part of the pole mass is negative, while the
imaginary part of $s$ is always positive.
One representation of the solution, which maintains manifest
gauge invariance at each order in perturbation theory, is
\beq
{\rm Det}\left [
(m_i^2 - s_k)\delta_{ij} + [\widetilde \Pi_k]_{ij} \right ] = 0,
\label{eq:expandpole}
\eeq
where, at one-loop order, the solution $s_k^{(1)}$ is obtained using
\beq
[\widetilde \Pi_k]_{ij} = \frac{1}{16 \pi^2} \Pi^{(1)}_{ij}(m_k^2) ,
\eeq
and then at two-loop order,
\beq
[\widetilde \Pi_k]_{ij} &=&  \frac{1}{16 \pi^2} \Pi^{(1)}_{ij}(m_k^2) 
+ \frac{1}{16 \pi^2} (s_k^{(1)} - m_k^2) \Pi^{(1)\prime}_{ij}(m_k^2) 
\nonumber \\ && + \frac{1}{(16 \pi^2)^2} \Pi^{(2)}_{ij}(m_k^2) .
\label{eq:taylorpole}
\eeq
Formally, the difference between this method and the method of iterating
eq.~(\ref{eq:iteratepole}) directly is of three-loop order. However, the
tree-level value of $m_{h^0}^2$ runs quite rapidly with the
renormalization scale $Q$, so performing a Taylor series expansion about
it is formally valid but numerically suspect. The difference between these
two methods for computing the pole masses of the Higgs scalars usually
turns out to be small for the real parts, but the procedure of iterating
eq.~(\ref{eq:iteratepole}) directly gives a result for the imaginary part
of the complex pole mass of $h^0$ that is much more stable with respect to
changes in the renormalization scale $Q$.

The calculations used in this paper neglect the Yukawa couplings of the
first two families, and the corresponding soft (scalar)$^3$ 
interactions. Thus, I use as inputs the following
33 \DRbarprime parameters at a specified renormalization scale $Q$:
\begin{eqnarray}
&&v_u,\> v_d,\\
&&g_3, \>g,\> \gp,\> y_t,\> y_b,\> y_\tau, \\
&&m_{Q_i}^2,\> m_{L_i}^2,\> m_{u_i}^2,\> m_{d_i}^2,\> m_{e_i}^2,\qquad
\!(i=1,2,3)\phantom{xxx}\\
&& m_{H_u}^2,\> m_{H_d}^2,\> b,\> \mu, \\
&& M_3,\> M_2,\> M_1,\> a_t,\> a_b,\> a_\tau,
\end{eqnarray}
in the notation of refs.~\cite{Martin:1997ns,effpotMSSM}.  
No assumptions regarding
CP-violating phases are made, so the last 7 parameters may be complex. The
other parameters are always real, either by definition or by convention,
without loss of generality.  
This means that the formulas below are valid for general CP violation in
the soft terms of the MSSM, but neglecting the usual
Cabibbo-Kobayashi-Maskawa CP violating parameter.  At tree-level, there is
no CP-violation in the Higgs sector, so one defines tree-level mass
eigenstates $\phi^0_i = (h^0, H^0, G^0, A^0)$ and $\phi^\pm_i = (G^\pm,
H^\pm)$ with the usual CP quantum number assignments. In general, the 
self-energy functions then consist of a 
$4 \times 4$ matrix for the neutral scalars $\phi^0_i$, and 
a $2\times 2$ matrix for $\phi_i^\pm$. The parameters $v_u$ and
$v_d$ are actually redundant; they are defined to be the Landau gauge
vacuum expectation values of the Higgs fields at the minimum of the
two-loop effective potential evaluated at $Q$.  In practice, they can be
taken as given and used to eliminate $b$ and $|\mu|$, or vice versa. 


In calculating the effective potential, and the self-energies below, I use
the Landau gauge for electroweak bosons, and a general covariant gauge for
gluon propagators. The fact that $v_u$ and $v_d$ minimize the Landau gauge
two-loop effective potential means that the sum of all tadpole diagrams,
including the tree-level contributions, vanishes identically through the
same order, so that they do not need to be included explicitly in
perturbative calculations.  As in ref. \cite{effpotMSSM}, the tree-level
neutral Higgs squared mass matrices are therefore given by: 
\begin{widetext}
\beq m^2_{\phi_R^0} &=& 
\begin{pmatrix} 
|\mu|^2 + m^2_{H_u} +
(g^2 + \gp^2 )(3 v_u^2 - v_d^2)/4 
& 
-b-({g^2 + \gp^2}) v_u v_d/2 
\cr
-b-({g^2 + \gp^2}) v_u v_d/2 
& 
|\mu|^2 + m^2_{H_d} + (g^2 + \gp^2)(3 v_d^2 - v_u^2)/4 
\end{pmatrix}; 
\label{eq:treescalarhiggs} 
\\
m^2_{\phi_I^0} &=& 
\begin{pmatrix} 
|\mu|^2 + m^2_{H_u} + (g^2 + \gp^2)( v_u^2 - v_d^2)/4 
&
b 
\cr 
b 
& 
|\mu|^2 + m^2_{H_d} + (g^2 + \gp^2)( v_d^2 - v_u^2)/4 
\end{pmatrix},
\label{eq:treepseudoscalarhiggs} 
\eeq 
in the $({\rm Re}[H_u^0],{\rm Re}[H_d^0])$ and 
$({\rm Im}[H_u^0],{\rm Im}[H_d^0])$ bases, respectively.
Note that even in the presence of arbitrary CP violation, the tree-level
squared mass matrices always seperate into $2\times 2$ blocks in this way,
because of the freedom to choose $b$ real and positive at any given
renormalization scale. The complex charge $\pm 1$ Higgs scalar tree-level
squared masses are obtained by diagonalizing the matrix 
\beq
m^2_{\phi^\pm} &=& 
\begin{pmatrix} 
|\mu|^2 + m^2_{H_u} + (g^2 + \gp^2)v_u^2/4 + (g^2 - \gp^2) v_d^2/4 
& 
b + {g^2} v_u v_d/2 
\cr 
b + {g^2} v_u v_d/2 
&
|\mu|^2 + m^2_{H_d} + (g^2 + \gp^2) v_d^2/4 + (g^2 - \gp^2) v_u^2/4 
\end{pmatrix},\phantom{xxxx} 
\label{eq:treechargedhiggs} 
\eeq 
\end{widetext}
in the $(H_u^+, H_d^{-*})$ basis.  

There is another
approach (see for example Appendix E of ref.~\cite{Pierce:1996zz}) in
which the condition of vanishing of the tadpoles is used to replace the
tree-level masses with different expressions, by eliminating $|\mu|^2 +
m^2_{H_u}$ and $|\mu|^2 + m^2_{H_d}$ in favor of combinations of $b$ and
$m_Z^2$. In that approach, tadpole terms do appear explicitly in the
loop-level part of the mass matrices.  Of course, both approaches must
agree in principle on their predictions for the physical masses, through
whatever loop order one is working.  In the approach followed here, the
tree-level eigenvalues of the lightest neutral Higgs and the Goldstone
bosons as obtained from
eqs.~(\ref{eq:treescalarhiggs})-(\ref{eq:treechargedhiggs})  are rather
strongly dependent on the choice of renormalization scale.  These
tree-level masses enter into the kinematic loop integral functions.
However, as we will see below, the resulting scale dependences of the
calculated physical Higgs scalar masses are very small.  A wide range of
renormalization scales gives consistent results for the physical masses,
within the uncertainties inherent in the two-loop approximation.  (It is
also possible to expand the analytical formulas presented here around any
choice of tree-level squared masses, treating the differences as 
perturbations. The one-loop integral functions are all known analytically,
so this does not present any technical difficulties, but will not be
explored in detail here.)


In this paper, I include all one-loop corrections to the Higgs scalar
boson self-energies. 
The two-loop corrections that are included are of two types. First, I
include all diagrams that involve the QCD coupling $g_3$. This includes
all effects of order:
\beq
g_3^2 y_t^2,\>\>
g_3^2 y_t y_b,\>\>
g_3^2 y_b^2,\>\>
g_3^2 g^2,\>\>
g_3^2 g g',\>\>
g_3^2 g^{\prime 2} ,
\eeq
and those related by replacing one or both powers of $y_t$ or $y_b$ by the
corresponding soft coupling $a_t$ or $a_b$. This means all diagrams
involving the gluon or the gluino, and also the diagrams involving the
four-squark interactions proportional to $g_3^2$. Second, I include all
diagrams that do not vanish when the electroweak gauge couplings are 
turned off. These include effects proportional to
\beq
y_t^4,\>\> 
y_t^3 y_b,\>\> 
y_t^2 y_b^2,\>\> 
y_t y_b^3,\>\> 
y_b^4,\>\> 
y_\tau^4,\>\> 
y_b^2 y_\tau^2,
\eeq
and those related by replacing one or more Yukawa coupling(s) by
the corresponding soft terms $a_t, a_b, a_\tau$.  Also, I include
electroweak effects whenever they contribute to the same Feynman diagrams 
as just mentioned. This includes both explicit factors of $g,g'$ in the
scalar couplings that also involve Yukawa couplings, and implicit factors
in the mixing angles of the Higgs scalars, squarks, sleptons, neutralinos
and charginos. (It would seem counterproductive to try to disentangle the 
latter anyway.) In the future when all of the two-loop self-energy
contributions become available, it will just be a matter of adding in
the contributions of the Feynman diagrams not considered here.  
It follows that some, but not all, effects of order e.g.~$y_t^2 g^2$ are
included in the present paper. Thus, the formal level of approximation is
to neglect electroweak effects not involving $g_3$ at two-loop order; but 
for future convenience some of them are included anyway.
In the case of the lightest Higgs scalar boson $h^0$, all other two-loop
corrections to the self-energy are included in the effective potential
approximation, as in refs.~\cite{effpot,effpotMSSM,Martin:2002wn}.

In much of the parameter space of the MSSM, including the decoupling limit
for the heavier Higgs scalars, the scalar $h^0$ is predominantly made out
of the gauge eigenstate field that couples to the top quark, 
while $H^0$, $A^0$, and $H^\pm$ are predominantly made out of the
gauge eigenstate field that has a Yukawa couplings to the bottom quark. 
Therefore, because of the large top mass compared to the other
quarks and leptons, the effects detailed above are generally more 
significant for $h^0$ than for the other Higgs scalars, at least when
$\tan\beta$ is moderate. 
 
\begin{figure*}[p]
\begin{picture}(108,65)(-54,-40)
\SetWidth{0.9}
\DashLine(-45,-22)(0,-22){4}
\DashLine(45,-22)(0,-22){4}
\DashCArc(0,0)(22,-90,270){4}
\Text(0,-32)[]{$\propA_{S}$}
\end{picture}
\hspace{0.35cm}
\begin{picture}(108,65)(-54,-40)
\SetWidth{0.9}
\DashLine(-45,0)(-22,0){4}
\DashLine(45,0)(22,0){4}
\DashCArc(0,0)(22,0,180){4}
\DashCArc(0,0)(22,180,360){4}
\Text(0,-32)[]{$\propB_{SS}$}
\end{picture}
\hspace{0.35cm}
\begin{picture}(108,65)(-54,-40)
\SetWidth{0.9}
\DashLine(-45,0)(-22,0){4}
\DashLine(45,0)(22,0){4}
\CArc(0,0)(22,0,180)
\CArc(0,0)(22,180,360)
\Text(0,-32)[]{$\propB_{FF}$}
\end{picture}
\begin{picture}(108,70)(-54,-40)
\SetWidth{0.9}
\DashLine(-45,-22)(0,-22){4}
\DashLine(45,-22)(0,-22){4}
\PhotonArc(0,0)(22,-90,270){-2}{14.5}
\Text(0,-32)[]{$\propA_{V}$}
\end{picture}
\hspace{0.35cm}
\begin{picture}(108,70)(-54,-40)
\SetWidth{0.9}
\DashLine(-45,0)(-22,0){4}
\DashLine(45,0)(22,0){4}
\PhotonArc(0,0)(22,0,180){2}{8.5}
\DashCArc(0,0)(22,180,360){4}
\Text(0,-32)[]{$\propB_{SV}$}
\end{picture}
\hspace{0.35cm}
\begin{picture}(108,70)(-54,-40)
\SetWidth{0.9}
\DashLine(-45,0)(-22,0){4}
\DashLine(45,0)(22,0){4}
\PhotonArc(0,0)(22,0,180){2}{8.5}
\PhotonArc(0,0)(22,180,360){2}{8.5}
\Text(0,-32)[]{$\propB_{VV}$}
\end{picture}

\vspace{.5cm}

\begin{picture}(108,80)(-54,-40)
\SetWidth{0.9}
\DashLine(-45,0)(-22,0){4}
\DashLine(45,0)(22,0){4}
\CArc(0,0)(22,0,180)
\CArc(0,0)(22,180,360)
\Gluon(0,-22)(0,22){3}{5}
\Text(0,-32)[]{$\propM_{FFFFV}$}
\end{picture}
\begin{picture}(108,80)(-54,-40)
\SetWidth{0.9}
\DashLine(-45,-18)(-22,-18){4}
\DashLine(45,-18)(22,-18){4}
\Line(-22,-18)(22,-18)
\GlueArc(0,18)(22,0,180){3}{7}
\Line(-22,18)(22,18)
\Line(-22,-18)(-22,18)
\Line(22,-18)(22,18)
\Text(0,-32)[]{$\propV_{FFFFV}$}
\end{picture}
\begin{picture}(108,80)(-54,-40)
\SetWidth{0.9}
\DashLine(-45,0)(-22,0){4}
\DashLine(45,0)(22,0){4}
\DashCArc(0,0)(22,0,180){4}
\DashCArc(0,0)(22,180,360){4}
\Gluon(0,-22)(0,22){3}{5}
\Text(0,-32)[]{$\propM_{SSSSV}$}
\end{picture}
\begin{picture}(108,80)(-54,-40)
\SetWidth{0.9}
\DashLine(-45,-18)(-22,-18){4}
\DashLine(45,-18)(22,-18){4}
\DashLine(-22,-18)(22,-18){4}
\GlueArc(0,18)(22,0,180){3}{7}
\DashLine(-22,18)(22,18){4}
\DashLine(-22,-18)(-22,18){4}
\DashLine(22,-18)(22,18){4}
\Text(0,-32)[]{$\propV_{SSSSV}$}
\end{picture}

\vspace{.5cm}

\begin{picture}(108,80)(-54,-40)
\SetWidth{0.9}
\DashLine(-45,-16)(0,-16){4}
\DashLine(45,-16)(0,-16){4}
\GlueArc(0,18)(22,0,180){3}{7}
\DashLine(-22,18)(22,18){4}
\DashLine(0,-16)(-22,18){4}
\DashLine(0,-16)(22,18){4}
\Text(0,-32)[]{$\propW_{SSSV}$}
\end{picture}
\begin{picture}(108,80)(-54,-40)
\SetWidth{0.9}
\DashLine(-45,0)(-22,0){4}
\DashLine(45,0)(22,0){4}
\CArc(0,0)(22,-90,90)
\DashCArc(0,0)(22,90,270){4}
\Line(0,-22)(0,22)
\Text(0,-32)[]{$\propM_{SFSFF}$}
\end{picture}
\begin{picture}(108,80)(-54,-40)
\SetWidth{0.9}
\DashLine(-45,-18)(-22,-18){4}
\DashLine(45,-18)(22,-18){4}
\Line(-22,-18)(22,-18)
\DashCArc(0,18)(22,0,180){4}
\Line(-22,18)(22,18)
\Line(-22,-18)(-22,18)
\Line(22,-18)(22,18)
\Text(0,-32)[]{$\propV_{FFFFS}$}
\end{picture}
\begin{picture}(108,80)(-54,-40)
\SetWidth{0.9}
\DashLine(-45,-16)(0,-16){4}
\DashLine(45,-16)(0,-16){4}
\CArc(0,18)(22,0,180)
\Line(-22,18)(22,18)
\DashLine(0,-16)(-22,18){4}
\DashLine(0,-16)(22,18){4}
\Text(0,-32)[]{$\propW_{SSFF}$}
\end{picture}

\vspace{0.46cm}

\begin{picture}(108,80)(-54,-40)
\SetWidth{0.9}
\DashLine(-45,-18)(-22,-18){4}
\DashLine(45,-18)(22,-18){4}
\DashLine(-22,-18)(22,-18){4}
\CArc(0,18)(22,0,180)
\Line(-22,18)(22,18)
\DashLine(-22,-18)(-22,18){4}
\DashLine(22,-18)(22,18){4}
\Text(0,-32)[]{$\propV_{SSSFF}$}
\end{picture}
\begin{picture}(108,80)(-54,-40)
\SetWidth{0.9}
\DashLine(-45,-22)(0,-22){4}
\DashLine(45,-22)(0,-22){4}
\DashCArc(0,-7)(15,-90,90){4}
\DashCArc(0,-7)(15,90,270){4}
\DashCArc(0,23)(15,-90,270){4}
\Text(0,-32)[]{$\propX_{SSS}$}
\end{picture}
\begin{picture}(108,80)(-54,-40)
\SetWidth{0.9}
\DashLine(-45,-22)(-20,-22){4}
\DashLine(45,-22)(20,-22){4}
\DashLine(-20,-22)(20,-22){4}
\DashLine(-20,-22)(0,8){4}
\DashLine(20,-22)(0,8){4}
\DashCArc(0,23)(15,-90,270){4}
\Text(0,-32)[]{$\propY_{SSSS}$}
\end{picture}
\begin{picture}(108,80)(-54,-40)
\SetWidth{0.9}
\DashLine(-48,0)(-30,0){4}
\DashLine(48,0)(30,0){4}
\DashCArc(-15,0)(15,0,180){4}
\DashCArc(-15,0)(15,180,360){4}
\DashCArc(15,0)(15,0,180){4}
\DashCArc(15,0)(15,180,360){4}
\Text(0,-32)[]{$\propZ_{SSSS}$}
\end{picture}

\vspace{0.395cm}

\begin{picture}(108,80)(-54,-40)
\SetWidth{0.9}
\DashLine(-45,0)(-22,0){4}
\DashLine(45,0)(22,0){4}
\DashCArc(0,0)(22,0,180){4}
\DashCArc(0,0)(22,180,360){4}
\DashLine(0,-22)(0,22){4}
\Text(0,-32)[]{$\propM_{SSSSS}$}
\end{picture}
\begin{picture}(108,80)(-54,-40)
\SetWidth{0.9}
\DashLine(-45,0)(-22,0){4}
\DashLine(45,0)(22,0){4}
\DashCArc(0,0)(22,0,180){4}
\DashCArc(0,0)(22,180,360){4}
\DashLine(-22,0)(22,0){4}
\Text(0,-32)[]{$\propS_{SSS}$}
\end{picture}
\begin{picture}(108,80)(-54,-40)
\SetWidth{0.9}
\DashLine(-45,-14)(-22,-14){4}
\DashLine(45,-14)(22,-14){4}
\DashLine(-22,-14)(0,14){4}
\DashLine(22,-14)(0,14){4}
\DashCArc(11,0)(17.8045,-51.843,128.157){4}
\DashLine(-22,-14)(22,-14){4}
\Text(0,-32)[]{$\propU_{SSSS}$}
\end{picture}
\begin{picture}(108,80)(-54,-40)
\SetWidth{0.9}
\DashLine(-45,-18)(-22,-18){4}
\DashLine(45,-18)(22,-18){4}
\DashLine(-22,-18)(22,-18){4}
\DashCArc(0,18)(22,0,180){4}
\DashLine(-22,18)(22,18){4}
\DashLine(-22,-18)(-22,18){4}
\DashLine(22,-18)(22,18){4}
\Text(0,-32)[]{$\propV_{SSSSS}$}
\end{picture}

\vspace{0.395cm}

\begin{picture}(108,80)(-54,-40)
\SetWidth{0.9}
\DashLine(-45,-16)(0,-16){4}
\DashLine(45,-16)(0,-16){4}
\DashCArc(0,18)(22,0,180){4}
\DashLine(-22,18)(22,18){4}
\DashLine(-22,18)(22,18){4}
\DashLine(0,-16)(-22,18){4}
\DashLine(0,-16)(22,18){4}
\Text(0,-32)[]{$\propW_{SSSS}$}
\end{picture}
\begin{picture}(108,80)(-54,-40)
\SetWidth{0.9}
\DashLine(-45,0)(-22,0){4}
\DashLine(45,0)(22,0){4}
\CArc(0,0)(22,0,180)
\CArc(0,0)(22,180,360)
\DashLine(0,-22)(0,22){4}
\Text(0,-32)[]{$\propM_{FFFFS}$}
\end{picture}

\caption{\label{fig:diagrams} The one-loop and two-loop Feynman diagram
topologies in this paper, by order of first appearance. Dashed lines are
for scalars, solid lines for fermions, wavy lines for electroweak vector
bosons and ghosts, and curly lines for gluons. The label on each diagram 
refers to a
corresponding renormalized integral function, as defined in
ref.~\cite{Martin:2003it}. There are 7 one-loop and 40 two-loop distinct
topologies here, accounting for fermion mass insertions (indicated
below with a bar over the appropriate subscript $F$), but not
counting separately diagrams obtained by exchanging external lines or
reversing all fermion chiralities.} 
\end{figure*}
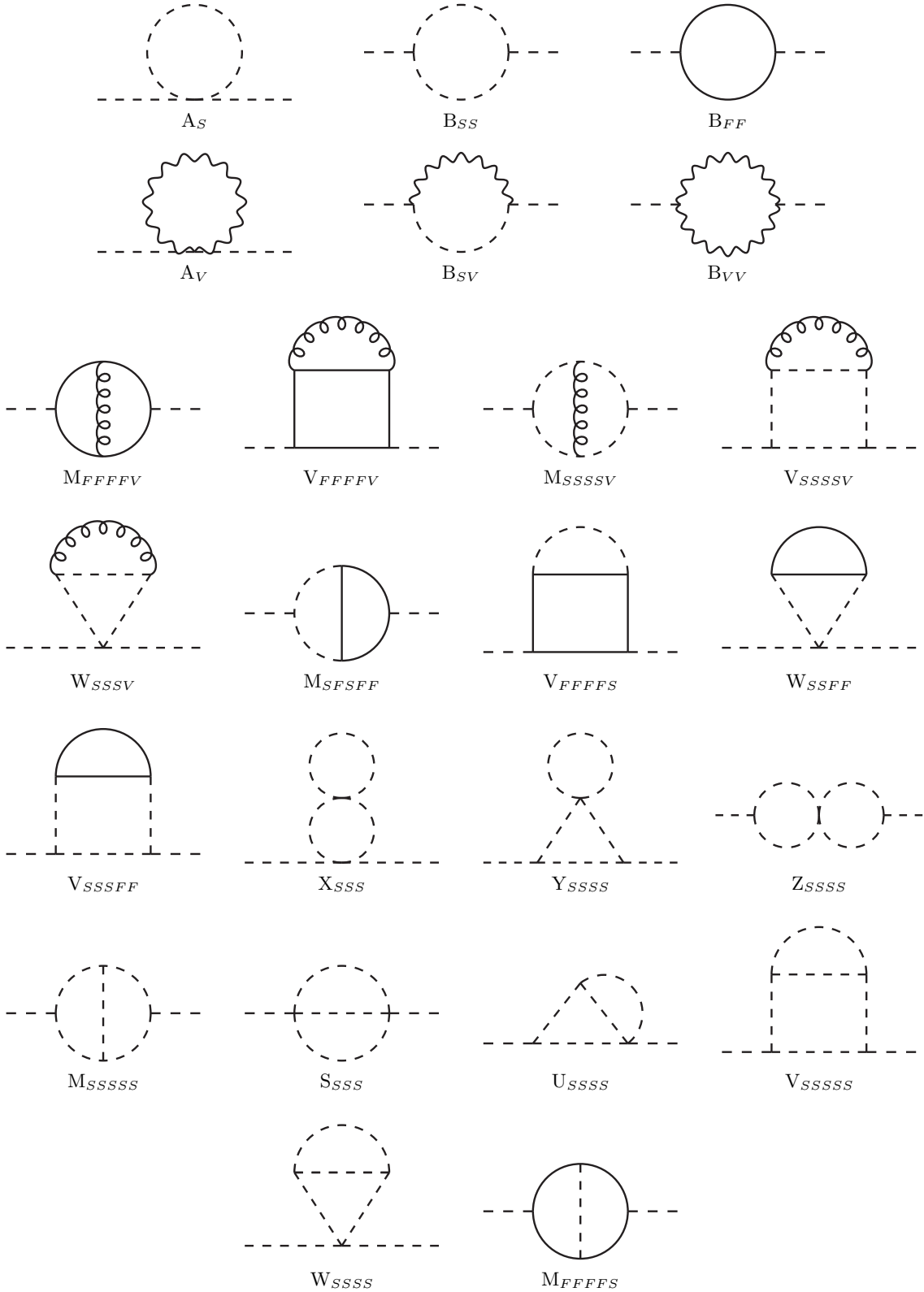

The Feynman diagram topologies that play a role in this paper are shown in
Figure 1. Each diagram shown that involves fermions actually refers to
several distinct ones, with chirality-reversing fermion mass insertions
inserted in all possible ways. For each diagram, there is a corresponding
finite loop integral function, which also includes the \DRbarprime
counterterms for that diagram. The label on each diagram refers to that
function, strictly following the notation and definitions found in
\cite{Martin:2003it}, which lists them in terms of the minimal set of
basis functions. The numerical evaluation of the basis functions is in
turn described in detail in ref.~\cite{evaluation}.

The rest of this paper is organized as follows. Section
\ref{sec:couplings} presents the complete list of three- and
four-particle couplings used in the calculations. The known one-loop
results for the Higgs scalar self-energies are reviewed in section
\ref{sec:oneloop}. The two-loop self-energy contributions described above
are given for the neutral Higgs scalars in \ref{sec:neutraltwoloop}, and
for the charged Higgs scalars in \ref{sec:chargedtwoloop}. Section
\ref{sec:examples} briefly recounts some consistency checks, and studies 
some numerical results for specific model
parameters.

\section{Couplings}\label{sec:couplings}
\setcounter{equation}{0}

In this section, I provide the list of three- and four-particle couplings
needed in the rest of the paper. The conventions and notations for the
MSSM Lagrangian parameters and mixing matrices strictly follow those 
given in section II of \cite{effpotMSSM}, which will not be repeated here
for brevity.

[The signs of some of the couplings listed here do differ from
those listed in section III of \cite{effpotMSSM}, namely equations
(3.6)-(3.9), (3.11)-(3.13), (3.27)-(3.33), (3.35)-(3.38), and
(3.44)-(3.47) of that paper. These sign conventions actually make no
difference at all for ref.~\cite{effpotMSSM}, because three-particle
couplings always appear squared in the two-loop effective potential.
However, the signs are important in the present paper, and have been
chosen to agree consistently with ref.~\cite{Martin:2003it}. To avoid 
confusion, all of
the relevant couplings are listed here.]

The couplings of fermions to the Higgs scalar bosons $\phi_i^0 = 
(h^0, H^0, G^0, A^0)$ and $\phi_i^\pm = (G^\pm, H^\pm)$ are:
\beq
Y_{t\tbar \phi_i^0} &=& y_t k_{u\phi_i^0}/\sqrt{2},\\
Y_{b\bbar \phi_i^0} &=& y_b k_{d\phi_i^0}/\sqrt{2},\\
Y_{\tau\taubar \phi_i^0} &=& y_\tau k_{d\phi_i^0}/\sqrt{2},\\
Y_{\tbar b\phi_i^+} &=& -y_t k_{u\phi_i^+},\\
Y_{\bbar t\phi_i^-} &=& -y_b k_{d\phi_i^+},\\
Y_{\taubar\nu_\tau\phi_i^-} &=& -y_\tau k_{d\phi_i^+}.
\eeq 
The fermion-neutralino-sfermion couplings are:
\beq
Y_{u\tilde N_i \tilde u^*_L}
     &=& (g N_{i2}^* + \gp N_{i1}^*/3)/\sqrt{2},
\\
Y_{\overline u\tilde N_i \tilde u_R}&=& -2 \sqrt{2} \gp N_{i1}^*/3,
\\
Y_{d\tilde N_i \tilde d^*_L}
     &=& (-g N_{i2}^* + \gp N_{i1}^*/3)/\sqrt{2},
\\
Y_{\overline d\tilde N_i \tilde d_R} &=& \sqrt{2} \gp N_{i1}^*/3,
\\
Y_{e\tilde N_i \tilde e^*_L}&=& -(g N_{i2}^* +\gp N_{i1}^*)/\sqrt{2},
\\
Y_{\overline e\tilde N_i \tilde e_R} &=& \sqrt{2} \gp N_{i1}^*,
\\
Y_{\nu\tilde N_i \tilde \nu^*} &=& (g N_{i2}^* -\gp N_{i1}^*)/\sqrt{2},
\\
Y_{t\tilde N_i \tilde t^*_j}
    &=& L_{\tilde t_j}^* Y_{u\tilde N_i \tilde u^*_L} 
     + R_{\tilde t_j}^* N_{i4}^* y_t,
\\
Y_{\overline t\tilde N_i \tilde t_j}
    &=& R_{\tilde t_j} Y_{\overline u\tilde N_i \tilde u_R}
    + L_{\tilde t_j} N_{i4}^* y_t,
\\
Y_{b\tilde N_i \tilde b^*_j}
    &=& L_{\tilde b_j}^* Y_{d\tilde N_i \tilde d^*_L}
    + R_{\tilde b_j}^* N_{i3}^* y_b,
\\
Y_{\overline b\tilde N_i \tilde b_j}
    &=& R_{\tilde b_j} Y_{\overline d\tilde N_i \tilde d_R}
    + L_{\tilde b_j} N_{i3}^* y_b,
\\
Y_{\tau\tilde N_i \tilde \tau^*_j}
    &=& L_{\tilde \tau_j}^* Y_{e\tilde N_i \tilde e^*_L}
    + R_{\tilde \tau_j}^* N_{i3}^* y_\tau,
\\
Y_{\overline \tau\tilde N_i \tilde \tau_j}
    &=& R_{\tilde \tau_j} Y_{\overline e\tilde N_i \tilde e_R}
    + L_{\tilde \tau_j} N_{i3}^* y_\tau.
\eeq
The fermion-chargino-sfermion couplings are:
\beq
Y_{d\tilde C_i \tilde u_L^*} &=& Y_{e\tilde C_i \tilde \nu_e^*}
\,=\, Y_{\tau\tilde C_i \tilde \nu_\tau^*} \,=\, g V_{i1}^* ,
\\
Y_{u\tilde C_i \tilde d_L^*} &=& Y_{\nu_e \tilde C_i \tilde e_L^*}
\,=\, g U_{i1}^* ,
\\
Y_{b\tilde C_i \tilde t_j^*} &=& 
  L^*_{\tilde t_j} g V_{i1}^* - R^*_{\tilde t_j} V_{i2}^* y_t ,
\\
Y_{\overline b\tilde C_i \tilde t_j} &=& -L_{\tilde t_j} U_{i2}^* y_b ,
\\
Y_{t\tilde C_i \tilde b_j^*} &=& L^*_{\tilde b_j} g U_{i1}^*
  -R^*_{\tilde b_j} U_{i2}^* y_b ,
\\
Y_{\overline t\tilde C_i \tilde b_j} &=& -L_{\tilde b_j} V_{i2}^* y_t ,
\\
Y_{\nu_\tau \tilde C_i \tilde \tau_j^*} &=&
  L^*_{\tilde \tau_j} g U_{i1}^* - R^*_{\tilde \tau_j} U_{i2}^* y_\tau ,
\\
Y_{\overline \tau\tilde C_i \tilde \nu_\tau} &=& -y_\tau U_{i2}^* .
\eeq
The neutralino and chargino couplings to Higgs scalar bosons are:
\beq
Y_{\tilde C_i^+ \tilde C_j^- \phiOk} &=& g (k_{d\phiOk}^* V_{i1}^* U_{j2}^*
       + k_{u\phiOk}^* V_{i2}^* U_{j1}^*)/\sqrt{2},\phantom{xxxx}
\\
Y_{\tilde N_i \tilde N_j \phiOk} &=& (g N_{i2}^* - g' N_{i1}^*)
       (k^*_{d\phiOk} N_{j3}^* 
\nonumber \\ &&
- k^*_{u\phiOk} N_{j4}^* )/2
        + (i \leftrightarrow j),
\\
Y_{\tilde C_i^+ \tilde N_j \phimk} &=&
       k_{u\phipk} [ g V_{i1}^* N_{j4}^*
\nonumber \\ &&
       + V_{i2}^* (g N_{j2}^* + g' N_{j1}^*)/\sqrt{2} ] ,
\\
Y_{\tilde C_i^- \tilde N_j \phipk} &=&
       k_{d\phipk} [g U_{i1}^* N_{j3}^*
\nonumber \\ &&
       -U_{i2}^* (g N_{j2}^* + g' N_{j1}^*)/\sqrt{2} ] .
\eeq
The couplings of electroweak gauge bosons to each other and to the
Higgs scalar bosons are:
\beq
&&\!\!\!\!\!\!\!\!\!\!\!\!\!\!\! g_{W\phiOj \phipk} = 
   i g (k_{d\phiOj} k_{d\phipk}- k_{u\phiOj}^* k_{u\phipk})/2,
 \\
&&\!\!\!\!\!\!\!\!\!\!\!\!\!\!\! g_{Z \phiOj \phiOk} = \sqrt{g^2 + 
g^{\prime 2}} 
{\rm Im}[
   k_{u\phiOj} k_{u\phiOk}^* - k_{d\phiOj} k_{d\phiOk}^*]/2,
 \\
&&\!\!\!\!\!\!\!\!\!\!\!\!\!\!\! g_{Z \phipj \phimk} =  
   i \delta_{jk} (g^2 - g^{\prime 2})/2 \sqrt{g^2 + g^{\prime 2}},
\phantom{\biggl [ \biggr ]}
\\
&&\!\!\!\!\!\!\!g_{WW\phiOi\phiOj} = \delta_{ij} g^2/2,
\\
&&\!\!\!\!\!\!\!g_{WW\phipi\phimj} = \delta_{ij} g^2/2,
 \\
&&\!\!\!\!\!\!\!g_{ZZ\phiOi\phiOj} = \delta_{ij} (g^2 + g^{\prime 2})/2,
 \\
&&\!\!\!\!\!\!\!g_{ZZ\phipi\phimj} = \delta_{ij}
  (g^2 - g^{\prime 2})^2/2(g^2 + g^{\prime 2}),
\\
&&\!\!\!\!\!\!\!\!\!\!g_{ZZ\phiOi} = (g^2 + g^{\prime 2}) {\rm Re} 
                 [v_u k_{u\phiOi} + v_d k_{d\phiOi}]/\sqrt{2},
\\
&&\!\!\!\!\!\!\!\!\!\!g_{WW\phiOi} = g^2  {\rm Re}
                 [v_u k_{u\phiOi} + v_d k_{d\phiOi}]/\sqrt{2} ,
\\
&&\!\!\!\!\!\!\!\!\!\!g_{WA\phipi} = e g (v_u k_{u\phipi} - v_d 
k_{d\phipi})/\sqrt{2} ,
\\
&&\!\!\!\!\!\!\!\!\!\!g_{WZ\phipi} = -e g' (v_u k_{u\phipi} - v_d 
k_{d\phipi})/\sqrt{2},
\eeq
where $e = \gp g/\sqrt{g^2+g^{\prime 2}}$ is the QED coupling.
\begin{widetext}
The couplings of four Higgs scalar bosons are given by:
\beq
\lambda_{\phiOi \phiOj \phiOk \phiOm} &=& 
   (g^2 + g^{\prime 2})
{\rm Re}[k_{u\phiOi} k_{u\phiOj}^* - k_{d\phiOi} k_{d\phiOj}^*]
{\rm Re}[k_{u\phiOk} k_{u\phiOm}^* - k_{d\phiOk} k_{d\phiOm}^*]/4
+ (i \leftrightarrow k)
+ (i \leftrightarrow m),
\\
\lambda_{\phiOi \phiOj \phipk \phimm} &=& 
\Bigl [
  g^2 (\delta_{ij}\delta_{km}
   + 2 k_{u\phiOi} k_{d\phiOj} k_{d\phipk} k_{u\phipm}
   + 2 k_{u\phiOi}^* k_{d\phiOj}^* k_{u\phipk} k_{d\phipm})
\nonumber \\ &&
  + g^{\prime 2}
    (k_{u\phiOi} k_{u\phiOj}^* -k_{d\phiOi} k_{d\phiOj}^*)
    (k_{u\phipk} k_{u\phipm} - k_{d\phipk} k_{d\phipm})
\Bigr ]/8+ (i \leftrightarrow j),
\\
\lambda_{\phipi \phipj \phimk \phimm} &=&
(g^2 + g^{\prime 2})
(2 k_{u\phipi} k_{u\phipj} k_{u\phipk} k_{u\phipm}
  -k_{u\phipi} k_{d\phipj} k_{u\phipk} k_{d\phipm}
  -k_{d\phipi} k_{u\phipj} k_{u\phipk} k_{d\phipm})/4
+ (u \leftrightarrow d). \phantom{xx}
\eeq
and the couplings of three Higgs scalars are:
\begin{eqnarray}
\lambda_{\phiOi\phiOj\phiOk} &=& (g^2 + g^{\prime 2})
{\rm Re}[k_{u\phiOi} k_{u\phiOj}^* - k_{d\phiOi} k_{d\phiOj}^*]
{\rm Re}[k_{u\phiOk} v_u - k_{d\phiOk} v_d]/2\sqrt{2}
+ (k\leftrightarrow i) + (k\leftrightarrow j)
\nonumber \\
\lambda_{\phiOi\phi^+_j\phi^-_k} &=&
\bigl \lbrace
g^2 \bigl (
[v_d k_{u\phiOi} + v_u k_{d\phiOi}]
k_{d\phi^+_j} k_{u\phi^+_k}
+
[v_d k_{u\phiOi}^* + v_u k_{d\phiOi}^*]
k_{u\phi^+_j} k_{d\phi^+_k}
+ \delta_{jk}{\rm Re}[v_d k_{d\phiOi} + v_u k_{u\phiOi}]
\bigr )
\nonumber \\ &&
+ g^{\prime 2} [k_{d\phi^+_j} k_{d\phi^+_k} -
k_{u\phi^+_j} k_{u\phi^+_k}]
{\rm Re}[v_d k_{d\phiOi} - v_u k_{u\phiOi}]
\bigr \rbrace/2\sqrt{2} .
\end{eqnarray}

The couplings involving sfermions are conveniently written using the
quantities
$I_{\tilde f}$ and $Y_{\tilde f}$, defined to be the
third component of weak isospin and the weak hypercharge of the
left-handed chiral superfield containing the squark or slepton 
$\sfermion$:
\renewcommand{\arraystretch}{1.3}
\begin{center}
\begin{tabular}{l|ccccccc}
\phantom{xxx}
& $\suL$ & $\sdL$ & $\snu_e$ & $\seL$ & $\suR$ & $\sdR$ & $\seR$
\\
\hline
$I_{\tilde f}$ &
\phantom{X}$1/2$\phantom{X}& 
\phantom{X}$-1/2$\phantom{X}& 
\phantom{X}$1/2$ \phantom{X}& 
\phantom{X}$-1/2$\phantom{X}& 
\phantom{X}$0$\phantom{X}&
\phantom{X}$0$\phantom{X}& 
\phantom{X}$0$\phantom{X}
\\
$Y_{\tilde f}$ & 
\phantom{X}$1/6$\phantom{X}& 
\phantom{X}$1/6$\phantom{X}& 
\phantom{X}$-1/2$\phantom{X}& 
\phantom{X}$-1/2$\phantom{X}& 
\phantom{X}$-2/3$\phantom{X}& 
\phantom{X}$1/3$ \phantom{X}& 
\phantom{X}$1$\phantom{X}
\end{tabular}
\renewcommand{\arraystretch}{1.0}
\end{center}
Then we have for the couplings of two neutral Higgs scalars to 
sfermion-antisfermion pairs
\beq
\lambda_{\phiOi \phiOj \tilde f \tilde f^*} &=& 
(I_{\tilde f} g^2 -Y_{\tilde f} g^{\prime 2})
{\rm Re}[k_{d\phiOi}k_{d\phiOj}^* -k_{u\phiOi}k_{u\phiOj}^*]/2
\eeq
for the sfermions $\tilde f$ of the first and second families
and $\tilde \nu_\tau$, 
and
\beq
\lambda_{\phiOi \phiOj \stopk \stopm^*} &=&
{\rm Re}[k_{u \phiOi} k_{u \phiOj}^*] y_t^2 \delta_{km}
+ L_{\stopk} L^*_{\stopm} \lambda_{\phiOi \phiOj \suL \suL^*}
+ R_{\stopk} R^*_{\stopm} \lambda_{\phiOi \phiOj \suR \suR^*} ,
\\
\lambda_{\phiOi \phiOj \sbotk \sbotm^*} &=&
{\rm Re}[k_{d \phiOi} k_{d \phiOj}^*] y_b^2 \delta_{km}
+ L_{\sbotk} L^*_{\sbotm} \lambda_{\phiOi \phiOj \sdL \sdL^*}
+ R_{\sbotk} R^*_{\sbotm} \lambda_{\phiOi \phiOj \sdR \sdR^*} ,
\\
\lambda_{\phiOi \phiOj \stauk \staum^*} &=&
{\rm Re}[k_{d \phiOi} k_{d \phiOj}^*] y_\tau^2 \delta_{km}
+ L_{\stauk} L^*_{\staum} \lambda_{\phiOi \phiOj \seL \seL^*}
+ R_{\stauk} R^*_{\staum} \lambda_{\phiOi \phiOj \seR \seR^*}
\eeq
for the other sfermions of the third family. The 
neutral Higgs-sfermion-antisfermion couplings are similarly given by
\beq
\lambda_{\phiOi \tilde f \tilde f^*} &=& 
   (I_{\tilde f} g^2 -Y_{\tilde f} g^{\prime 2}) 
   {\rm Re}[k_{d \phiOi} v_d - k_{u \phiOi} v_u]/\sqrt{2}
\eeq
for the first two families and $\tilde \nu_\tau$, and by
\beq
\lambda_{\phiOi \stopk \stopm^*} &=&
  \sqrt{2} v_u y_t^2 {\rm Re}[k_{u \phiOi}] \delta_{km} 
  +L_{\stopk} R^*_{\stopm} (k_{u \phiOi} a_t 
                        - k_{d \phiOi}^* \mu^* y_t)/\sqrt{2} 
  +R_{\stopk} L^*_{\stopm}(k_{u \phiOi}^* a_t^* 
                        - k_{d \phiOi} \mu y_t)/\sqrt{2} 
  \nonumber \\ &&
  + L_{\stopk} L^*_{\stopm} \lambda_{\phiOi \suL \suL^*} 
  + R_{\stopk} R^*_{\stopm} \lambda_{\phiOi \suR \suR^*} ,
\\
\lambda_{\phiOi \sbotk \sbotm^*} &=&
  \sqrt{2} v_d y_b^2 {\rm Re}[k_{d \phiOi}] \delta_{km} 
  +L_{\sbotk} R^*_{\sbotm} (k_{d \phiOi} a_b 
                        -k_{u \phiOi}^* \mu^* y_b)/\sqrt{2} 
  +R_{\sbotk} L^*_{\sbotm}(k_{d \phiOi}^* a_b^* 
                        -k_{u \phiOi} \mu y_b)/\sqrt{2} 
  \nonumber \\ &&
  + L_{\sbotk} L^*_{\sbotm} \lambda_{\phiOi \sdL \sdL^*} 
  + R_{\sbotk} R^*_{\sbotm} \lambda_{\phiOi \sdR \sdR^*} ,
\\
\lambda_{\phiOi \stauk \staum^*} &=&
  \sqrt{2} v_d y_\tau^2 {\rm Re}[k_{d \phiOi}] \delta_{km} 
  +L_{\stauk} R^*_{\staum} (k_{d \phiOi} a_\tau 
                        -k_{u \phiOi}^* \mu^* y_\tau)/\sqrt{2} 
  +R_{\stauk} L^*_{\staum}(k_{d \phiOi}^* a_\tau^* 
                        -k_{u \phiOi} \mu y_\tau)/\sqrt{2} 
  \nonumber \\ &&
  + L_{\stauk} L^*_{\staum} \lambda_{\phiOi \seL \seL^*} 
  + R_{\stauk} R^*_{\staum} \lambda_{\phiOi \seR \seR^*} 
\eeq
for the other third family sfermions.
The couplings of pairs of charged Higgs scalars 
to sfermions of the first two families are
\beq
\lambda_{\phipi \phimj \tilde f \tilde f^*} = 
   (I_{\tilde f} g^2 + Y_{\tilde f} g^{\prime 2}) 
   (k_{u\phipi} k_{u\phipj} - k_{d\phipi} k_{d\phipj})/2 .
\eeq
For the sfermions of the third family,
\beq 
\lambda_{\phipi \phimj \stopk \stopm^*} &=&
  R_{\stopk} R_{\stopm}^* (y_t^2 k_{u\phipi} k_{u\phipj} 
  +\lambda_{\phipi \phimj \suR \suR^*})
  + L_{\stopk} L_{\stopm}^* (y_b^2 k_{d\phipi} k_{d\phipj} 
  +\lambda_{\phipi \phimj \suL \suL^*}) ,
\\ 
\lambda_{\phipi \phimj \sbotk \sbotm^*} &=&
  L_{\sbotk} L_{\sbotm}^* (y_t^2 k_{u\phipi} k_{u\phipj} 
  +\lambda_{\phipi \phimj \sdL \sdL^*})
  + R_{\sbotk} R_{\sbotm}^* (y_b^2 k_{d\phipi} k_{d\phipj} 
  +\lambda_{\phipi \phimj \sdR \sdR^*}) ,
\\ 
\lambda_{\phipi \phimj \tilde \nu_\tau \tilde \nu_\tau^*} &=&
  (y_\tau^2 k_{d\phipi} k_{d\phipj} 
  +\lambda_{\phipi \phimj \tilde \nu \tilde \nu^*}) ,
\\ 
\lambda_{\phipi \phimj \stauk \staum^*} &=&
  L_{\stauk} L_{\staum}^* \lambda_{\phipi \phimj \seL \seL^*}
  + R_{\stauk} R_{\staum}^* (y_\tau^2 k_{d\phipi} k_{d\phipj} 
  +\lambda_{\phipi \phimj \seR \seR^*}) .
\eeq
The charged Higgs-sfermion-antisfermion couplings are
\beq
\lambda_{\phipi \sdL \suL^*} = \lambda_{\phipi \seL \tilde \nu_e^*} 
      &=& g^2 (k_{u\phipi} v_u + k_{d\phipi} v_d)/2 
\eeq
for the first two families, and
\beq
\lambda_{\phipi \sbotk \stopm^*} &=& L_{\sbotk} L^*_{\stopm} (
    \lambda_{\phipi \sdL \suL^*}
   -y_t^2 v_u k_{u\phipi} -y_b^2 v_d k_{d\phipi} )
   -R_{\sbotk} R^*_{\stopm} y_t y_b (k_{d\phipi} v_u + k_{u \phipi}v_d )
\nonumber \\
&&
   -L_{\sbotk} R^*_{\stopm} (k_{u\phipi} a_t + k_{d\phipi} \mu^* y_t)
   -R_{\sbotk} L^*_{\stopm} (k_{d\phipi} a_b^* + k_{u\phipi} \mu y_b),
\\
\lambda_{\phipi \stauk \tilde \nu_\tau^*} &=&
   L_{\stauk} (\lambda_{\phipi \seL \tilde \nu_e^*}
   -y_\tau^2 v_d k_{d\phipi} )
   -R_{\stauk} (k_{d\phipi} a_\tau^* +k_{u\phipi} \mu y_\tau) .
\eeq
for the third family.
The charged Higgs-neutral Higgs-sfermion-antisfermion couplings are
\beq
\lambda_{\phiOi\phipj \sdL \suL^*} =
\lambda_{\phiOi\phipj \seL \snu^*} =
   g^2 (k_{u\phiOi}^* k_{u\phipj} + k_{d\phiOi} k_{d\phipj})/2\sqrt{2}
\eeq
for the first two families, and
\beq
\lambda_{\phiOi\phipj \sbotk \stopm^*} &=&
   L_{\sbotk} L_{\stopm}^* (\lambda_{\phiOi\phipj \sdL \suL^*}
                        -[y_t^2 k_{u\phiOi}^* k_{u\phipj}
                         +y_b^2 k_{d\phiOi} k_{d\phipj}]/\sqrt{2}) 
\nonumber \\ &&
-R_{\sbotk} R_{\stopm}^* y_t y_b (k_{u \phiOi} k_{d\phipj} +
                                  k_{d \phiOi}^* k_{u\phipj})/\sqrt{2} ,
\\
\lambda_{\phiOi\phipj \stauk \snutau^*} &=&
     L_{\stauk} (\lambda_{\phiOi\phipj \seL \snu^*}
     -y_\tau^2 k_{d\phiOi} k_{d\phipj}/\sqrt{2})
\eeq
for the third family.

The sfermion-antisfermion-sfermion-antisfermion couplings in the 
Lagrangian are written as 
\beq
-{\cal L} = \frac{1}{2} 
\lambda_{\sfermioni\sfermionj^*\sfermionk\sfermionm^*}
(\sfermioni\sfermionj^*)(\sfermionk\sfermionm^*) ,
\eeq
where each combination in parentheses forms a color singlet.
Then
\beq
\lambda_{\sfermioni\sfermionj^*\sfermionk\sfermionm^*}
= 
\xx_{\sfermioni\sfermionj^*\sfermionk\sfermionm^*}
+ g^2 \sum_{n=1}^3 
  x^{(n)}_{\sfermioni\sfermionj^*} x^{(n)}_{\sfermionk\sfermionm^*}
+ g^{\prime 2} x'_{\sfermioni\sfermionj^*} 
x'_{\sfermionk\sfermionm^*}
,
\label{eq:sfsfcsfsfc}
\eeq
where the non-zero Yukawa $F$-term contributions are:
\beq
&&\xx_{\stopi\stopj^*\stopk\stopm^*} =
  y_t^2 (L_{\stopi} R_{\stopj}^* R_{\stopk} L_{\stopm}^*
  +R_{\stopi} L_{\stopj}^* L_{\stopk} R_{\stopm}^*) ,
\\
&&\xx_{\sboti\sbotj^*\sbotk\sbotm^*} =
  y_b^2 (L_{\sboti} R_{\sbotj}^* R_{\sbotk} L_{\sbotm}^*
  +R_{\sboti} L_{\sbotj}^* L_{\sbotk} R_{\sbotm}^*) ,
\\
&&\xx_{\staui\stauj^*\stauk\staum^*} =
  y_\tau^2 (L_{\staui} R_{\stauj}^* R_{\stauk} L_{\staum}^*
  +R_{\staui} L_{\stauj}^* L_{\stauk} R_{\staum}^*) ,
\\
&&\xx_{\stopi\sbotj^*\sbotk\stopm^*} =
  \xx_{\sbotk\stopm^*\stopi\sbotj^*} =
  y_t^2 R_{\stopi} L_{\sbotj}^* L_{\sbotk} R_{\stopm}^*
  + y_b^2 L_{\stopi} R_{\sbotj}^* R_{\sbotk} L_{\stopm}^* ,
\\
&&\xx_{\snutau\stauj^*\stauk\snutau^*} =
  \xx_{\stauk\snutau^*\snutau\stauj^*} =
  y_\tau^2 R_{\stauj}^* R_{\stauk} ,
\\
&&\xx_{\sboti\sbotj^*\stauk\staum^*} = \xx_{\stauk\staum^*\sboti\sbotj^*} 
= 
    y_b y_\tau (L_{\sboti} R_{\sbotj}^* R_{\stauk} L_{\staum}^*
    +R_{\sboti} L_{\sbotj}^* L_{\stauk} R_{\staum}^*) ,
\\
&&\xx_{\stopi\sbotj^*\stauk\snutau^*} =
\xx_{\stauk\snutau^*\stopi\sbotj^*} =
(\xx_{\sbotj\stopi^*\snutau\stauk^*})^* =
(\xx_{\snutau\stauk^*\sbotj\stopi^*})^* =
    y_b y_\tau L_{\stopi} R_{\sbotj}^* R_{\stauk} .
\eeq
\end{widetext}
The electroweak $U(1)_Y$ $D$-term contributions to 
eq.~(\ref{eq:sfsfcsfsfc}) are:
\beq
x'_{\sfermion\sfermion^*} &=& Y_{\sfermion} 
\eeq
for the sfermions of the first two families, and
\beq
x'_{\stopj\stopk^*} &=& 
L_{\stopj} L_{\stopk}^*/6 -2 R_{\stopj} R_{\stopk}^*/3 ,
\\ 
x'_{\sbotj\sbotk^*} &=& 
L_{\sbotj} L_{\sbotk}^*/6 + R_{\sbotj} R_{\sbotk}^*/3 ,
\\ 
x'_{\stauj\stauk^*} &=& 
-L_{\stauj} L_{\stauk}^*/2 + R_{\stauj} R_{\stauk}^* ,
\eeq
for the third family sfermions. The $SU(2)_L$ $D$-term
contributions to eq.~(\ref{eq:sfsfcsfsfc}) are 
\beq
x^{(1)}_{\suL\sdL^*} = 
x^{(1)}_{\sdL\suL^*} = 
x^{(1)}_{\snu_e\seL^*} = 
x^{(1)}_{\seL\snu_e^*} &=& 1/2 ,\phantom{xx}
\\
x^{(2)}_{\suL\sdL^*} = 
-x^{(2)}_{\sdL\suL^*} = 
x^{(2)}_{\snu_e\seL^*} = 
-x^{(2)}_{\seL\snu_e^*} &=& i/2 ,
\\
x^{(3)}_{\suL\suL^*} = 
-x^{(3)}_{\sdL\sdL^*} = 
x^{(3)}_{\snu_e\snu_e^*} = 
-x^{(3)}_{\seL\seL^*} &=& 1/2
\eeq
for the first two family sfermions, and
\beq
x^{(1)}_{\stopj\sbotk^*}  = (x^{(1)}_{\sbotk\stopj^*})^* 
   &=& L_{\stopj} L_{\sbotk}^*/2 ,
\\
x^{(1)}_{\snutau\stauj^*}  = (x^{(1)}_{\stauj\snutau^*})^* 
   &=& L_{\stauj}^*/2 ,
\\
x^{(2)}_{\stopj\sbotk^*}  = (x^{(2)}_{\sbotk\stopj^*})^* 
   &=& i L_{\stopj} L_{\sbotk}^*/2 ,
\\
x^{(2)}_{\snutau\stauj^*}  = (x^{(2)}_{\stauj\snutau^*})^* 
   &=& i L_{\stauj}^*/2 ,
\\
x^{(3)}_{\stopj\stopk^*} &=& L_{\stopj} L_{\stopk}^*/2 ,
\\
x^{(3)}_{\snutau\snutau^*} &=& 1/2  ,
\\
x^{(3)}_{\sbotj\sbotk^*} &=& -L_{\sbotj} L_{\sbotk}^*/2 , 
\\
x^{(3)}_{\stauj\stauk^*} &=& -L_{\stauj} L_{\stauk}^*/2 .
\eeq
for the third family sfermions.
The $SU(3)_c$ $D$-term contributions to 
squark-antisquark-squark-antisquark couplings
are not included above, and will be kept track of separately in the
following.

I conclude this section with a few other important conventions to be 
observed throughout this paper. 
The symbol $\tilde f$ refers to a generic sfermion mass eigenstate. 
The symbol $\squark$ refers only to
the 8 first and second family squarks, 
$(\tilde u_L, \tilde d_L, \tilde u_R, \tilde d_R,
\tilde c_L, \tilde s_L, \tilde c_R, \tilde s_R)$, which are always
assumed to be mass eigenstates. Indices $i,j$ are used for the
external Higgs scalars.
Indices $k,m,n,p,\ldots$ for virtual particles are always
implicitly summed over all possible values, namely $1,2,3,4$ for
neutral Higgs scalars and neutralinos, or $1,2$ for charged Higgs 
scalars, charginos, top squarks, bottom squarks, and tau sleptons,
or over the 21 distinct sfermion mass eigenstates $\tilde f_k$.
The symbol $n_f$ or $n_{\tilde f}$ refers to the 
number of colors, and is always equal to 3 or 
1 in the obvious way. The name of a particle is always used in
place of its renormalized, tree-level squared mass when appearing as the 
argument of a loop function, so for example
$\propM_{SFS\Fbar F}(\stopk, t , \stopm, t, \gluino )$
means
$\propM_{SFS\Fbar F}(m^2_{\stopk}, m^2_t , m^2_{\stopm}, m^2_t, 
m^2_{\gluino} )$.
Each of the integral functions also has an implicit dependence on $s$ and $Q$,
as in ref.~\cite{Martin:2003it}. All of the couplings and masses appearing
below are tree-level running \DRbarprime~ parameters in the MSSM with no
particles decoupled.

\begin{widetext}
\section{One-loop contributions to Higgs scalar boson 
self-energies}\label{sec:oneloop}
\setcounter{equation}{0}

In this section, I review the known results for the one-loop self-energies 
of the Higgs scalar bosons. The Feynman gauge versions of these formulas
can be found in ref.~\cite{Pierce:1996zz}, but here we use the Landau
gauge results in order to agree with the two-loop calculation of the
effective potential. 

For the neutral Higgs scalar bosons $\phi_i^0 = (h^0,H^0,G^0,A^0)$,
\beq
\Pi^{(1)}_{\phiOi\phiOj} &=& 
\frac{1}{2} \lambda_{\phiOi \phiOj \phiOk \phiOk} \propA_S(\phiOk)
+ \lambda_{\phiOi \phiOj \phipk \phimk} \propA_S(\phipk)
+ \sum_{\tilde f} n_{\tilde f} 
  \lambda_{\phiOi \phiOj \tilde f \tilde f^*} \propA_S(\tilde f)
+ \lambda_{\phiOi \phipk \phimm} \lambda_{\phiOj \phipm \phimk}
  \propB_{SS}(\phipk,\phipm)
\nonumber \\ &&
+ \frac{1}{2} \lambda_{\phiOi \phiOk \phiOm} \lambda_{\phiOj \phiOk 
\phiOm}
  \propB_{SS}(\phiOk,\phiOm)
+ \sum_{\tilde f,\tilde f'} n_{\tilde f} 
  \lambda_{\phiOi \tilde f \tilde f^{\prime *}}
  \lambda_{\phiOj \tilde f' \tilde f^*} 
  \propB_{SS}(\tilde f, \tilde f')
\nonumber \\ && 
+ 2{\rm Re}[ Y_{\tilde C_k^+ \tilde C_m^- \phiOi} 
             Y^*_{\tilde C_k^+ \tilde C_m^- \phiOj}] 
             \propB_{FF}(\tilde C_k, \tilde C_m) 
+ 2{\rm Re}[ Y_{\tilde C_k^+ \tilde C_m^- \phiOi} 
             Y_{\tilde C_m^+ \tilde C_k^- \phiOj}] 
             m_{\tilde C_k}m_{\tilde C_m} 
             \propB_{\Fbar\Fbar}(\tilde C_k, \tilde C_m) 
\nonumber \\ && 
+ {\rm Re}[ Y_{\tilde N_k \tilde N_m \phiOi} 
            Y^*_{\tilde N_k \tilde N_m \phiOj}] 
             \propB_{FF}(\tilde N_k, \tilde N_m) 
+ {\rm Re}[ Y_{\tilde N_k \tilde N_m \phiOi} 
            Y_{\tilde N_k \tilde N_m \phiOj}] 
       m_{\tilde N_k} m_{\tilde N_m} 
       \propB_{\Fbar\Fbar}(\tilde N_k, \tilde N_m) 
\nonumber \\ && 
+ 6 {\rm Re}[Y_{t\tbar \phiOi} Y_{t\tbar \phiOj}^*] \propB_{FF}(t,t)
+ 6 {\rm Re}[Y_{t\tbar \phiOi} Y_{t\tbar \phiOj}] m_t^2 
            \propB_{\Fbar\Fbar}(t,t)
+ 6 {\rm Re}[Y_{b\bbar \phiOi} Y_{b\bbar \phiOj}^*] \propB_{FF}(b,b)
\nonumber \\ &&
+ 6 {\rm Re}[Y_{b\bbar \phiOi} Y_{b\bbar \phiOj}] m_b^2 
            \propB_{\Fbar\Fbar}(b,b)
+ 2 {\rm Re}[Y_{\tau\taubar \phiOi} Y_{\tau\taubar \phiOj}^*] 
             \propB_{\Fbar\Fbar}(\tau,\tau)
+ 2 {\rm Re}[Y_{\tau\taubar \phiOi} Y_{\tau\taubar \phiOj}] 
             m_\tau^2 \propB_{FF}(\tau,\tau)
\nonumber \\ &&
+ \frac{1}{2} g_{ZZ\phiOi\phiOj} \propA_V(Z) 
+g_{WW\phiOi\phiOj} \propA_V(W) 
+ 2 {\rm Re}[g_{W\phiOi \phipk} g_{W\phiOj \phipk}^*] 
             \propB_{SV} (\phipk, W)
\nonumber \\ && 
+g_{Z\phiOi\phiOk} g_{Z \phiOj \phiOk} \propB_{SV} (\phiOk, Z)
+ \frac{1}{2} g_{ZZ\phiOi} g_{ZZ\phiOj} \propB_{VV}(Z,Z)
+ g_{WW\phiOi} g_{WW\phiOj} \propB_{VV}(W,W) .
\eeq
In general, this is a $4\times 4$ matrix. It has the form of two
block-diagonal $2\times 2$ matrices in the special case of no CP
violation. The couplings here are as defined in section 
\ref{sec:couplings}. 
The renormalized and finite loop-integral functions $\propA_S(x)$,
$\propA_{SS}(x,y)$, $\propB_{SS}(x,y)$, $\propB_{FF}(x,y)$, 
$\propB_{\Fbar\Fbar}(x,y)$, $\propA_V (x,y)$, $\propB_{SV}(x,y)$,
and $\propB_{VV}(x,y)$ are explicitly functions of the tree-level squared 
masses of the virtual particles in the loops, and they are all 
also implicitly functions of $s$. They can be found in section
III of ref.~\cite{Martin:2003it}.

For the charged Higgs scalar bosons $\phi_i^\pm = (G^\pm, H^\pm)$, the 
result is a $2\times 2$ matrix:
\beq
\Pi^{(1)}_{\phipi\phimj} &=& 
\frac{1}{2} \lambda_{\phiOk \phiOk \phipi \phimj} \propA_S(\phiOk)
+ \lambda_{\phipi \phipk \phimj \phimk} \propA_S(\phipk)
+ \sum_{\tilde f} n_{\tilde f} 
  \lambda_{\phipi \phimj \tilde f \tilde f^*} \propA_S(\tilde f)
\nonumber \\ &&
+ \lambda_{\phiOm \phipi \phimk} \lambda_{\phiOm \phipk \phimj}
  \propB_{SS}(\phipk,\phiOm)
+ \sum_{\tilde f,\tilde f'} n_{\tilde f}
  \lambda_{\phipi \tilde f \tilde f^{\prime *}}
  \lambda^*_{\phipj \tilde f \tilde f^{\prime *}} 
  \propB_{SS}(\tilde f, \tilde f')
\nonumber \\ &&
+ ( Y_{\tilde C_k^- \tilde N_m \phipi} Y^*_{\tilde C_k^- \tilde N_m \phipj} 
+ Y_{\tilde C_k^+ \tilde N_m\phimi}^* Y_{\tilde C_k^+ \tilde N_m \phimj} 
) \propB_{FF}(\tilde C_k, \tilde N_m) 
\nonumber \\ && 
+ ( Y_{\tilde C_k^- \tilde N_m \phipi} Y_{\tilde C_k^+ \tilde N_m \phimj} 
+ Y_{\tilde C_k^+ \tilde N_m \phimi}^* Y_{\tilde C_k^- \tilde N_m \phipj}^* 
) m_{\tilde C_k} m_{\tilde N_m} \propB_{\Fbar\Fbar}(\tilde C_k, \tilde N_m)
\nonumber \\ && 
+ 3 (Y_{\tbar b\phipi} Y_{\tbar b\phipj} +
     Y_{\bbar t\phimi} Y_{\bbar t\phimj}) \propB_{FF}(t,b)
+ 3 (Y_{\tbar b\phipi} Y_{\bbar t\phimj} +
     Y_{\bbar t\phimi} Y_{\tbar b\phipj}) m_b m_t 
     \propB_{\Fbar\Fbar}(t,b)
\nonumber \\ &&
+ Y_{\taubar\nu_\tau \phimi} Y_{\taubar\nu_\tau \phimj} 
     \propB_{FF}(0,\tau) 
+ \frac{1}{2} g_{ZZ\phipi\phimj} \propA_V(Z) 
+g_{WW\phipi\phimj} \propA_V(W) 
\nonumber \\ && 
+ e^2 \delta_{ij} \propB_{SV}(\phipi,0) 
+g_{W\phiOk \phipi} g_{W\phiOk \phipj}^* \propB_{SV} (\phiOk, W)
-g_{Z\phipi\phimk} g_{Z \phipk \phimj} \propB_{SV} (\phipk, Z)
\nonumber \\ && 
+ g_{WA\phipi} g_{WA\phipj} \propB_{VV}(0,W)
+ g_{WZ\phipi} g_{WZ\phipj} \propB_{VV}(Z,W) .
\eeq

\section{Two-loop contributions to neutral Higgs scalar boson 
self-energies}\label{sec:neutraltwoloop}
\setcounter{equation}{0}

In this section, I present analytical formulas for the contributions to 
the two-loop self-energies of the neutral Higgs scalars. These are 
labeled in the form $\Pi^{(2),N}_{\phiOi\phiOj}$, where $N$ is 
used to distinguish the various contributions and will be equal to the
equation number.

\subsection{Strong contributions}\label{subsec:neutralstrong}

The contributions to the neutral Higgs scalar boson self-energy 
matrix involving the gluon are:
\begin{eqnarray}
\Pi^{(2),1}_{\phiOi\phiOj}
& = & 
4 g_3^2 \Bigl \lbrace \Bigl (
2 {\rm Re}[Y_{t\tbar \phiOi} Y_{t\tbar \phiOj}^*] \propgaugeFF(t,t)
+2 {\rm Re}[Y_{t\tbar \phiOi} Y_{t\tbar \phiOj}] m_t^2 
\propgaugeFbarFbar(t,t)
\nonumber \\ &&
+ \lambda_{\phiOi \stopk \stopm^*} \lambda_{\phiOj \stopm \stopk^*}
\propgaugeSS (\stopk,\stopm)
+  \lambda_{\phiOi \phiOj \stopk \stopk^*}
\propW_{SSSV}(\stopk, \stopk , \stopk , 0)
\Bigr )
+ (t \rightarrow b) 
\nonumber \\ &&
+ \sum_{\squark}
\lambda_{\phiOi \squark\squark^*} \lambda_{\phiOj \squark\squark^*}
\propgaugeSS (\squark,\squark)
+ \sum_{\squark}
\lambda_{\phiOi \phiOj \squark \squark^*}
\propW_{SSSV}(\squark,\squark,\squark , 0)
\Bigr \rbrace .
\label{eq:phiOgluon}
\end{eqnarray}
The functions 
$\propgaugeFF (x,y)$,
$\propgaugeFbarFbar (x,y)$, and
$\propgaugeSS (x,y)$ are defined in section V of \cite{Martin:2003it};
they follow from the Feynman diagrams labeled $\propM_{FFFFV}$,
$\propV_{FFFFV}$ (with fermion mass insertions in all possible ways)
and  $\propM_{SSSSV}$, $\propV_{SSSSV}$ in figure
\ref{fig:diagrams} of the present paper.
The contributions involving the gluino are given by:
\begin{eqnarray}
\Pi^{(2),2}_{\phiOi\phiOj}
&= & 16 g_3^2 \Bigl \lbrace
\Bigr [
{\rm Re}[ ( Y_{t\tbar \phiOi} L_{\stopk} L_{\stopm}^* +
Y^*_{t\tbar \phiOi} R_{\stopk} R_{\stopm}^* )
\lambda_{\phiOj \stopm \stopk^*} ] m_t
\propM_{SFS\Fbar F}(\stopk, t , \stopm, t, \gluino )
\nonumber \\ &&
-{\rm Re}[Y_{t\tbar \phiOi} L_{\stopm} R^*_{\stopk}
  \lambda_{\phiOj \stopk \stopm^*} ] m_{\gluino}
  \propM_{SFSF \Fbar} (\stopk, t, \stopm , t, \gluino )
\nonumber \\ &&
-{\rm Re}[Y^*_{t\tbar \phiOi} L_{\stopm} R^*_{\stopk}
  \lambda_{\phiOj \stopk \stopm^*}]
  m_t^2 m_{\gluino} \propM_{S\Fbar S\Fbar \Fbar}
(\stopk, t, \stopm, t, \gluino)
\Bigl ]
+ \bigl (i \leftrightarrow j \bigr ) \Bigr \rbrace
+ \bigl (t \rightarrow b \bigr ) ,
\label{eq:phiOgluinoa}
\\
\Pi^{(2),3}_{\phiOi\phiOj}
& = &
16 g_3^2 \Bigl \lbrace
{\rm Re}[Y_{t\tbar \phiOi} Y_{t\tbar \phiOj }^*] \left [
\propV_{FFFFS}(t,t,t,\gluino, \stopk)
+ m_t^2 \propV_{F\Fbar \Fbar FS}(t,t,t,\gluino, \stopk)
\right ]
\nonumber \\ &&
+2 {\rm Re}[Y_{t\tbar \phiOi} Y_{t\tbar \phiOj}] m_t^2
   \propV_{\Fbar F \Fbar FS}(t,t,t,\gluino, \stopk )
-4 {\rm Re}[Y_{t\tbar \phiOi} Y_{t\tbar \phiOj}^*] {\rm Re} [
  L_{\stopk} R^*_{\stopk}] m_t m_{\gluino}
  \propV_{FF\Fbar\Fbar S}(t,t,t,\gluino, \stopk )
\nonumber \\ &&
-2 {\rm Re}[Y_{t\tbar \phiOi} Y_{t\tbar \phiOj }
  L_{\stopk} R^*_{\stopk}] m_t m_{\gluino}
  \propV_{\Fbar FF\Fbar S}(t,t,t,\gluino, \stopk )
\nonumber \\ &&
-2 {\rm Re}[Y_{t\tbar \phiOi} Y_{t\tbar \phiOj}
  L^*_{\stopk} R_{\stopk}] m_t^3 m_{\gluino}
  \propV_{\Fbar\Fbar\Fbar\Fbar S}(t,t,t,\gluino , \stopk )
\Bigr \rbrace
+ (t \rightarrow b) , 
\\
\Pi^{(2),4}_{\phiOi\phiOj}
& = & 8 g_3^2
\Bigl \lbrace
\Bigl [
\lambda_{\phiOi \phiOj \stopk \stopk^*}
\propW_{SSFF} (\stopk, \stopk, t, \gluino)
- 2
{\rm Re} [
\lambda_{\phiOi \phiOj \stopk \stopm^*} L_{\stopk}^* R_{\stopm} ]
m_t m_{\gluino} \propW_{SS\Fbar \Fbar} (\stopk, \stopm, t, \gluino)
\nonumber \\ &&
+ 2
{\rm Re} [
\lambda_{\phiOi \stopk \stopm^*}\lambda_{\phiOj \stopm \stopk^*}]
\propV_{SSSFF} (\stopk,\stopm,\stopm,t,\gluino)
\nonumber \\ &&
- 2
{\rm Re} [
\lambda_{\phiOi \stopk \stopm^*}\lambda_{\phiOj \stopn \stopk^*}
(L_{\stopm} R_{\stopn}^* +R_{\stopm} L_{\stopn}^* )]
m_t m_{\gluino} \propV_{SSS\Fbar \Fbar} (\stopk,\stopm,\stopn,t,\gluino)
\Bigr ] + (t \rightarrow b)
\nonumber \\ &&
+ \sum_{\squark}
\lambda_{\phiOi \phiOj \squark \squark^*}
\propW_{SSFF}(\squark, \squark,0,\gluino)
+ 2 \sum_{\squark} {\rm Re}[
\lambda_{\phiOi \squark \squark^*}
\lambda_{\phiOj \squark \squark^*}] \propV_{SSSFF} (
\squark,\squark,\squark,0,\gluino) 
\Bigr \rbrace .
\label{eq:phiOgluinob}
\end{eqnarray}
Finally, the contributions from squark-antisquark-squark-antisquark
interactions proportional to $g_3^2$ are:
\begin{eqnarray}
\Pi^{(2),5}_{\phiOi\phiOj}
&= & 
4 g_3^2 \Bigl\lbrace\Bigl [
\lambda_{\phiOi \phiOj \stopk \stopm^*}
(L^*_{\stopk} L_{\stopn} -R^*_{\stopk} R_{\stopn})
(L_{\stopm} L^*_{\stopn}-R_{\stopm} R^*_{\stopn})
\propX_{SSS}(\stopk,\stopm,\stopn)
\nonumber \\ &&
+ 2
{\rm Re}[
\lambda_{\phiOi \stopk \stopm^*}
\lambda_{\phiOj \stopn \stopk^*}
(L_{\stopm} L^*_{\stopp}-R_{\stopm} R^*_{\stopp})
(L_{\stopp} L^*_{\stopn}-R_{\stopp} R^*_{\stopn})]
\propY_{SSSS} (\stopk,\stopm,\stopn,\stopp)
\nonumber \\ &&
+
\lambda_{\phiOi \stopk \stopm^*}
\lambda_{\phiOj \stopn \stopp^*}
(L_{\stopm} L^*_{\stopn}-R_{\stopm} R^*_{\stopn})
(L_{\stopp} L^*_{\stopk}-R_{\stopp} R^*_{\stopk})
\propZ_{SSSS} (\stopk,\stopm,\stopn,\stopp)
\Bigr ] + (t \rightarrow b)
\nonumber \\ &&
+
\sum_{\tilde q}
\left (
\lambda_{\phiOi \phiOj \tilde q \tilde q^*}
\propX_{SSS} (\tilde q,\tilde q, \tilde q)
+
\lambda_{\phiOi \tilde q \tilde q^*}
\lambda_{\phiOj \tilde q \tilde q^*}
[2 \propY_{SSSS} (\tilde q,\tilde q,\tilde q,\tilde q)
+ \propZ_{SSSS} (\tilde q,\tilde q,\tilde q,\tilde q)]
\right )
\Bigr \rbrace .
\end{eqnarray}

\subsection{Yukawa and related contributions}\label{subsec:neutralYukawa}

In this section, I present contributions to the neutral Higgs scalar 
boson two-loop self energy that involve Yukawa couplings and the
corresponding soft (scalar)$^3$ interactions, as specified in the 
Introduction.

The contributions involving charginos and neutralinos are given by:
\begin{eqnarray}
\Pi^{(2),6}_{\phiOi\phiOj} &=& 
2 n_t \Bigl [
\Bigl (
{\rm Re} [ Y_{t\tbar\phiOi} \lambda_{\phiOj\stopm\stopn^*}
           Y_{t\neutk\stopn^*}^* Y_{\overline t\neutk\stopm}^*]
           m_{\neutk} \propM_{SFSF \Fbar} (\stopm,t,\stopn,t,\neutk)
\nonumber \\ &&
+{\rm Re} [Y_{t\tbar\phiOi} \lambda_{\phiOj\stopm\stopn^*}
    (Y_{t\neutk\stopn^*}^* Y_{t\neutk\stopm^*}
    +Y_{\overline t\neutk\stopm}^* Y_{\overline t\neutk\stopn} )]
    m_t \propM_{SFS \Fbar F} (\stopn,t,\stopm,t,\neutk)
\nonumber \\ &&
+ {\rm Re} [Y_{t\tbar\phiOi} \lambda_{\phiOj\stopm\stopn^*}
            Y_{t\neutk\stopm^*} Y_{\overline t\neutk\stopn}]
            m_t^2 m_{\neutk}
    \propM_{S\Fbar S\Fbar \Fbar} (\stopm,t,\stopn,t,\neutk)
\Bigr )
+ (i \leftrightarrow j) \Bigr ]
+ \bigl (t \rightarrow b \bigr )
+ \bigl (t \rightarrow \tau \bigr ),\phantom{xxx}
\\
\Pi^{(2),7}_{\phiOi\phiOj} &=& 2 \Bigl [
3 \Bigl ({\rm Re} [Y_{t\tbar\phiOi} \lambda_{\phiOj\sbotm\sbotn^*}
          Y_{t\chark\sbotn^*}^* Y_{\overline t\chark\sbotm}^*]
          m_{\chark} \propM_{SFSF \Fbar} (\sbotm,t,\sbotn,t,\chark)
\nonumber \\ &&
+{\rm Re} [Y_{t\tbar\phiOi} \lambda_{\phiOj\sbotm\sbotn^*}
          (Y_{t\chark\sbotn^*}^* Y_{t\chark\sbotm^*}
          +Y_{\overline t\chark\sbotm}^* Y_{\overline t\chark\sbotn} )]
           m_t \propM_{SFS \Fbar F} (\sbotn,t,\sbotm,t,\chark)
\nonumber \\ &&
+ {\rm Re} [Y_{t\tbar\phiOi} \lambda_{\phiOj\sbotm\sbotn^*}
            Y_{t\chark\sbotm^*} Y_{\overline t\chark\sbotn}]
            m_t^2 m_{\chark}
  \propM_{S\Fbar S \Fbar \Fbar} (\sbotm,t,\sbotn,t,\chark) \Bigr )
+ (t \leftrightarrow b)
\nonumber \\ &&
+ {\rm Re} [Y_{\tau\taubar\phiOi} \lambda_{\phiOj\snu\snu^*}
        Y_{\tau\chark\snutau^*}^* Y_{\overline \tau\chark\snutau}^*]
        m_{\chark} \propM_{SFSF\Fbar} (\snutau,\tau,\snutau,\tau,\chark)
\nonumber \\ &&
+{\rm Re} [Y_{\tau\taubar\phiOi} \lambda_{\phiOj\snu\snu^*}]
          (|Y_{\tau\chark\snutau^*}|^2
          +|Y_{\overline \tau\chark\snutau}|^2 )
   m_\tau \propM_{SFS \Fbar F} (\snutau,\tau,\snutau,\tau,\chark)
\nonumber \\ &&
+ {\rm Re} [Y_{\tau\taubar\phiOi} \lambda_{\phiOj\snu\snu^*}
            Y_{\tau\chark\snutau^*} Y_{\overline \tau\chark\snutau}^*]
            m_\tau^2 m_{\chark}
  \propM_{S\Fbar S\Fbar\Fbar} (\snutau,\tau,\snutau,\tau,\chark) 
\Bigr ]
+ (i \leftrightarrow j), 
\phantom{xxx}
\\
\Pi^{(2),8}_{\phiOi\phiOj} &=&
2 n_t \Bigl \lbrace
(|Y_{t \neutk \stopm^*}|^2 + |Y_{\overline t \neutk \stopm}|^2)
    \Bigl [
    {\rm Re} [ Y_{t\tbar\phiOi} Y_{t\tbar \phiOj}^*] \lbrace
    \propV_{FFFFS}(t,t,t,\neutk,\stopm) 
+m_t^2 \propV_{F\Fbar \Fbar FS}(t,t,t,\neutk,\stopm) \rbrace
\nonumber \\ &&
+ 2 {\rm Re} [ Y_{t\tbar\phiOi} Y_{t\tbar \phiOj}] m_t^2
     \propV_{\Fbar F\Fbar FS}(t,t,t,\neutk,\stopm) \Bigr ]
\nonumber \\ &&
+2 m_t m_{\neutk} \Bigl [
    {\rm Re} [ Y_{t\tbar\phiOi}^* Y_{t\tbar \phiOj}^*
    Y_{t \neutk \stopm^*}Y_{\overline t \neutk \stopm}]
    \propV_{\Fbar FF \Fbar S}(t,t,t,\neutk,\stopm)
\nonumber \\ &&
+2 {\rm Re}[Y_{t\tbar\phiOi} Y_{t\tbar \phiOj}^*]
   {\rm Re} [ Y_{t \neutk \stopm^*}Y_{\overline t \neutk \stopm}]
   \propV_{F F \Fbar \Fbar S}(t,t,t,\neutk,\stopm)
\nonumber \\ &&
+ {\rm Re} [ Y_{t\tbar\phiOi} Y_{t\tbar \phiOj}
   Y_{t \neutk \stopm^*}Y_{\overline t \neutk \stopm}]
   m_t^2 \propV_{\Fbar \Fbar \Fbar \Fbar S}(t,t,t,\neutk,\stopm)\Bigr ] 
\Bigr \rbrace 
+ (t \rightarrow b) 
+ (t \rightarrow \tau) ,
\\
\Pi^{(2),9}_{\phiOi\phiOj} &=& 2 n_t \Bigl [
(|Y_{t \chark \sbotm^*}|^2 + |Y_{\overline t \chark \sbotm}|^2)
      \Bigl ( {\rm Re} [ Y_{t\tbar\phiOi} Y_{t\tbar \phiOj}^*] 
      \lbrace \propV_{FFFFS}(t,t,t,\chark,\sbotm ) 
        +m_t^2 \propV_{F\Fbar \Fbar FS}(t,t,t,\chark,\sbotm ) \rbrace
\nonumber \\ &&
+2 {\rm Re} [ Y_{t\tbar\phiOi} Y_{t\tbar \phiOj}]
           m_t^2 \propV_{\Fbar F\Fbar FS}(t,t,t,\chark,\sbotm)
\Bigr ) 
\nonumber \\ &&
+ 2 m_t m_{\chark} \Bigl (
{\rm Re} [ Y_{t\tbar\phiOi}^* Y_{t\tbar \phiOj}^*
           Y_{t \chark \sbotm^*}Y_{\overline t \chark \sbotm}]
           \propV_{\Fbar FF \Fbar S}(t,t,t,\chark,\sbotm)
\nonumber \\ &&
+2 {\rm Re}[Y_{t\tbar\phiOi} Y_{t\tbar \phiOj}^*]
   {\rm Re} [Y_{t \chark \sbotm^*}Y_{\overline t \chark \sbotm}]
       \propV_{F F \Fbar \Fbar S}(t,t,t,\chark,\sbotm)
\nonumber \\ &&
+{\rm Re} [ Y_{t\tbar\phiOi} Y_{t\tbar \phiOj}
            Y_{t \chark \sbotm^*}Y_{\overline t \chark \sbotm}]
            m_t^2 \propV_{\Fbar \Fbar \Fbar \Fbar S}(t,t,t,\chark,\sbotm)
\Bigr ) \Bigr ]
+ \bigl (t \leftrightarrow b \bigr )
+ \bigl (t \leftrightarrow \tau \bigr ) ,
\\
\Pi^{(2),10}_{\phiOi \phiOj} &=& 
n_t \Bigl \lbrace 
2 {\rm Re} [\lambda_{\phiOi \stopm \stopn^*}
    \lambda_{\phiOj \stopp \stopm^*}
    (Y_{t \neutk \stopn^*}^* Y_{t \neutk \stopp^*}
   + Y_{\overline t \neutk \stopn} Y_{\overline t \neutk \stopp}^*) ]
   \propV_{SSSFF} (\stopm, \stopn, \stopp, t, \neutk)
\nonumber \\ && 
+2 {\rm Re} [
    \lambda_{\phiOi \stopm \stopn^*} \lambda_{\phiOj \stopp \stopm^*}
    (Y_{\overline t \neutk \stopn} Y_{t \neutk \stopp^*}
   + Y_{\overline t \neutk \stopp}^* Y_{t \neutk \stopn^*}^*) ]
    m_t m_{\neutk}
    \propV_{SSS\Fbar \Fbar} (\stopm, \stopn, \stopp, t, \neutk)
\nonumber \\ &&
+ \lambda_{\phiOi \phiOj \stopm \stopn^*}
\bigl [\bigl (Y_{t \neutk \stopm^*} Y_{t \neutk \stopn^*}^*
      + Y_{\overline t \neutk \stopm}^* Y_{\overline t \neutk \stopn}
      \bigr ) \propW_{SSFF}(\stopm,\stopn,t,\neutk)
\nonumber \\ &&
+\bigl ( Y_{t \neutk \stopm^*} Y_{\overline t \neutk \stopn}
       + Y_{\overline t \neutk \stopm}^* Y_{t \neutk \stopn^*}^* \bigr )
       m_t m_{\neutk} \propW_{SS\Fbar \Fbar}(\stopm,\stopn,t,\neutk)
\bigr ]
\Bigr \rbrace 
+ (t \rightarrow b) 
+ (t \rightarrow \tau) 
\nonumber \\ &&
+ |Y_{\nu \neutk \snu^*}|^2 \bigl [
2 \lambda_{\phiOi \snu \snu^*} \lambda_{\phiOj \snutau \snu^*}
   \propV_{SSSFF} (\snutau, \snutau, \snutau, 0, \neutk)
+\lambda_{\phiOi\phiOj \snu \snu^*} 
  \propW_{SSFF}(\snutau,\snutau,0,\neutk) 
\bigr ] ,
\\
\Pi^{(2),11}_{\phiOi\phiOj} &=& 
3 \Bigl \lbrace 
2 {\rm Re} [\lambda_{\phiOi \sbotm \sbotn^*}
              \lambda_{\phiOj \sbotp \sbotm^*}
             (Y_{t \chark \sbotn^*}^* Y_{t \chark \sbotp^*}
            + Y_{\overline t \chark \sbotn} 
              Y_{\overline t \chark \sbotp}^*) ]
           \propV_{SSSFF} (\sbotm, \sbotn, \sbotp, t, \chark)
\nonumber \\ &&
+ 2 {\rm Re} [\lambda_{\phiOi \sbotm \sbotn^*} 
              \lambda_{\phiOj \sbotp \sbotm^*}
             (Y_{\overline t \chark \sbotn} Y_{t \chark \sbotp^*}
            + Y_{\overline t \chark \sbotp}^* 
              Y_{t \chark \sbotn^*}^*) ] m_t m_{\chark}
             \propV_{SSS\Fbar \Fbar} (\sbotm, \sbotn, \sbotp, t, \chark )
\nonumber \\ &&
+ \lambda_{\phiOi \phiOj \sbotm \sbotn^*}
\Bigl [ \bigl (Y_{t \chark \sbotm^*} Y_{t \chark \sbotn^*}^*
         + Y_{\overline t \chark \sbotm}^* Y_{\overline t \chark \sbotn}
        \bigr ) \propW_{SSFF}(\sbotm,\sbotn,t,\chark)
\nonumber \\ &&
+ \bigl ( Y_{t \chark \sbotm^*} Y_{\overline t \chark \sbotn}
        + Y_{\overline t \chark \sbotm}^* Y_{t \chark \sbotn^*}^* \bigr )
         m_t m_{\chark} \propW_{SS\Fbar \Fbar}(\sbotm,\sbotn,t,\chark)
\Bigr ] 
\Bigr \rbrace
+ (t \leftrightarrow b)
\nonumber \\ &&
+ 2 \lambda_{\phiOi \snu \snu^*} \lambda_{\phiOj \snu \snu^*}
             (|Y_{\tau \chark \snutau^*}|^2 
            + |Y_{\overline \tau \chark \snutau}|^2 ) 
           \propV_{SSSFF} (\snutau, \snutau, \snutau,\tau, \chark)
\nonumber \\ &&
+ 4 \lambda_{\phiOi \snu \snu^*} \lambda_{\phiOj \snu \snu^*}
    {\rm Re} [Y_{\overline \tau \chark \snutau} Y_{\tau \chark \snutau^*}
            ] m_\tau m_{\chark}
    \propV_{SSS\Fbar \Fbar} (\snutau, \snutau, \snutau, \tau, \chark )
\nonumber \\ &&
+ \lambda_{\phiOi \phiOj \snu \snu^*}
    \Bigl [ \bigl (|Y_{\tau \chark \snutau^*}|^2
         + |Y_{\overline \tau  \chark \snutau}|^2 
        \bigr ) \propW_{SSFF}(\snutau,\snutau,\tau,\chark)
\nonumber \\ &&
+ 2 {\rm Re}[Y_{\tau \chark \snutau^*} Y_{\overline \tau \chark \snutau}]
         m_\tau m_{\chark} 
         \propW_{SS\Fbar\Fbar}(\snutau,\snutau,\tau,\chark)
\Bigr ] 
\nonumber \\ &&
+ 2 {\rm Re} [\lambda_{\phiOi \staum \staun^*}
              \lambda_{\phiOj \staup \staum^*}
             Y_{\nu_\tau \chark \staun^*}^* Y_{\nu_\tau \chark \staup^*}]
           \propV_{SSSFF} (\staum, \staun, \staup, 0, \chark)
\nonumber \\ &&
+ \lambda_{\phiOi \phiOj \staum \staun^*}
           Y_{\nu_\tau \chark \staum^*} Y_{\nu_\tau \chark \staun^*}^*
           \propW_{SSFF}(\staum,\staun,0,\chark) .
\end{eqnarray}
The contributions involving virtual Higgs scalar bosons and third-family 
fermions are:
\begin{eqnarray}
\Pi^{(2),12}_{\phiOi\phiOj} &=& 
4 n_{t} \Bigl [
{\rm Re} [Y_{t\tbar\phiOi} Y_{t\tbar\phiOj}^*]  |Y_{t\tbar\phiOk}|^2
\lbrace \propV_{FFFFS}(t,t,t,t,\phiOk)
+ m_t^2 \propV_{F\Fbar \Fbar FS}(t,t,t,t,\phiOk) \rbrace
\nonumber \\ &&
+ 2 {\rm Re} [Y_{t\tbar\phiOi} Y_{t\tbar\phiOj}]  |Y_{t\tbar\phiOk}|^2
m_t^2 \propV_{\Fbar F \Fbar FS}(t,t,t,t,\phiOk)
+ {\rm Re} [Y_{t\tbar\phiOi} Y_{t\tbar\phiOj}  (Y_{t\tbar\phiOk}^*)^2]
m_t^2 \propV_{\Fbar F F \Fbar S}(t,t,t,t,\phiOk)
\nonumber \\ &&
+ 2 {\rm Re} [Y_{t\tbar\phiOi} Y_{t\tbar\phiOj}^*]  
    {\rm Re}[(Y_{t\tbar\phiOk})^2]
m_t^2 \propV_{F F \Fbar \Fbar S}(t,t,t,t,\phiOk)
+ {\rm Re} [Y_{t\tbar\phiOi} Y_{t\tbar\phiOj}  (Y_{t\tbar\phiOk})^2]
m_t^4 \propV_{\Fbar \Fbar \Fbar \Fbar S}(t,t,t,t,\phiOk)
\nonumber \\ &&
+ {\rm Re}[Y_{t\tbar\phiOi} Y_{t\tbar\phiOj} (Y_{t\tbar\phiOk}^*)^2]
  \propM_{FFFFS} (t,t,t,t,\phiOk)/2
+ {\rm Re}[Y_{t\tbar\phiOi} Y_{t\tbar\phiOj}^*]
  {\rm Re}[(Y_{t\tbar\phiOk})^2]
  m_t^2 \propM_{F\Fbar F \Fbar S} (t,t,t,t,\phiOk)
\nonumber \\ &&
+ {\rm Re}[Y_{t\tbar\phiOi} Y_{t\tbar\phiOj}^*] |Y_{t\tbar\phiOk}|^2
  m_t^2 \propM_{F\Fbar\Fbar FS} (t,t,t,t,\phiOk)
  + {\rm Re}[Y_{t\tbar\phiOi} Y_{t\tbar\phiOj}] |Y_{t\tbar\phiOk}|^2
  m_t^2 \propM_{FF\Fbar\Fbar S} (t,t,t,t,\phiOk)
\nonumber \\ &&
+ {\rm Re}[Y_{t\tbar\phiOi} Y_{t\tbar\phiOj} (Y_{t\tbar\phiOk})^2]
  m_t^4 \propM_{\Fbar\Fbar\Fbar\Fbar S} (t,t,t,t,\phiOk)/2
\Bigr ]
+ (t \rightarrow b )
+ (t \rightarrow \tau ) ,
\\
\Pi^{(2),13}_{\phiOi\phiOj} &=& 
6 \Bigl [(Y_{\tbar b \phipk}^2 + Y_{\bbar t\phimk}^2)
\bigl \lbrace
   {\rm Re}[Y_{t\tbar\phiOi} Y_{t\tbar\phiOj}^*]
   [\propV_{FFFFS}(t,t,t,b,\phipk) +
   m_t^2 \propV_{F\Fbar \Fbar FS}(t,t,t,b,\phipk)]
\nonumber \\ &&
+2 {\rm Re}[Y_{t\tbar\phiOi} Y_{t\tbar\phiOj}] m_t^2
   \propV_{\Fbar F\Fbar FS}(t,t,t,b,\phipk)
+ (t\leftrightarrow b) 
\bigr \rbrace
\nonumber \\ &&
+ 2 Y_{\tbar b\phipk} Y_{\bbar t\phimk} m_t m_b
\bigl \lbrace
{\rm Re}[Y_{t\tbar\phiOi} Y_{t\tbar\phiOj}]
[\propV_{\Fbar FF\Fbar S}(t,t,t,b,\phipk)
+ m_t^2 \propV_{\Fbar \Fbar \Fbar \Fbar S}(t,t,t,b,\phipk)]
\nonumber \\ &&
+ 2 {\rm Re}[Y_{t\tbar\phiOi} Y_{t\tbar\phiOj}^*]
\propV_{FF \Fbar\Fbar S}(t,t,t,b,\phipk)
+ (t\leftrightarrow b) 
\bigr \rbrace \Bigl ]
\nonumber \\ &&
+ 2 
Y_{\taubar\nu_\tau \phimk}^2
\bigl \lbrace
{\rm Re}[Y_{\tau\taubar\phiOi} Y_{\tau\taubar\phiOj}^*]
[\propV_{FFFFS}(\tau,\tau,\tau,0,\phipk) +
  m_\tau^2 \propV_{F\Fbar \Fbar FS}(\tau,\tau,\tau,0,\phipk)]
\nonumber \\ &&
+2 {\rm Re}[Y_{\tau\taubar\phiOi} Y_{\tau\taubar\phiOj}] m_\tau^2
  \propV_{\Fbar F\Fbar FS}(\tau,\tau,\tau,0,\phipk)
\bigr \rbrace ,
\\
\Pi^{(2),14}_{\phiOi\phiOj} &=& 6 \Bigl [
(Y_{\tbar b \phipk}^2 + Y_{\bbar t\phimk}^2) m_t m_b
  \bigl \lbrace
  {\rm Re}[Y_{t\tbar\phiOi} Y_{b\bbar\phiOj}^*]
  \propM_{F\Fbar\Fbar FS}(b,t,b,t,\phipk)
+ {\rm Re}[Y_{t\tbar\phiOi} Y_{b\bbar\phiOj}]
  \propM_{FF \Fbar \Fbar S}(t,b,t,b,\phipk)
  \bigr \rbrace
\nonumber \\ &&
+ Y_{\tbar b \phipk} Y_{\bbar t\phimk} \bigl \lbrace
  {\rm Re}[Y_{t\tbar\phiOi} Y_{b\bbar\phiOj}]
  [\propM_{FFFF S}(t,b,t,b,\phipk)
+ m_t^2 m_b^2 \propM_{\Fbar \Fbar \Fbar \Fbar S}(t,b,t,b,\phipk)]
\nonumber \\ &&
+ {\rm Re}[Y_{t\tbar\phiOi} Y_{b\bbar\phiOj}^*]
  [m_t^2 \propM_{F \Fbar F \Fbar S}(b,t,b,t,\phipk)
  + m_b^2 \propM_{F \Fbar F \Fbar S}(t,b,t,b,\phipk)]
\bigr \rbrace \Bigr ]
+ (i \leftrightarrow j) .
\end{eqnarray}
Contributions involving virtual Higgs scalar bosons and third-family
sfermions are:
\begin{eqnarray}
\Pi^{(2),15}_{\phiOi\phiOj} &=& n_{\tilde t} \Bigl [
\lambda_{\phiOi\stopm\stopp^*} \lambda_{\phiOj\stopq\stopn^*}
  \lambda_{\phiOk\stopp\stopq^*} \lambda_{\phiOk\stopn\stopm^*}
  \propM_{SSSSS}(\stopm,\stopn,\stopp,\stopq,\phiOk)
+ \lambda_{\phiOi\phiOk \stopm\stopn^*}
  \lambda_{\phiOj\phiOk \stopn\stopm^*}
  \propS_{SSS} (\phiOk,\stopm,\stopn)
\nonumber \\ &&
+ 2 {\rm Re}[ \lambda_{\phiOi\stopm\stopn^*}\lambda_{\phiOj\stopp\stopm^*}
            \lambda_{\phiOk\stopn\stopq^*} \lambda_{\phiOk\stopq\stopp^*}]
    \propV_{SSSSS}(\stopm,\stopn,\stopp,\stopq,\phiOk)
\nonumber \\ &&
+2{\rm Re}[(\lambda_{\phiOi\stopm\stopn^*}
           \lambda_{\phiOj \phiOk \stopp \stopm^*}
          +\lambda_{\phiOj\stopm\stopn^*}
           \lambda_{\phiOi \phiOk \stopp \stopm^*})
           \lambda_{\phiOk\stopn\stopp^*}]
   \propU_{SSSS}(\stopm,\stopn,\stopp,\phiOk)
\nonumber \\ &&
+ \lambda_{\phiOi \phiOj \stopm \stopn^*}
    \lambda_{\phiOk\stopn\stopp^*} \lambda_{\phiOk\stopp\stopm^*}
    \propW_{SSSS} (\stopm,\stopn,\stopp,\phiOk)
+ \frac{1}{2} \lambda_{\phiOi\phiOj\stopm\stopn^*}
              \lambda_{\phiOk\phiOk\stopn\stopm^*}
              \propX_{SSS}(\stopm,\stopn,\phiOk)
\nonumber \\ &&
+ {\rm Re}[\lambda_{\phiOi\stopm\stopn^*}
             \lambda_{\phiOj\stopp\stopm^*}
             \lambda_{\phiOk\phiOk\stopn\stopp^*}]
             \propY_{SSSS}(\stopm,\stopn,\stopp,\phiOk)
\Bigr ]
+ (\tilde t \rightarrow \tilde b) 
+ (\tilde t \rightarrow \tilde \tau) 
+ (\tilde t \rightarrow \tilde \nu_\tau) ,
\\
\Pi^{(2),16}_{\phiOi\phiOj} &=& n_{\tilde t} \Bigl [
  (\lambda_{\phiOi\stopm\stopp^*} \lambda_{\phiOj\sbotq\sbotn^*}
  +\lambda_{\phiOj\stopm\stopp^*} \lambda_{\phiOi\sbotq\sbotn^*})
  \lambda_{\phipk\sbotn\stopm^*}
  \lambda_{\phipk\sbotq\stopp^*}^*
  \propM_{SSSSS}(\stopm,\sbotn,\stopp,\sbotq,\phipk)
\nonumber \\ &&
+ 2 {\rm Re}[\lambda_{\phiOi\stopm\stopn^*}
           \lambda_{\phiOj\stopp\stopm^*}
           \lambda_{\phipk\sbotq\stopn^*}^*
           \lambda_{\phipk\sbotq\stopp^*}]
           \propV_{SSSSS}(\stopm,\stopn,\stopp,\sbotq,\phipk)
\nonumber \\ &&
+ 2 {\rm Re}[\lambda_{\phiOi\sbotm\sbotn^*}
           \lambda_{\phiOj\sbotp\sbotm^*}
           \lambda_{\phipk\sbotp\stopq^*}^*
           \lambda_{\phipk\sbotn\stopq^*}]
           \propV_{SSSSS}(\sbotm,\sbotn,\sbotp,\stopq,\phipk)
\nonumber \\ &&
+2 {\rm Re}[(\lambda_{\phiOi\stopm\stopn^*}
             \lambda_{\phiOj \phipk \sbotp \stopm^*}
            +\lambda_{\phiOj\stopm\stopn^*}
             \lambda_{\phiOi \phipk \sbotp \stopm^*})
             \lambda_{\phipk\sbotp\stopn^*}^*]
             \propU_{SSSS}(\stopm,\stopn,\sbotp,\phipk)
\nonumber \\ &&
+2 {\rm Re}[(\lambda_{\phiOi\sbotm\sbotn^*}
             \lambda_{\phiOj \phipk \sbotm \stopp^*}^*
            +\lambda_{\phiOj\sbotm\sbotn^*}
             \lambda_{\phiOi \phipk \sbotm \stopp^*}^*)
             \lambda_{\phipk\sbotn\stopp^*}]
             \propU_{SSSS}(\sbotm,\sbotn,\stopp,\phipk)
\nonumber \\ &&
+ \lambda_{\phiOi \phiOj \stopm \stopn^*}
  \lambda_{\phipk\sbotp\stopm^*}
  \lambda_{\phipk\sbotp\stopn^*}^*
  \propW_{SSSS} (\stopm,\stopn,\sbotp,\phipk)
+ \lambda_{\phiOi \phiOj \sbotm \sbotn^*}
  \lambda_{\phipk\sbotn\stopp^*}
  \lambda_{\phipk\sbotm\stopp^*}^*
  \propW_{SSSS} (\sbotm,\sbotn,\stopp,\phipk)
\nonumber \\ &&
+ \lambda_{\phiOi\phiOj\stopm\stopn^*} 
  \lambda_{\phipk\phimk\stopn\stopm^*}
  \propX_{SSS}(\stopm,\stopn,\phipk)
+2 {\rm Re}[\lambda_{\phiOi\stopm\stopn^*}
           \lambda_{\phiOj\stopp\stopm^*}
           \lambda_{\phipk\phimk\stopn\stopp^*}]
           \propY_{SSSS}(\stopm,\stopn,\stopp,\phipk)
\nonumber \\ &&
+ \lambda_{\phiOi\phiOj\sbotm\sbotn^*} 
  \lambda_{\phipk\phimk\sbotn\sbotm^*}
  \propX_{SSS}(\sbotm,\sbotn,\phipk)
+2 {\rm Re}[\lambda_{\phiOi\sbotm\sbotn^*}
           \lambda_{\phiOj\sbotp\sbotm^*}
           \lambda_{\phipk\phimk\sbotn\sbotp^*}]
           \propY_{SSSS}(\sbotm,\sbotn,\sbotp,\phipk)
\nonumber \\ &&
+ 2 {\rm Re}[\lambda_{\phiOi\phipk \sbotm\stopn^*}
  \lambda_{\phiOj\phipk \sbotm\stopn^*}^*]
  \propS_{SSS} (\phipk,\sbotm,\stopn)
\Bigr ] + (\tilde t\rightarrow \tilde \nu_\tau, 
            \tilde b\rightarrow \tilde \tau) .
\end{eqnarray}
Finally, the contributions involving only virtual sfermions are
given by:
\begin{eqnarray}
\Pi^{(2),17}_{\phiOi\phiOj} &=& 
\lambda_{\phiOi\phiOj\sfermionk\sfermionm^*} [
          n_{\sfermionk} n_{\sfermionn}
          \lambda_{\sfermionm \sfermionk^* \sfermionn \sfermionn^*}
        + n_{\sfermionk} 
          \lambda_{\sfermionm \sfermionn^* \sfermionn \sfermionk^*}]
          \propX_{SSS}(\sfermionk,\sfermionm,\sfermionn)
\nonumber \\ &&
+2 {\rm Re}[ \lambda_{\phiOi \sfermionk \sfermionm^*}
            \lambda_{\phiOj \sfermionn \sfermionk^*}
            ( n_{\sfermionk} n_{\sfermionp}
            \lambda_{\sfermionm \sfermionn^* \sfermionp \sfermionp^*}
            + n_{\sfermionk}
            \lambda_{\sfermionm \sfermionp^* \sfermionp \sfermionn^*})]
          \propY_{SSSS}(\sfermionk,\sfermionm,\sfermionn,\sfermionp)
\nonumber \\ &&
+ \lambda_{\phiOi \sfermionk \sfermionm^*}
  \lambda_{\phiOj \sfermionn \sfermionp^*} 
  (n_{\sfermionk} n_{\sfermionn} 
  \lambda_{\sfermionm \sfermionk^* \sfermionp \sfermionn^*}
  + n_{\sfermionk} 
  \lambda_{\sfermionm \sfermionn^* \sfermionp \sfermionk^*})
  \propZ_{SSSS}(\sfermionk,\sfermionm,\sfermionn,\sfermionp) .
\label{eq:Hneutallsfermions}
\end{eqnarray}
This expression includes the contributions for the sfermions of the first 
two families, which only have gauge interactions. In the numerical
results of section \ref{sec:examples}, only the third-family 
sfermion contributions from eq.~(\ref{eq:Hneutallsfermions}) are included.

\section{Two-loop contributions to charged Higgs scalar boson 
self-energies}\label{sec:chargedtwoloop}
\setcounter{equation}{0}

In this section, I present analytical formulas for two-loop contributions
to the charged Higgs scalar boson self-energies, as specified in the 
Introduction. They are labeled in the
form $\Pi^{(2),N}_{\phipi\phimj}$, where $N$ is the equation number.

\subsection{Strong contributions}\label{subsec:chargedstrong}

The contributions to the two-loop charged Higgs scalar boson self-energy
involving the gluon are:
\begin{eqnarray}
\Pi^{(2),1}_{\phipi\phimj}
& = & 
4 g_3^2 \Bigl ([Y_{\tbar b\phipi} Y_{\tbar b\phipj} + 
                Y_{\bbar t\phimi} Y_{\bbar t\phimj}] \propgaugeFF(t,b)
+[Y_{\tbar b\phipi} Y_{\bbar t\phimj} + 
  Y_{\bbar t\phimi} Y_{\tbar b\phipj}]
  m_t m_b \propgaugeFbarFbar(t,b)
\nonumber \\ &&
+ \lambda_{\phipi \sbotk \stopm^*} \lambda^*_{\phipj \sbotk \stopm^*}
\propgaugeSS (\sbotk,\stopm)
+  \lambda_{\phipi \phimj \stopk \stopk^*}
\propW_{SSSV}(\stopk, \stopk , \stopk , 0)
+  \lambda_{\phipi \phimj \sbotk \sbotk^*}
\propW_{SSSV}(\sbotk, \sbotk , \sbotk , 0)
\nonumber \\ &&
+\lambda_{\phipi \sdL\suL^*} \lambda_{\phipj \sdL \suL^*}
[\propgaugeSS (\sdL,\suL) + \propgaugeSS (\ssL,\scL)]
+ 
\sum_{\tilde q} \lambda_{\phipi \phimj \tilde q \tilde q^*}
\propW_{SSSV}(\squark,\squark,\squark , 0)
\Bigr ) .
\label{eq:phipmgluon}
\end{eqnarray}
The contributions involving the gluino are:
\begin{eqnarray}
\Pi^{(2),2}_{\phipi\phimj}
& = & 8 g_3^2 \lambda_{\phipi \sbotk \stopm^*} \Bigl \lbrace
(Y_{\bbar t\phimj} R_{\stopm} R_{\sbotk}^*
  +Y_{\tbar b\phipj} L_{\stopm} L_{\sbotk}^*)
   m_t \propM_{SFS\Fbar F}(\sbotk,b,\stopm,t,\tilde g)
\nonumber \\ &&
+(Y_{\bbar t\phimj} L_{\stopm} L_{\sbotk}^*
  +Y_{\tbar b\phipj} R_{\stopm} R_{\sbotk}^*)
  m_b \propM_{SFS\Fbar F}(\stopm,t,\sbotk,b,\tilde g)
\nonumber \\ &&
-(Y_{\bbar t\phimj} L_{\stopm} R_{\sbotk}^*
  + Y_{\tbar b \phipj} R_{\stopm} L_{\sbotk}^*) 
  m_{\tilde g} \propM_{SFSF\Fbar}(\sbotk,b,\stopm,t,\tilde g)
\nonumber \\ &&
-(Y_{\bbar t\phimj} R_{\stopm} L_{\sbotk}^* 
  + Y_{\tbar b\phipj} L_{\stopm} R_{\sbotk}^*)
  m_b m_t m_{\tilde g} \propM_{S\Fbar S\Fbar\Fbar}(\sbotk,b,\stopm,t,\tilde g)
\Bigr \rbrace
+ (i \leftrightarrow j)^* ,
\label{eq:mentionijstar}
\\
\Pi^{(2),3}_{\phipi\phimj} &=& 8 g_3^2 \Bigl \lbrace
(Y_{\tbar b\phipi} Y_{\tbar b\phipj} |L_{\sbotk}|^2
  +Y_{\bbar t\phimi} Y_{\bbar t\phimj} |R_{\sbotk}|^2)
  \propV_{FFFFS} (t,b,b,\tilde g,\sbotk)
\nonumber \\ &&
+ (Y_{\tbar b\phipi} Y_{\tbar b\phipj} |R_{\sbotk}|^2
  +Y_{\bbar t\phimi} Y_{\bbar t\phimj} |L_{\sbotk}|^2)
  m_b^2 \propV_{F\Fbar\Fbar FS} (t,b,b,\tilde g,\sbotk)
\nonumber \\ &&
+ (Y_{\tbar b\phipi}Y_{\bbar t\phimj} + 
   Y_{\bbar t\phimi}Y_{\tbar b\phipj}) m_t m_b 
   \propV_{\Fbar F\Fbar FS} (t,b,b,\tilde g,\sbotk)  
\nonumber \\ &&
-2 (Y_{\tbar b\phipi}Y_{\tbar b\phipj} + 
    Y_{\bbar t\phimi} Y_{\bbar t\phimj})  
   {\rm Re}[L_{\sbotk}R_{\sbotk}^*] m_b  m_{\tilde g} 
   \propV_{FF\Fbar\Fbar S} (t,b,b,\tilde g,\sbotk) 
\nonumber \\ &&
-(Y_{\tbar b\phipi}Y_{\bbar t\phimj} L_{\sbotk} R_{\sbotk}^*+
   Y_{\bbar t\phimi}Y_{\tbar b\phipj} R_{\sbotk} L_{\sbotk}^*)
  m_t m_{\tilde g} \propV_{\Fbar FF\Fbar S} (t,b,b,\tilde g,\sbotk)
\nonumber \\ &&
-(Y_{\tbar b\phipi} Y_{\bbar t\phimj} R_{\sbotk} L_{\sbotk}^*+
  Y_{\bbar t\phimi} Y_{\tbar b\phipj} L_{\sbotk} R_{\sbotk}^*)
  m_b^2 m_t m_{\tilde g} 
  \propV_{\Fbar\Fbar\Fbar\Fbar S}(t,b,b,\tilde g,\sbotk)
\Bigr \rbrace
+ (t\leftrightarrow b,\> \phi^+ \leftrightarrow \phi^-)
,
\\
\Pi^{(2),4}_{\phipi\phimj}
& = & 8 g_3^2
\Bigl \lbrace
\Bigl [\lambda_{\phipi \phimj \stopk \stopk^*}
       \propW_{SSFF} (\stopk, \stopk, t, \gluino)
-\lambda_{\phipi \phimj \stopk \stopm^*} 
       [L_{\stopk}^* R_{\stopm} + R_{\stopk}^* L_{\stopm} ]
    m_t m_{\gluino} \propW_{SS\Fbar \Fbar} (\stopk, \stopm, t, \gluino)
\Bigr ] 
\nonumber \\ && 
+ (t \rightarrow b)
\,+ \lambda_{\phipi \sbotk \stopm^*} \lambda^*_{\phipj \sbotk \stopm^*}
[\propV_{SSSFF} (\sbotk,\stopm,\stopm,t,\gluino)
+\propV_{SSSFF} (\stopm,\sbotk,\sbotk,b,\gluino)]
\nonumber \\ &&
-\lambda_{\phipi \sbotk \stopm^*}\lambda^*_{\phipj \sbotk \stopn^*}
(L_{\stopm} R_{\stopn}^* +R_{\stopm} L_{\stopn}^* )
m_t m_{\gluino} \propV_{SSS\Fbar \Fbar} (\sbotk,\stopm,\stopn,t,\gluino)
\nonumber \\ &&
-\lambda_{\phipi \sbotm \stopk^*}\lambda^*_{\phipj \sbotn \stopk^*}
(L^*_{\sbotm} R_{\sbotn} +R^*_{\sbotm} L_{\sbotn} )
m_b m_{\gluino} \propV_{SSS\Fbar \Fbar} (\stopk,\sbotm,\sbotn,b,\gluino)
\nonumber \\ &&
+ \sum_{\tilde q} \lambda_{\phipi \phimj \tilde q \tilde q^*}
\propW_{SSFF} (\tilde q, \tilde q, 0, \gluino) 
+ \lambda_{\phipi \sdL \suL^*} \lambda_{\phipj \sdL \suL^*}
[\propV_{SSSFF} (\suL,\sdL,\sdL,0,\gluino) 
\nonumber \\ &&
+\propV_{SSSFF} (\sdL,\suL,\suL,0,\gluino)
+\propV_{SSSFF} (\scL,\ssL,\ssL,0,\gluino) 
+\propV_{SSSFF} (\ssL,\scL,\scL,0,\gluino)
] 
\Bigr \rbrace .
\end{eqnarray}
In eq.~(\ref{eq:mentionijstar}) and in the following, the symbol
$(i\leftrightarrow j)^*$ means the preceding expression with $i$ and $j$
interchanged, and with complex conjugation applied to all of the 
couplings but not to the loop-integral functions.

The contributions involving squark-antisquark-squark-antisquark couplings
proportional to $g_3^2$ are:
\begin{eqnarray}
\Pi^{(2),5}_{\phipi\phimj}
&=&
4 g_3^2 \Bigl [\lambda_{\phipi \phimj \stopk \stopm^*}
    (L^*_{\stopk} L_{\stopn} -R^*_{\stopk} R_{\stopn})
    (L_{\stopm} L^*_{\stopn}-R_{\stopm} R^*_{\stopn})
    \propX_{SSS}(\stopk,\stopm,\stopn)
+ (t \rightarrow b) 
\nonumber \\ &&
+ \lambda_{\phipi \sbotm \stopk^*} \lambda^*_{\phipj \sbotn \stopk^*}
        (L^*_{\sbotm} L_{\sbotp} -R^*_{\sbotm} R_{\sbotp})
        (L_{\sbotn} L^*_{\sbotp}-R_{\sbotn} R^*_{\sbotp})
        \propY_{SSSS}(\stopk,\sbotm,\sbotn,\sbotp)
\nonumber \\ &&
+\lambda_{\phipi \sbotk \stopm^*} \lambda^*_{\phipj \sbotk \stopn^*}
        (L_{\stopm} L^*_{\stopp} -R_{\stopm} R^*_{\stopp})
        (L^*_{\stopn} L_{\stopp}-R^*_{\stopn} R_{\stopp})
        \propY_{SSSS}(\sbotk,\stopm,\stopn,\stopp)
\nonumber \\ &&
+\lambda_{\phipi \sbotk \stopm^*} \lambda_{\phipj \sbotn \stopp^*}^*
       (L_{\sbotk}^* L_{\sbotn} - R_{\sbotk}^* R_{\sbotn})
       (L_{\stopp}^* L_{\stopm} - R_{\stopp}^* R_{\stopm})
       \propZ_{SSSS}(\sbotk,\stopm,\sbotn,\stopp) 
\nonumber \\ &&
+\sum_{\tilde q} \lambda_{\phipi \phimj \tilde q \tilde q^*}
                 \propX_{SSS}(\tilde q,\tilde q,\tilde q)
+\lambda_{\phipi \sdL \suL^*} \lambda_{\phipj \sdL \suL^*}
    [\propY_{SSSS}(\suL,\sdL,\sdL,\sdL)
    +\propY_{SSSS}(\sdL,\suL,\suL,\suL)
\nonumber \\ &&
    +\propY_{SSSS}(\scL,\ssL,\ssL,\ssL)
    +\propY_{SSSS}(\ssL,\scL,\scL,\scL) 
    +\propZ_{SSSS}(\sdL,\suL,\sdL,\suL)
    +\propZ_{SSSS}(\ssL,\scL,\ssL,\scL)]
\Bigr ]. \phantom{xx}
\end{eqnarray}

\subsection{Yukawa and related contributions}\label{subsec:chargedYukawa}

In this subsection, I present two-loop contributions to the charged Higgs
scalar boson self-energies that involve Yukawa couplings and (scalar)$^3$
couplings, as specified in the Introduction.

Contributions involving neutralinos and charginos are given by:
\beq
\Pi^{(2),6}_{\phipi\phimj} &=& 
3 \lambda_{\phipi \sbotm \stopk^*} \Bigl \lbrace
(Y_{\bbar t \phimj} Y_{t\neutn\stopk^*}^* Y_{\bbar \neutn \sbotm}^*
   +Y_{\tbar b \phipj} Y_{\tbar\neutn\stopk} Y_{b \neutn \sbotm^*})
    m_{\neutn} \propM_{SFSF\Fbar} (\stopk,t,\sbotm,b,\neutn)
\nonumber \\ &&
+ (Y_{\bbar t \phimj} Y_{t\neutn\stopk^*}^* Y_{b\neutn\sbotm^*}
  +Y_{\tbar b \phipj} Y_{\tbar\neutn\stopk} Y_{\bbar\neutn\sbotm}^*)
    m_b \propM_{SFS\Fbar F} (\stopk,t,\sbotm,b,\neutn)
\nonumber \\ &&
+ (Y_{\bbar t \phimj} Y_{\tbar\neutn\stopk} Y_{\bbar \neutn \sbotm}^* 
  +Y_{\tbar b \phipj} Y_{t\neutn\stopk^*}^* Y_{b\neutn\sbotm^*})
    m_t  \propM_{SFS\Fbar F} (\sbotm,b,\stopk,t,\neutn)
\nonumber \\ &&
+ (Y_{\bbar t \phimj} Y_{b\neutn\sbotm^*} Y_{\tbar \neutn \stopk}
   +Y_{\tbar b \phipj} Y_{t\neutn\stopk^*}^* Y_{\bbar\neutn\sbotm}^*)
    m_b m_t m_{\neutn} 
    \propM_{S\Fbar S\Fbar\Fbar} (\stopk,t,\sbotm,b,\neutn)
\Bigr \rbrace 
\nonumber \\ &&
+ \lambda_{\phipi \staum \nutau^*} 
Y_{\taubar \nutau \phimj} Y_{\nutau\neutn\snutau^*}^* 
    \bigl [Y_{\taubar \neutn \staum}^*
    m_{\neutn} \propM_{SFSF\Fbar} (\snutau,0,\staum,\tau,\neutn)
\nonumber \\ &&
    + Y_{\tau\neutn\staum^*}
    m_\tau \propM_{SFS\Fbar F} (\snutau,0,\staum,\tau,\neutn) \bigr ]
+ (i\leftrightarrow j)^*
,
\\
\Pi^{(2),7}_{\phipi\phimj} &=& 3 \Bigl \lbrace
(Y_{\tbar b\phipi} Y_{\tbar b\phipj} |Y_{b\neutk\sbotm^*}|^2
  +Y_{\bbar t\phimi} Y_{\bbar t\phimj} |Y_{\bbar\neutk\sbotm}|^2)
  \propV_{FFFFS}(t,b,b,\neutk,\sbotm)
\nonumber \\ &&
+ (Y_{\tbar b\phipi} Y_{\tbar b\phipj} |Y_{\bbar\neutk\sbotm}|^2 
  +Y_{\bbar t\phimi} Y_{\bbar t\phimj} |Y_{b\neutk\sbotm^*}|^2)
  m_b^2 \propV_{F\Fbar\Fbar FS}(t,b,b,\neutk,\sbotm)
\nonumber \\ &&
+ (Y_{\tbar b \phipi} Y_{\bbar t\phimj} +
   Y_{\bbar t\phimi} Y_{\tbar b \phipj})
   (|Y_{b\neutk\sbotm^*}|^2 + |Y_{\bbar\neutk\sbotm}|^2)
   m_b m_t \propV_{\Fbar F\Fbar FS}(t,b,b,\neutk,\sbotm)
\nonumber \\ &&
+ (Y_{\tbar b \phipi} Y_{\bbar t\phimj} 
   Y_{b\neutk\sbotm^*}^* Y_{\bbar\neutk\sbotm}^*
  +Y_{\bbar t\phimi} Y_{\tbar b \phipj} 
   Y_{\bbar\neutk\sbotm} Y_{b\neutk\sbotm^*})
   m_t m_{\neutk} \propV_{\Fbar FF\Fbar S}(t,b,b,\neutk,\sbotm)
\nonumber \\ &&
+ 2 (Y_{\tbar b\phipi} Y_{\tbar b\phipj} +
     Y_{\bbar t\phimi} Y_{\bbar t\phimj})
     {\rm Re}[Y_{b\neutk\sbotm^*} Y_{\bbar\neutk\sbotm}]
   m_b m_{\neutk} \propV_{FF\Fbar\Fbar S}(t,b,b,\neutk,\sbotm)
\nonumber \\ &&
+ (Y_{\tbar b \phipi} Y_{\bbar t\phimj} 
   Y_{b\neutk\sbotm^*} Y_{\bbar\neutk\sbotm}
  +Y_{\bbar t\phimi} Y_{\tbar b \phipj} 
   Y_{\bbar\neutk\sbotm}^* Y_{b\neutk\sbotm^*}^*)
   m_b^2 m_t m_{\neutk} \propV_{\Fbar\Fbar\Fbar\Fbar S}(t,b,b,\neutk,\sbotm)
\Bigr \rbrace 
\nonumber \\ &&
+ (t \leftrightarrow b, \> \phi^+ \leftrightarrow \phi^-)^*
\>
+  Y_{\taubar \nutau\phimi} Y_{\taubar \nutau\phimj} \Bigl \lbrace
  |Y_{\taubar\neutk\staum}|^2
  \propV_{FFFFS}(0,\tau,\tau,\neutk,\staum)
\nonumber \\ &&
+|Y_{\tau\neutk\staum^*}|^2
  m_\tau^2 \propV_{F\Fbar\Fbar FS}(0,\tau,\tau,\neutk,\staum)
+ 2 {\rm Re}[Y_{\tau\neutk\staum^*} Y_{\taubar\neutk\staum}]
   m_\tau m_{\neutk} \propV_{FF\Fbar\Fbar S}(0,\tau,\tau,\neutk,\staum)
\nonumber \\ &&
+ |Y_{\nutau\neutk\snutau^*}|^2
  \propV_{FFFFS}(\tau,0,0,\neutk,\snutau) \Bigr \rbrace
,
\\
\Pi^{(2),8}_{\phipi\phimj} &=& 3 \Bigl \lbrace
(Y_{\tbar b\phipi} Y_{\tbar b\phipj} |Y_{b\chark\stopm^*}|^2
  +Y_{\bbar t\phimi} Y_{\bbar t\phimj} |Y_{\bbar\chark\stopm}|^2)
  \propV_{FFFFS}(t,b,b,\chark,\stopm)
\nonumber \\ &&
+ (Y_{\tbar b\phipi} Y_{\tbar b\phipj} |Y_{\bbar\chark\stopm}|^2 
  +Y_{\bbar t\phimi} Y_{\bbar t\phimj} |Y_{b\chark\stopm^*}|^2)
  m_b^2 \propV_{F\Fbar\Fbar FS}(t,b,b,\chark,\stopm)
\nonumber \\ &&
+ (Y_{\tbar b \phipi} Y_{\bbar t\phimj} +
   Y_{\bbar t\phimi} Y_{\tbar b \phipj})
   (|Y_{b\chark\stopm^*}|^2 + |Y_{\bbar\chark\stopm}|^2)
   m_b m_t \propV_{\Fbar F\Fbar FS}(t,b,b,\chark,\stopm)
\nonumber \\ &&
+ (Y_{\tbar b \phipi} Y_{\bbar t\phimj} 
   Y_{b\chark\stopm^*}^* Y_{\bbar\chark\stopm}^*
  +Y_{\bbar t\phimi} Y_{\tbar b \phipj} 
   Y_{\bbar\chark\stopm} Y_{b\chark\stopm^*})
   m_t m_{\chark} \propV_{\Fbar FF\Fbar S}(t,b,b,\chark,\stopm)
\nonumber \\ &&
+ 2 (Y_{\tbar b\phipi} Y_{\tbar b\phipj} +
     Y_{\bbar t\phimi} Y_{\bbar t\phimj})
     {\rm Re}[Y_{b\chark\stopm^*} Y_{\bbar\chark\stopm}]
   m_b m_{\chark} \propV_{FF\Fbar\Fbar S}(t,b,b,\chark,\stopm)
\nonumber \\ &&
+ (Y_{\tbar b \phipi} Y_{\bbar t\phimj} 
   Y_{b\chark\stopm^*} Y_{\bbar\chark\stopm}
  +Y_{\bbar t\phimi} Y_{\tbar b \phipj} 
   Y_{\bbar\chark\stopm}^* Y_{b\chark\stopm^*}^*)
   m_b^2 m_t m_{\chark} \propV_{\Fbar\Fbar\Fbar\Fbar S}(t,b,b,\chark,\stopm)
\Bigr \rbrace 
\nonumber \\ &&
+ (t \leftrightarrow b, \> \phi^+ \leftrightarrow \phi^-)^*
\>
+ Y_{\taubar \nutau\phimi} Y_{\taubar \nutau\phimj} \Bigl \lbrace
  |Y_{\taubar\chark\snutau}|^2
  \propV_{FFFFS}(0,\tau,\tau,\chark,\snutau)
\nonumber \\ &&
+|Y_{\tau\chark\snutau^*}|^2
  m_\tau^2 \propV_{F\Fbar\Fbar FS}(0,\tau,\tau,\chark,\snutau)
+ 2 {\rm Re}[Y_{\tau\chark\snutau^*} Y_{\taubar\chark\snutau}]
   m_\tau m_{\chark} \propV_{FF\Fbar\Fbar S}(0,\tau,\tau,\chark,\snutau)
\nonumber \\ &&
+ |Y_{\nutau\chark\staum^*}|^2 \propV_{FFFFS}(\tau,0,0,\chark,\staum) 
\Bigr \rbrace
,
\\
\Pi^{(2),9}_{\phipi \phimj} &=& n_t \lambda_{\phipi \phimj \stopm 
\stopn^*} 
  \Bigl [\bigl (Y_{t \neutk \stopm^*} Y_{t \neutk \stopn^*}^*
      + Y_{\overline t \neutk \stopm}^* Y_{\overline t \neutk \stopn}
      \bigr ) \propW_{SSFF}(\stopm,\stopn,t,\neutk)
\nonumber \\ && 
  +\bigl ( Y_{t \neutk \stopm^*} Y_{\overline t \neutk \stopn}
       + Y_{\overline t \neutk \stopm}^* Y_{t \neutk \stopn^*}^* \bigr )
       m_t m_{\neutk} \propW_{SS\Fbar \Fbar}(\stopm,\stopn,t,\neutk)
  \Bigr ] + (t \rightarrow b) + (t \rightarrow \tau)
\nonumber \\ &&
+ \lambda_{\phipi \phimj \snutau \snutau^*} |Y_{\nu_\tau \neutk \snutau^*}|^2
      \propW_{SSFF}(\snutau,\snutau,0,\neutk)
,
\\
\Pi^{(2),10}_{\phipi \phimj} &=&
3 \Bigl \lbrace
  \lambda_{\phipi \sbotm\stopn^*} \lambda_{\phipj \sbotm \stopp^*}^*
   [(Y_{t\neutk\stopn^*}^* Y_{t\neutk\stopp^*}   
   +Y_{\tbar\neutk\stopn} Y_{\tbar\neutk\stopp}^*) 
   \propV_{SSSFF}(\sbotm,\stopn,\stopp,t,\neutk) 
\nonumber \\ &&
   +(Y_{t\neutk\stopn^*}^* Y_{\tbar\neutk\stopp}^*   
   +Y_{\tbar\neutk\stopn} Y_{t\neutk\stopp^*}) 
   m_t m_{\neutk} \propV_{SSS\Fbar\Fbar}(\sbotm,\stopn,\stopp,t,\neutk)] 
\nonumber \\ &&
+  \lambda_{\phipi \sbotn\stopm^*} \lambda_{\phipj \sbotp \stopm^*}^*
   [(Y_{b\neutk\sbotn^*} Y_{b\neutk\sbotp^*}^*   
   +Y_{\bbar\neutk\sbotn}^* Y_{\bbar\neutk\sbotp}) 
   \propV_{SSSFF}(\stopm,\sbotn,\sbotp,b,\neutk) 
\nonumber \\ &&
   +(Y_{b\neutk\sbotn^*} Y_{\bbar\neutk\sbotp}   
   +Y_{\bbar\neutk\sbotn}^* Y_{b\neutk\sbotp^*}^*) 
   m_b m_{\neutk} \propV_{SSS\Fbar\Fbar}(\stopm,\sbotn,\sbotp,b,\neutk)] 
\Bigr \rbrace
\nonumber \\ &&
+  \lambda_{\phipi \staun\snutau^*} \lambda_{\phipj \staup \snutau^*}^*
   [(Y_{\tau\neutk\staun^*} Y_{\tau\neutk\staup^*}^*   
   +Y_{\taubar\neutk\staun}^* Y_{\taubar\neutk\staup}) 
   \propV_{SSSFF}(\snutau,\staun,\staup,\tau,\neutk) 
\nonumber \\ &&
   +(Y_{\tau\neutk\staun^*} Y_{\taubar\neutk\staup}   
   +Y_{\taubar\neutk\staun}^* Y_{\tau\neutk\staup^*}^*) 
   m_\tau m_{\neutk} 
   \propV_{SSS\Fbar\Fbar}(\snutau,\staun,\staup,\tau,\neutk)] 
\nonumber \\ &&
+\lambda_{\phipi \staum\snutau^*} \lambda_{\phipj \staum \snutau^*}^*
   |Y_{\nu_\tau\neutk\snutau^*}|^2  
   \propV_{SSSFF}(\staum,\snutau,\snutau,0,\neutk) 
,
\\
\Pi^{(2),11}_{\phipi\phimj} &=&
3 \Bigl \lbrace
\lambda_{\phipi \phimj \sbotm \sbotn^*}
\Bigl [ \bigl (Y_{t \chark \sbotm^*} Y_{t \chark \sbotn^*}^*
         + Y_{\overline t \chark \sbotm}^* Y_{\overline t \chark \sbotn}
        \bigr ) \propW_{SSFF}(\sbotm,\sbotn,t,\chark)
\nonumber \\ &&
+ \bigl ( Y_{t \chark \sbotm^*} Y_{\overline t \chark \sbotn}
        + Y_{\overline t \chark \sbotm}^* Y_{t \chark \sbotn^*}^* \bigr )
         m_t m_{\chark} \propW_{SS\Fbar \Fbar}(\sbotm,\sbotn,t,\chark)
\Bigr ]
\Bigr \rbrace
+ (t \leftrightarrow b)
\nonumber \\ &&
+ \lambda_{\phipi \phimj \snutau \snutau^*}
    \Bigl [ \bigl (|Y_{\tau \chark \snutau^*}|^2
         + |Y_{\overline \tau  \chark \snutau}|^2
        \bigr ) \propW_{SSFF}(\snutau,\snutau,\tau,\chark)
\nonumber \\ &&
+ 2 {\rm Re}[Y_{\tau \chark \snutau^*} Y_{\overline \tau \chark \snutau}]
         m_\tau m_{\chark}
         \propW_{SS\Fbar\Fbar}(\snutau,\snutau,\tau,\chark)
\Bigr ]
\nonumber \\ &&
+ \lambda_{\phipi \phimj \staum \staun^*}
           Y_{\nu_\tau \chark \staum^*} Y_{\nu_\tau \chark \staun^*}^*
           \propW_{SSFF}(\staum,\staun,0,\chark) 
,
\\
\Pi^{(2),12}_{\phipi\phimj} &=& 3 \Bigl \lbrace
\lambda_{\phipi \sbotm \stopn^*} \lambda_{\phipj \sbotm\stopp^*}^* [
    (Y_{b\chark\stopn^*}^* Y_{b\chark \stopp^*} 
    +Y_{\bbar\chark\stopn} Y_{\bbar\chark \stopp}^*)
    \propV_{SSSFF}(\sbotm,\stopn,\stopp,b,\chark)
\nonumber \\ &&
   +(Y_{\bbar\chark\stopn} Y_{b\chark \stopp^*} 
    +Y_{b\chark\stopn^*}^* Y_{\bbar\chark \stopp}^*)
    m_b m_{\chark} \propV_{SSS\Fbar\Fbar} (\sbotm,\stopn,\stopp,b,\chark)]
\nonumber \\ &&
+ \lambda_{\phipi\sbotn\stopm^*} \lambda_{\phipj\sbotp\stopm^*}^* [
  (Y_{t\chark\sbotn^*} Y_{t\chark\sbotp^*}^* 
  +Y_{\tbar\chark\sbotn}^* Y_{\tbar\chark\sbotp})
   \propV_{SSSFF}(\stopm,\sbotn,\sbotp,t,\chark)
\nonumber \\ &&
 +(Y_{t\chark\sbotn^*} Y_{\tbar\chark\sbotp} 
  +Y_{\tbar\chark\sbotn}^* Y_{t\chark\sbotp^*}^*)
   m_t m_{\chark} \propV_{SSS\Fbar\Fbar}(\stopm,\sbotn,\sbotp,t,\chark)]
\Bigr \rbrace
\nonumber \\ &&
+ \lambda_{\phipi \staum \snutau^*} \lambda_{\phipj \staum\snutau^*}^* [
    (|Y_{\tau\chark\snutau^*}|^2 +|Y_{\taubar\chark\snutau}|^2 )
    \propV_{SSSFF}(\staum,\snutau,\snutau,\tau,\chark)
\nonumber \\ &&
   +2 {\rm Re}[Y_{\taubar\chark\snutau} Y_{\tau\chark \snutau^*}]
    m_\tau m_{\chark} 
    \propV_{SSS\Fbar\Fbar} (\staum,\snutau,\snutau,\tau,\chark)]
\nonumber \\ &&
+ \lambda_{\phipi\staun\snutau^*} \lambda_{\phipj\staup\snutau^*}^* 
  Y_{\nu_\tau\chark\staun^*} Y_{\nu_\tau\chark\staup^*}^*
   \propV_{SSSFF}(\snutau,\staun,\staup,0,\chark) .
\end{eqnarray}
Contributions involving virtual Higgs scalar bosons and third-family 
fermions are:
\beq
\Pi^{(2),13}_{\phipi\phimj} &=& 3 \Bigl \lbrace
(Y_{\tbar b\phipi} Y_{\tbar b\phipj} +
     Y_{\bbar t\phimi} Y_{\bbar t\phimj}) |Y_{t\tbar\phiOk}|^2
     [\propV_{FFFFS}(b,t,t,t,\phiOk)
     + m_t^2 \propV_{F\Fbar\Fbar FS}(b,t,t,t,\phiOk)]
\nonumber \\ &&
+2 (Y_{\tbar b\phipi} Y_{\bbar t\phimj} +
     Y_{\bbar t\phimi} Y_{\tbar b \phipj})
     |Y_{t\tbar\phiOk}|^2 m_b m_t
     \propV_{\Fbar F\Fbar FS}(b,t,t,t,\phiOk)
\nonumber \\ &&
+[ Y_{\tbar b\phipi} Y_{\bbar t\phimj} (Y_{t\tbar\phiOk}^*)^2 
    + Y_{\bbar t\phimi} Y_{\tbar b\phipj} (Y_{t\tbar\phiOk})^2]
    m_b m_t \propV_{\Fbar FF\Fbar S}(b,t,t,t,\phiOk)
\nonumber \\ &&
+2 [ Y_{\tbar b\phipi} Y_{\tbar b\phipj}  
    + Y_{\bbar t\phimi} Y_{\bbar t\phimj}] 
    {\rm Re}[(Y_{t\tbar\phiOk})^2]
    m_t^2 \propV_{FF\Fbar\Fbar S}(b,t,t,t,\phiOk)
\nonumber \\ &&
+[ Y_{\tbar b\phipi} Y_{\bbar t\phimj} (Y_{t\tbar\phiOk})^2 
    + Y_{\bbar t\phimi} Y_{\tbar b\phipj} (Y_{t\tbar\phiOk}^*)^2]
    m_b m_t^3 \propV_{\Fbar\Fbar\Fbar\Fbar S}(b,t,t,t,\phiOk)
\Bigr \rbrace
+ (t \leftrightarrow b, \> \phi^+ \leftrightarrow \phi^-)^*
\nonumber \\ &&
+ Y_{\taubar \nutau\phimi} Y_{\taubar \nutau\phimj}  
     |Y_{\tau\taubar\phiOk}|^2
     [\propV_{FFFFS}(0,\tau,\tau,\tau,\phiOk)
     + m_\tau^2 \propV_{F\Fbar\Fbar FS}(0,\tau,\tau,\tau,\phiOk)]
\nonumber \\ &&
+2 Y_{\taubar \nutau\phimi} Y_{\taubar \nutau\phimj} 
    {\rm Re}[(Y_{\tau\taubar\phiOk})^2] 
    m_\tau^2 \propV_{FF\Fbar\Fbar S}(0,\tau,\tau,\tau,\phiOk)
,
\\
\Pi^{(2),14}_{\phipi\phimj} &=& 3 \Bigl \lbrace
[Y_{\tbar b\phipi} Y_{\tbar b\phipj} Y_{\tbar b\phipk}^2
   + Y_{\bbar t\phimi} Y_{\bbar t\phimj} Y_{\bbar t\phimk}^2]
   \propV_{FFFFS}(t,b,b,t,\phipk)
\nonumber \\ &&
+ [Y_{\tbar b\phipi} Y_{\tbar b\phipj} Y_{\bbar t\phimk}^2
  +Y_{\bbar t\phimi} Y_{\bbar t\phimj} Y_{\tbar b\phipk}^2]
  m_b^2 \propV_{F\Fbar\Fbar FS}(t,b,b,t,\phipk)
\nonumber \\ &&
+ [Y_{\tbar b\phipi} Y_{\bbar t\phimj} +
   Y_{\bbar t\phimi} Y_{\tbar b\phipj}][
   Y_{\tbar b\phipk}^2 + Y_{\bbar t\phimk}^2]
   m_b m_t \propV_{\Fbar F\Fbar FS}(t,b,b,t,\phipk)
\nonumber \\ &&
+ [Y_{\tbar b\phipi} Y_{\bbar t\phimj} +
   Y_{\bbar t\phimi} Y_{\tbar b\phipj}] 
   Y_{\tbar b\phipk} Y_{\bbar t\phimk} 
   m_t^2 [\propV_{\Fbar FF\Fbar S}(t,b,b,t,\phipk)
   +m_b^2 \propV_{\Fbar\Fbar\Fbar\Fbar S}(t,b,b,t,\phipk)]
\nonumber \\ &&
+ 2[Y_{\tbar b\phipi} Y_{\tbar b\phipj} +
    Y_{\bbar t\phimi} Y_{\bbar t\phimj}] 
    Y_{\tbar b\phipk} Y_{\bbar t\phimk}
    m_b m_t \propV_{FF\Fbar\Fbar S}(t,b,b,t,\phipk)
\Bigr \rbrace + (t\leftrightarrow b,\,\phi^+\leftrightarrow\phi^-)
\nonumber \\ &&
+ Y_{\taubar \nutau \phimi} Y_{\taubar \nutau\phimj} 
   Y_{\taubar \nutau\phimk}^2
   [\propV_{FFFFS}(0,\tau,\tau,0,\phipk) +
    \propV_{FFFFS}(\tau,0,0,\tau,\phipk)]
,
\\
\Pi^{(2),15}_{\phipi\phimj} &=& 3 \Bigl \lbrace
(Y_{\tbar b\phipi} Y_{\bbar t\phimj} 
       Y_{t\tbar \phiOk}^* Y_{b\bbar \phiOk}^*
      +Y_{\bbar t\phimi} Y_{\tbar b\phipj} 
       Y_{t\tbar \phiOk} Y_{b\bbar \phiOk})
       \propM_{FFFFS}(t,t,b,b,\phiOk)
\nonumber \\ &&
+2(Y_{\tbar b\phipi} Y_{\tbar b\phipj} 
      +Y_{\bbar t\phimi} Y_{\bbar t\phimj}) 
       {\rm Re}[Y_{t\tbar \phiOk} Y_{b\bbar \phiOk}]
       m_b m_t \propM_{F\Fbar F\Fbar S}(t,t,b,b,\phiOk)
\nonumber \\ &&
+2(Y_{\tbar b\phipi} Y_{\tbar b\phipj} 
      +Y_{\bbar t\phimi} Y_{\bbar t\phimj}) 
       {\rm Re}[Y_{t\tbar \phiOk} Y_{b\bbar \phiOk}^*]
       m_b m_t \propM_{F\Fbar\Fbar FS}(t,t,b,b,\phiOk)
\nonumber \\ &&
+(Y_{\tbar b\phipi} Y_{\bbar t\phimj} 
       Y_{t\tbar \phiOk}^* Y_{b\bbar \phiOk}
      +Y_{\bbar t\phimi} Y_{\tbar b\phipj} 
       Y_{t\tbar \phiOk} Y_{b\bbar \phiOk}^*)
       m_b^2 \propM_{FF\Fbar\Fbar S}(t,t,b,b,\phiOk)
\nonumber \\ &&
+(Y_{\tbar b\phipi} Y_{\bbar t\phimj} 
       Y_{t\tbar \phiOk} Y_{b\bbar \phiOk}^*
      +Y_{\bbar t\phimi} Y_{\tbar b\phipj} 
       Y_{t\tbar \phiOk}^* Y_{b\bbar \phiOk})
       m_t^2 \propM_{FF\Fbar\Fbar S}(b,b,t,t,\phiOk)
\nonumber \\ &&
+(Y_{\tbar b\phipi} Y_{\bbar t\phimj} 
       Y_{t\tbar \phiOk} Y_{b\bbar \phiOk}
      +Y_{\bbar t\phimi} Y_{\tbar b\phipj} 
       Y_{t\tbar \phiOk}^* Y_{b\bbar \phiOk}^*)
       m_b^2 m_t^2 \propM_{\Fbar\Fbar\Fbar\Fbar S}(t,t,b,b,\phiOk)
\Bigr \rbrace
.
\eeq
Contributions involving virtual Higgs scalars and third-family sfermions
are:
\beq
\Pi^{(2),16}_{\phipi\phimj} &=& n_{\tilde t} \Bigl \lbrace
\Bigr (
\lambda_{\phipi\sbotn\stopm^*} [
  \lambda_{\phipk\phimj\stopm\stopp^*}
  \lambda_{\phipk\sbotn\stopp^*}^* 
  \propU_{SSSS}(\stopm,\sbotn,\stopp,\phipk)
+ \lambda_{\phiOk\phipj\sbotp\stopm^*}^*
  \lambda_{\phiOk\sbotp\sbotn^*} 
  \propU_{SSSS}(\stopm,\sbotn,\sbotp,\phiOk)]
\nonumber \\ &&
+ \lambda_{\phipi\sbotm\stopn^*} [
  \lambda_{\phipk\phimj\sbotp\sbotm^*}
  \lambda_{\phipk\sbotp\stopn^*}^*
   \propU_{SSSS}(\sbotm,\stopn,\sbotp,\phipk)
+  \lambda_{\phiOk\phipj\sbotm\stopp^*}^*
  \lambda_{\phiOk\stopn\stopp^*}
   \propU_{SSSS}(\sbotm,\stopn,\stopp,\phiOk)] \Bigl )
\nonumber \\ &&
+ (i\leftrightarrow j)^*
\>
+\lambda_{\phipi\phimk\stopm\stopn^*} \lambda_{\phipk\phimj\stopn\stopm^*}
  \propS_{SSS}(\phipk,\stopm,\stopn)
+\lambda_{\phipi\phimk\sbotm\sbotn^*} \lambda_{\phipk\phimj\sbotn\sbotm^*}
  \propS_{SSS}(\phipk,\sbotm,\sbotn)
\nonumber \\ &&
+ \lambda_{\phiOk\phipi\sbotm\stopn^*} 
  \lambda_{\phiOk\phipj\sbotm\stopn^*}^*
  \propS_{SSS}(\phiOk,\sbotm,\stopn)
+ \lambda_{\phipi\sbotm\stopp^*} \lambda_{\phipj\sbotn\stopq^*}^*
  \lambda_{\phiOk\sbotn\sbotm^*} \lambda_{\phiOk\stopp\stopq^*}
  \propM_{SSSSS}(\sbotm,\sbotn,\stopp,\stopq,\phiOk)
\nonumber \\ &&
+ \lambda_{\phipi\phimj\stopm\stopn^*} \bigl [
  \lambda_{\phiOk\stopn\stopp^*}
  \lambda_{\phiOk\stopp\stopm^*} 
  \propW_{SSSS}(\stopm,\stopn,\stopp,\phiOk)
+ \lambda_{\phipk\sbotp\stopm^*}
  \lambda_{\phipk\sbotp\stopn^*}^* 
  \propW_{SSSS}(\stopm,\stopn,\sbotp,\phipk)
\nonumber \\ &&
+ \lambda_{\phiOk\phiOk\stopn\stopm^*} 
  \propX_{SSS}(\stopm,\stopn,\phiOk)/2 
+ \lambda_{\phipk\phimk\stopn\stopm^*} 
  \propX_{SSS}(\stopm,\stopn,\phipk) \bigr ]
\nonumber \\ &&
+ \lambda_{\phipi\phimj\sbotm\sbotn^*} \bigl [
  \lambda_{\phiOk\sbotn\sbotp^*}
  \lambda_{\phiOk\sbotp\sbotm^*} 
  \propW_{SSSS}(\sbotm,\sbotn,\sbotp,\phiOk)
+ \lambda_{\phipk\sbotn\stopp^*}
  \lambda_{\phipk\sbotm\stopp^*}^* 
  \propW_{SSSS}(\sbotm,\sbotn,\stopp,\phipk)
\nonumber \\ &&
+ \lambda_{\phiOk\phiOk\sbotn\sbotm^*} 
  \propX_{SSS}(\sbotm,\sbotn,\phiOk)/2 +
  \lambda_{\phipk\phimk\sbotn\sbotm^*} 
  \propX_{SSS}(\sbotm,\sbotn,\phipk)\bigr ]
\nonumber \\ &&
+ \lambda_{\phipi\sbotn\stopm^*} \lambda_{\phipj\sbotp\stopm^*}^*
  \bigl [\lambda_{\phiOk\sbotp\sbotq^*}\lambda_{\phiOk\sbotq\sbotn^*}
  \propV_{SSSSS}(\stopm,\sbotn,\sbotp,\sbotq,\phiOk)
 +\lambda_{\phipk\sbotp\stopq^*}\lambda_{\phipk\sbotn\stopq^*}^*
  \propV_{SSSSS}(\stopm,\sbotn,\sbotp,\stopq,\phipk)
\nonumber \\ &&
+ \lambda_{\phiOk\phiOk\sbotp\sbotn^*}
  \propY_{SSSS}(\stopm,\sbotn,\sbotp,\phiOk)/2
  +\lambda_{\phipk\phimk\sbotp\sbotn^*}
  \propY_{SSSS}(\stopm,\sbotn,\sbotp,\phipk) \bigr ]
\nonumber \\ &&
+ \lambda_{\phipi\sbotm\stopn^*} \lambda_{\phipj\sbotm\stopp^*}^*
  \bigl [\lambda_{\phiOk\stopn\stopq^*}\lambda_{\phiOk\stopq\stopp^*}
  \propV_{SSSSS}(\sbotm,\stopn,\stopp,\stopq,\phiOk)
 +\lambda_{\phipk\sbotq\stopp^*}\lambda_{\phipk\sbotq\stopn^*}^*
  \propV_{SSSSS}(\sbotm,\stopn,\stopp,\sbotq,\phipk)
\nonumber \\ &&
  +\lambda_{\phiOk\phiOk\stopn\stopp^*}
  \propY_{SSSS}(\sbotm,\stopn,\stopp,\phiOk)/2
  +\lambda_{\phipk\phimk\stopn\stopp^*}
  \propY_{SSSS}(\sbotm,\stopn,\stopp,\phipk) \bigr ]
\Bigr\rbrace
+ (\tilde t \rightarrow \snutau, \tilde b \rightarrow \tilde \tau)
.
\eeq
Finally, contributions involving only virtual sfermions are given by:
\begin{eqnarray}
\Pi^{(2),17}_{\phipi\phimj} &=& 
\lambda_{\phipi\phimj\sfermionk\sfermionm^*} (
          n_{\sfermionk} n_{\sfermionn}
          \lambda_{\sfermionm \sfermionk^* \sfermionn \sfermionn^*}
        + n_{\sfermionk} 
          \lambda_{\sfermionm \sfermionn^* \sfermionn \sfermionk^*})
          \propX_{SSS}(\sfermionk,\sfermionm,\sfermionn)
\nonumber \\ &&
+(
\lambda_{\phipi\sfermionk\sfermionm^*}
       \lambda_{\phipj\sfermionk\sfermionn^*}^* 
   +\lambda_{\phipi\sfermionn\sfermionk^*} 
       \lambda_{\phipj\sfermionm\sfermionk^*}^* )(
       n_{\sfermionk} n_{\sfermionp} 
       \lambda_{\sfermionm\sfermionn^*\sfermionp\sfermionp^*} 
       +n_{\sfermionk}
       \lambda_{\sfermionm\sfermionp^*\sfermionp\sfermionn^*})
          \propY_{SSSS}(\sfermionk,\sfermionm,\sfermionn,\sfermionp)
\nonumber \\ &&
+ \lambda_{\phipi \sfermionk \sfermionm^*}
  \lambda_{\phipj \sfermionn \sfermionp^*}^* 
  (n_{\sfermionk} n_{\sfermionn} 
  \lambda_{\sfermionm \sfermionk^* \sfermionn \sfermionp^*}
  + n_{\sfermionk} 
  \lambda_{\sfermionm \sfermionp^* \sfermionn \sfermionk^*})
  \propZ_{SSSS}(\sfermionk,\sfermionm,\sfermionn,\sfermionp) .
\label{eq:Hcharallsfermions}
\end{eqnarray}
This expression includes the contributions for the sfermions of the first 
two families, which only have gauge interactions. In the numerical
results of section \ref{sec:examples}, only the third-family 
sfermion contributions from eq.~(\ref{eq:Hcharallsfermions}) are included.

\section{Discussion and numerical examples}\label{sec:examples}
\setcounter{equation}{0}

I have carried out several checks on the above expressions. First, the
quark/gluon two-loop contributions to the Higgs scalar boson self-energies
had already been computed in refs.~\cite{Kniehl:1994ph,Djouadi:1994gf}. I
have checked that my corresponding results, namely the $G_{FF}$ and
$G_{\Fbar\Fbar}$ terms in eqs.~(\ref{eq:phiOgluon}) and
(\ref{eq:phipmgluon}) of the present paper, agree with these, by
converting from the on-shell scheme used there into the \DRbarprime
scheme used here.

Second, I have verified that the self-energy contributions listed above
for $h^0, H^0$, when evaluated in the limit $s\rightarrow 0$, do
correspond precisely to the second derivatives with respect to $v_u, 
v_d$ of the appropriate terms in the
two-loop effective potential \cite{effpotMSSM}, according to:
\beq
\Pi^{(2)}(0) = \frac{1}{2}
\begin{pmatrix} c_\alpha & -s_\alpha \cr s_\alpha & c_\alpha
\end{pmatrix}
\begin{pmatrix}
\partial^2 V^{(2)}/\partial v_u^2 
&
\partial^2 V^{(2)}/\partial v_u \partial v_d 
\cr
\partial^2 V^{(2)}/\partial v_u \partial v_d 
&
\partial^2 V^{(2)}/\partial v_d^2 
\end{pmatrix}
\begin{pmatrix} c_\alpha & s_\alpha \cr -s_\alpha & c_\alpha
\end{pmatrix}
,
\label{eq:PIEP}
\eeq
\end{widetext}
where 
$c_\alpha = \cos\alpha$, $s_\alpha = \sin\alpha$, and
the two-loop effective potential is
\beq
V_{\rm eff} = V^{(0)} 
+ \frac {1}{16 \pi^2} V^{(1)}
+ \frac {1}{(16 \pi^2)^2} V^{(2)}
+ \ldots .
\eeq

Third, I have checked that the renormalization group scale
invariance of the pole masses is consistent with the known two-loop
renormalization group equations for the Lagrangian
parameters \cite{Martin:1993zk,Yamada:1994id,Jack:1994kd,DRbarprime} 
and VEVs \cite{effpotMSSM}.
These checks are quite involved, but follow the same pattern as given
explicitly in the toy model of section VI of \cite{Martin:2003it}.

Finally, there are non-realistic limits of the MSSM in which a 
global 
$SU(2)$ symmetry implies the equality of masses and
self-energies of the charged Higgs scalar bosons $G^\pm, H^\pm$ with two 
of the
neutral scalars. This occurs for $y_t=y_b$, $a_t=a_b$, $m_{H_u}^2 =
m_{H_d}^2$, $m_{u_i}^2 = m_{d_i}^2$, and either:
\beq
\mbox{case 1:}&\qquad &v_u=v_d\not=0,\qquad\> g = g'=0; \nonumber\\
\mbox{case 2:}&\qquad &v_u=v_d=0, \qquad \>g,g'\not = 0, \nonumber
\eeq
and neglecting all slepton contributions.
I have checked that in each case, the required equality between neutral
and charged Higgs scalars for the self-energy contributions of sections
\ref{sec:neutraltwoloop} and \ref{sec:chargedtwoloop} indeed occurs.

The most important application of the results above is probably to the
calculation of the ``momentum-dependent" contributions to the pole mass of 
the lightest scalar Higgs boson, $h^0$. 
Before reporting some numerical examples, it seems worthwhile to 
illustrate the role and rough size of the effects with a simple limiting 
case that can be 
treated analytically. Consider the degenerate decoupling limit in which
the top squarks and the gluino have the same mass $M$, with $s \ll m_t^2 
\ll M^2$, and with all bottom, tau, and electroweak
effects neglected. 
Then, at one 
loop order:
\beq
\Pi^{(1)}_{h^0h^0}(s) = y_t^2 
c^2_\alpha
\left [P_1 + s P_1' + \ldots \right ]
\eeq
where
\beq
P_1 &=& 
6 M^2 \bigl [\lnbar M^2 -1 \bigr ] 
-6 m_t^2 \bigl [\lnbar m_t^2 -1 \bigr ]
\nonumber \\ &&
+ 12 {\rm ln}(M^2/m_t^2)
\\
P_1' &=& 3 \lnbar m_t^2 + 2 ,
\eeq
and 
\beq
\lnbar{X} \equiv \mbox{ln}(X/Q^2) ,
\eeq
and we consistently neglect terms of order $m_t^2/M^2$.
Similarly, at two-loop order, we obtain from the results of section
\ref{subsec:neutralstrong}, and the analytical expressions of section 
VI of ref.~\cite{evaluation}, and for simplicity keeping only
terms of order $g_3^2$:
\beq
\Pi^{(2)}_{h^0h^0}(s) = g_3^2 y_t^2 
c^2_\alpha
\left [P_2 + s P_2' + \ldots \right ]
\eeq
where
\beq
P_2 &=& 32 M^2 \bigl [-(\lnbar M^2)^2 + 3 \lnbar M^2 -3\bigr ] 
\nonumber \\ &&
+ 16 m_t^2 
\bigl [
9(\lnbar m_t^2)^2 -9 \lnbar m_t^2 + 5 
\nonumber \\ &&
-2 (\lnbar M^2)^2
-6 \lnbar M^2 \lnbar m_t^2 + 5 \lnbar M^2 \bigr ],
\\
P_2' &=& -12 (\lnbar m_t^2)^2 -12 \lnbar m_t^2 + \frac{44}{3} 
\nonumber \\ &&
-4 (\lnbar M^2)^2 + \frac{4}{3} \lnbar M^2 
+ 8 \lnbar M^2 \lnbar m_t^2 .
\eeq
Now, including the tree-level contribution to the squared mass, one 
can use the condition
\beq
{\partial V_{\rm eff}}/{\partial v_u}  = 0
\eeq
to eliminate the terms proportional to $M^2$ in the expression for the
pole squared-mass. One then finds:
\beq
m^2_{h^0,{\rm pole}} &=& m_Z^2 \cos^2(2\beta) 
+ \frac{y_t^2}{16\pi^2} c_\alpha^2 \left [
m_t^2 \Delta_1 + m_{h^0}^2 \Delta_1' \right ]
\nonumber \\ &&
+ \frac{g_3^2 y_t^2}{(16\pi^2)^2} c_\alpha^2 \left [
m_t^2 \Delta_2 + m_{h^0}^2 \Delta_2' \right ],
\eeq
neglecting terms of order
$y_t^4$ and $m^4_{h^0}/m_t^2$, 
with 
\beq
\Delta_1 &=& 12 {\rm ln}(M^2/m_t^2),
\\
\Delta_1' &=& P_1',
\\
\Delta_2 &=& 32[ 3 (\lnbar m_t^2)^2 - \lnbar m_t^2 -1
- (\lnbar M^2)^2 
\nonumber \\ &&
- 2 \lnbar M^2 \lnbar m_t^2 + \lnbar M^2 ] ,
\\
\Delta_2' &=& P_2' .
\eeq
Choosing the renormalization scale $Q=M$, 
\beq
\Delta_1 &=& 12 L,\\
\Delta_1' &=&  2 - 3 L,\\
\Delta_2 &=& 96 L^2 + 32 L -32,\\
\Delta_2' &=& -12 L^2 + 12 L +44/3 ,
\eeq
where $L = {\rm ln}(M^2/m_t^2)$, and as usual the masses and $y_t$ and 
$g_3$
are \DRbarprime~ couplings in the MSSM (with the top quark and the 
superpartners not decoupled).
The terms $\Delta_1$ and $\Delta_2$ agree with the results obtained in 
eq.~(21) of ref.~\cite{Espinosa:1999zm}.
The last term, $\Delta_2'$, is a consequence of the new result obtained 
under much more 
general circumstances in this paper. However, even in this crude limit 
(which neglects the important ingredients of top squark mixing and mass 
hierarchy), we can see 
that it is smaller than one might perhaps have expected. This is both 
because the dimensionless number coefficients in the
$\Delta_2'$ term are smaller than those 
in the $\Delta_2$ term, and because there is a significant
cancellation between the leading logarithm squared term and the sub-leading
logarithm and constant term in $\Delta_2'$. Indeed, the 
leading-logarithm approximation to $\Delta_2'$ is clearly quite poor
unless $M$ is over 1 TeV.

For more precise results in realistic models, it is necessary to 
keep all of the terms in the two-loop self-energy, and evaluate
the integrals numerically. When computing the pole mass of
$h^0$, it is best to use the following trick for approximating
the full two-loop self-energy. 
Denote by 
$\Pi^{(2)}_{\rm par}(s)$ the sum of the $2\times 2$ matrix
self-energy contributions for the 
neutral Higgs scalars $h^0, H^0$ found in section 
\ref{sec:neutraltwoloop}.
(From here on I only apply the general results above to specific
examples without CP violation.)
Then we use the following expression for the two-loop self-energy:
\beq
\Pi^{(2)}(s) \approx  \Pi^{(2)}_{\rm par}(s) - \Pi^{(2)}_{\rm par}(0)
+ \Pi^{(2)}(0),
\label{eq:approxPI}
\eeq
where the last term is given exactly by eq.~(\ref{eq:PIEP}).
In this way, we include all other two-loop self-energy effects within
the effective potential approximation, 
while avoiding any possibility of
double-counting. Eventually, when all of the remaining diagrams are
calculated, this procedure will not be necessary, of course.

For a specific quasi-realistic numerical example, consider 
the model defined by the following \DRbarprime parameters at a
renormalization group scale
$Q_0 = 640$ GeV:
\begin{eqnarray}
&&
g' = 0.36,\>\> g=0.65,\>\> g_3 = 1.06, \>\>
\nonumber \\  &&
y_t = 0.90,\>\> y_b = 0.13,\>\> y_\tau = 0.10,
\end{eqnarray}
and, in GeV,
\begin{eqnarray}
&&
M_1=150,\>\> M_2 = 280,\>\> M_3 = 800,\>\>\>
\nonumber \\  &&
a_t = -600,\>\> a_b = -150,\>\> a_\tau=-40
\nonumber
\end{eqnarray}
and, in GeV$^2$,
\begin{eqnarray}
&&
m^2_{Q_{1,2}} = (780)^2, \>\,
m^2_{u_{1,2}} = (740)^2, \>\,
m^2_{d_{1,2}} = (735)^2, \>\,
\nonumber \\  &&
m^2_{L_{1,2}} = (280)^2, \>\,
m^2_{e_{1,2}} = (200)^2,
\phantom{xxx}
\nonumber \\
&&
m^2_{Q_3} = (700)^2,\>\>
m^2_{u_3} = (580)^2,\>\>
m^2_{d_3}= (725)^2,\>\>
\nonumber \\  &&
m^2_{L_3}= (270)^2,\>\>
m^2_{e_3} = (195)^2,
\nonumber \\
&&
m^2_{H_u}= -(500)^2,\>\>
m^2_{H_d} = (270)^2.
\label{templateparams}
\end{eqnarray}
The two-loop effective potential is then minimized by:
\begin{equation}
v_u(Q_0) = 172\>\,{\rm GeV};
\qquad\>\>\>
v_d(Q_0) = 17.2\>\,{\rm GeV},
\end{equation}
provided the remaining parameters are:
\begin{equation}
\mu = 504.18112\>\,{\rm GeV},
\qquad b = (184.22026\>\,{\rm GeV})^2 .
\label{templatemub}
\end{equation}

\begin{figure}[tb]
\includegraphics[width=8.6cm]{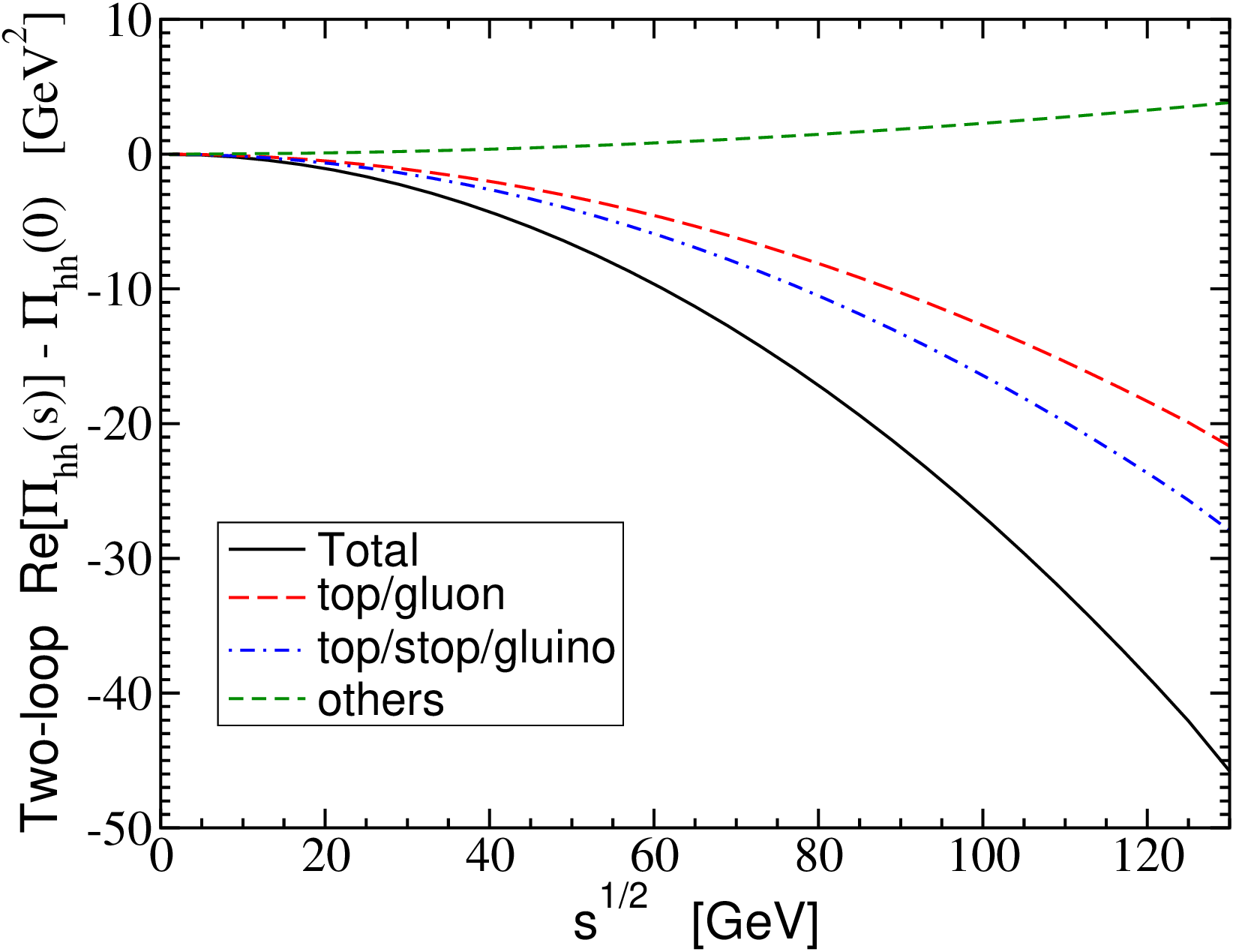}
\caption{\label{fig:Pihh} The two-loop contributions to
Re$[\Pi_{h^0h^0}(s)] - \Pi_{h^0h^0}(0)$ found in section 
\ref{sec:neutraltwoloop},
for the model described in the text, as a function of the 
momentum invariant $s$.}
\end{figure}

Figure \ref{fig:Pihh} shows the two-loop contribution to the quantity 
Re$[\Pi_{h^0h^0}(s)] - \Pi_{h^0h^0}(0)$ in this model, as a function of
$s$.
The solid line is the total calculated in section \ref{sec:neutraltwoloop}
of this paper. Various contributions to this are also shown separately: 
the part 
coming from diagrams involving a top quark loop and a gluon [the $G_{FF}$
and $G_{\Fbar\Fbar}$ terms in
eq.~(\ref{eq:phiOgluon})] 
are shown as the long-dashed line, the part from other diagrams involving 
top (s)quarks and gluinos 
are shown as the dot-dashed line, and all of 
the remaining contributions are lumped together as the short-dashed line.
This shows that, at least for the subset of contributions found in this 
paper, the deviation from the effective potential approximation comes 
mostly from top quark loops involving the strong interactions, as one
might expect. 
The relative proportions from different diagrams varies rather strongly
with the choice of renormalization scale, but the total has only a
small $Q$-dependence.
Diagrams involving only squarks contribute less to the
quantity Re$[\Pi_{h^0h^0}(s)] - \Pi_{h^0h^0}(0)$, because $s\ll m_{\tilde 
q}^2$.

\begin{figure}[tb]
\includegraphics[width=8.6cm]{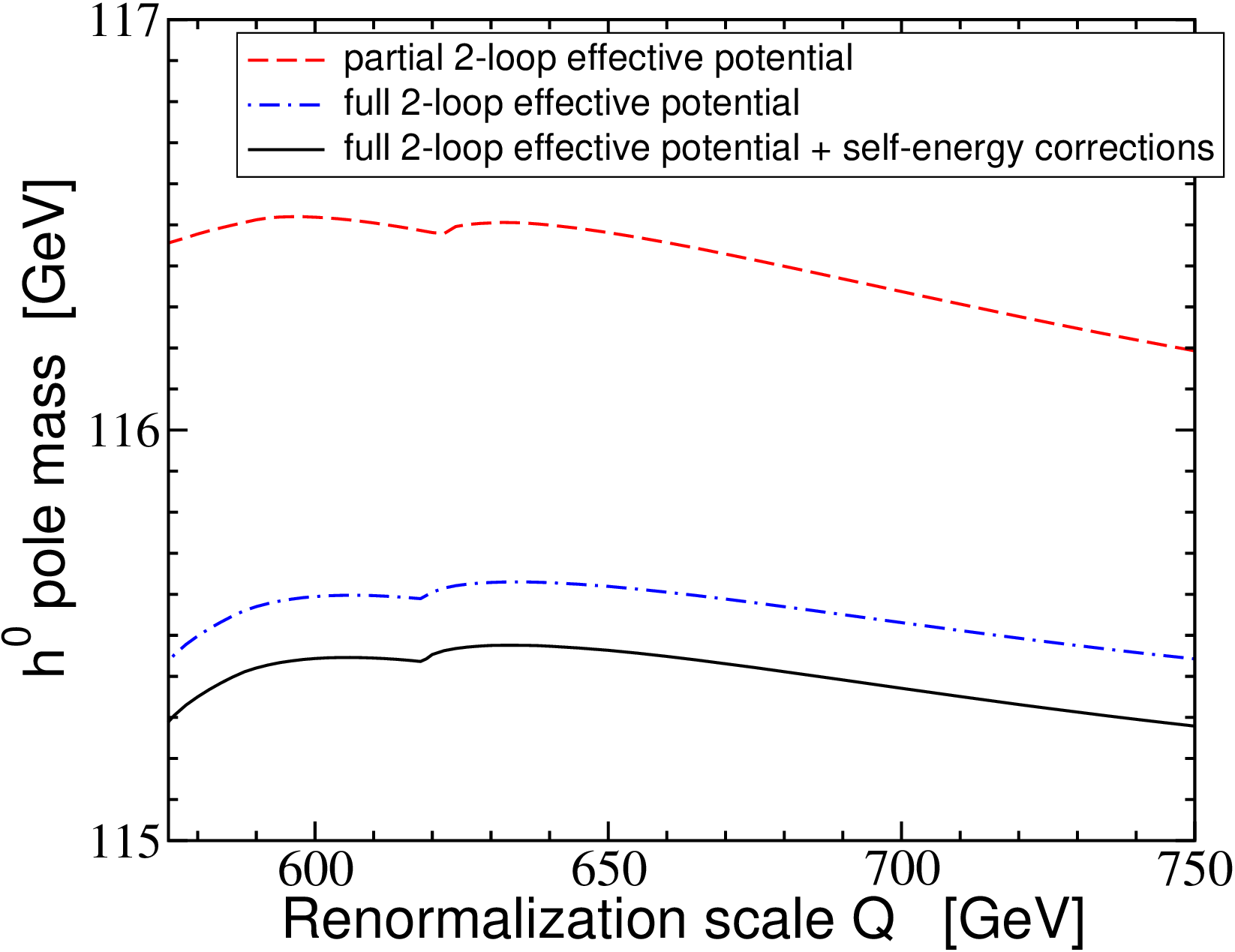}
\caption{\label{fig:hpole} The pole mass of $h^0$, computed in various 
approximations, for the model described in the test, as a function of the
renormalization scale $Q$. 
In each case, the 
full one-loop self-energy is used in
the computation. The dashed line also includes the contributions of the
two-loop self-energy in the effective potential approximation, neglecting 
electroweak couplings. The
dot-dashed line includes the contributions of the full
two-loop self-energy in the effective potential approximation. The
solid line also includes momentum-dependent contributions to the 
self-energy, as found in section \ref{sec:neutraltwoloop}.}
\end{figure}
The resulting pole mass of $h^0$ is shown in figure \ref{fig:hpole},
as a function of the choice\footnote{To
avoid instabilities in the effective potential approximation to 
the self-energy \cite{effpotMSSM}, only choices of $Q$ leading to positive 
Goldstone 
boson tree-level squared
masses are shown; in this model, that requirement limits us to 
$Q > 568$ GeV. This includes the geometric mean 
of the top squark masses, and also the scale where the 
sum of the one-loop and two-loop corrections to $m_{h^0}$ vanishes.}
of renormalization scale $Q$. To 
make this 
graph, all of
the model parameters including the VEVs are evolved using the two-loop 
renormalization group 
equations \cite{Martin:1993zk} from the
defining scale $Q_0 = 640$ GeV to the scale $Q$. 
The 
two-loop effective 
potential is then required to be minimized, determining the values of
$\mu$ and $b$ at that scale. Using these parameters as inputs, the 
dot-dashed line shows the pole mass as calculated in the full 
effective potential approximation, as in ref.~\cite{Martin:2002wn}.
The solid line shows the improved calculation of this paper, using
eq.~(\ref{eq:approxPI}) for the momentum-dependent self-energy. 
(For comparison, the dashed line shows the result within a partial 
two-loop 
effective potential approximation 
\cite{Zhang:1998bm,Espinosa:1999zm,Espinosa:2000df,Degrassi:2001yf}, 
in which all electroweak effects involving $g,g'$
are neglected in the two-loop effective potential.)
We see that including the $s$-dependence in the self-energy lowers the 
prediction for the
pole mass, by only about 160 MeV in this model, and nearly independently
of the choice of renormalization scale.

\begin{figure}[tb!p]
\includegraphics[width=8.6cm]{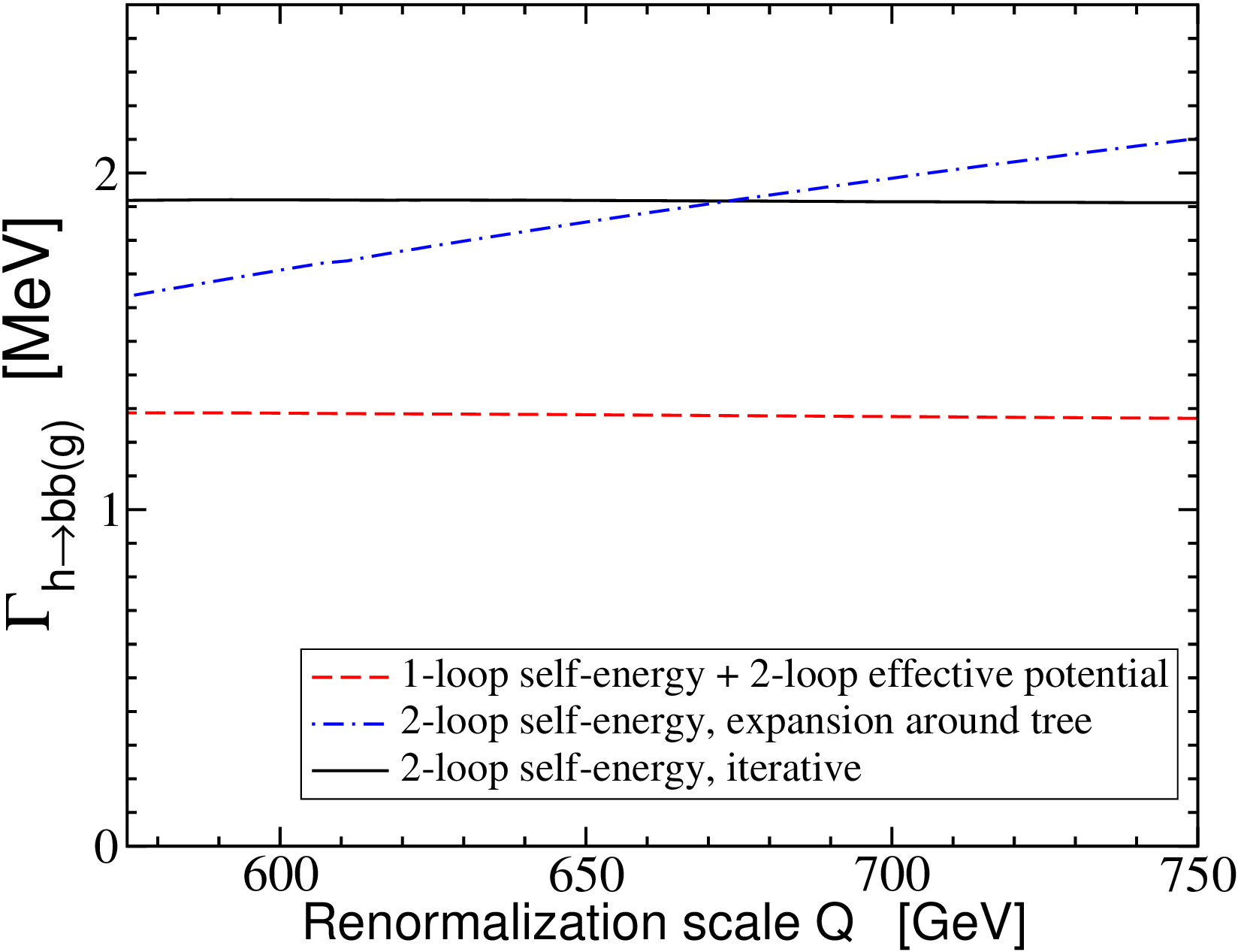}
\caption{\label{fig:gammahpole} The dependence of the $h^0 \rightarrow b 
\overline b (g)$ width, obtained from the corresponding 
contributions to the imaginary part of the pole
mass, as a function of the renormalization scale $Q$, in various
approximations.}
\end{figure}
\begin{figure*}[t]
\includegraphics[width=8.6cm]{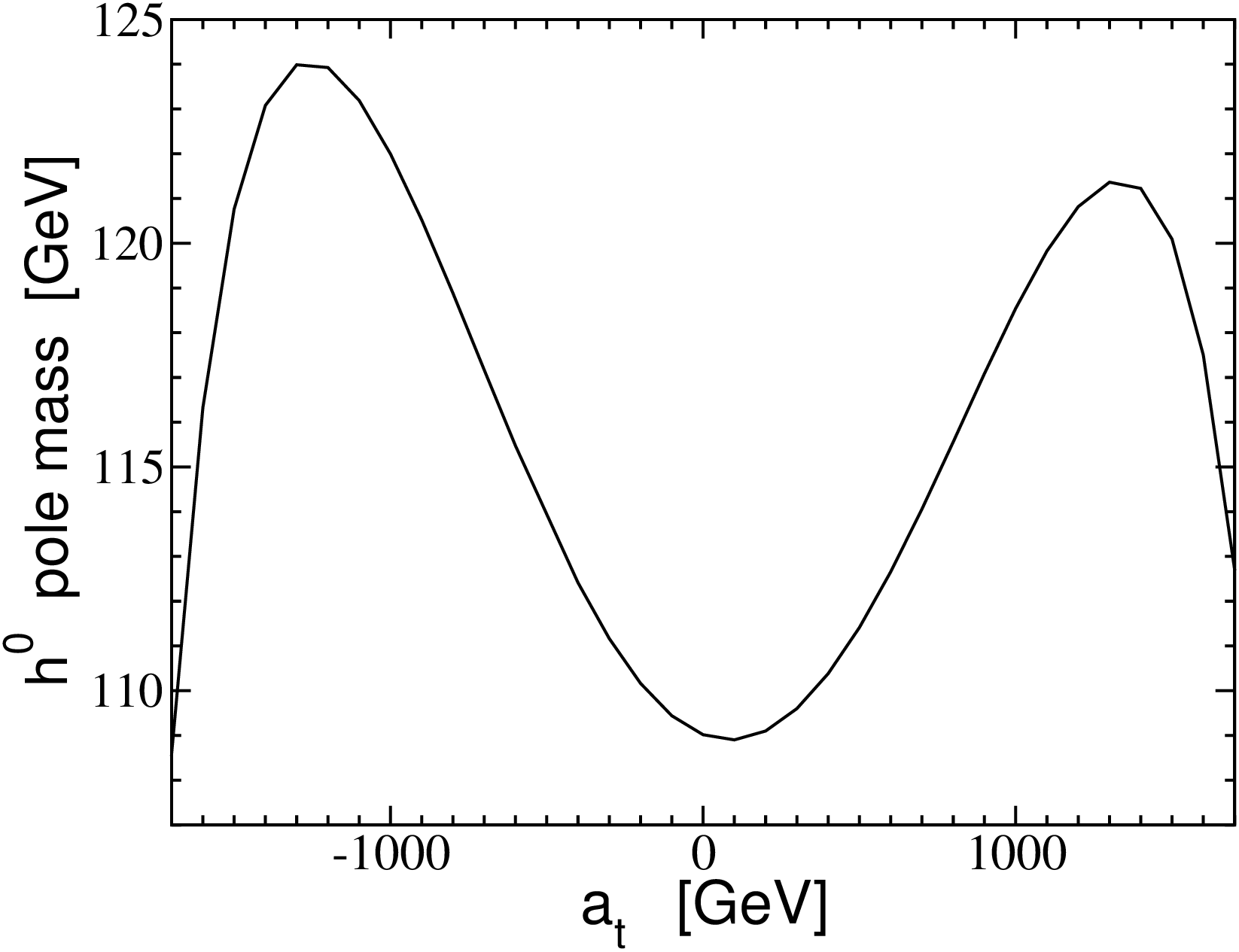}~~~~%
\includegraphics[width=8.6cm]{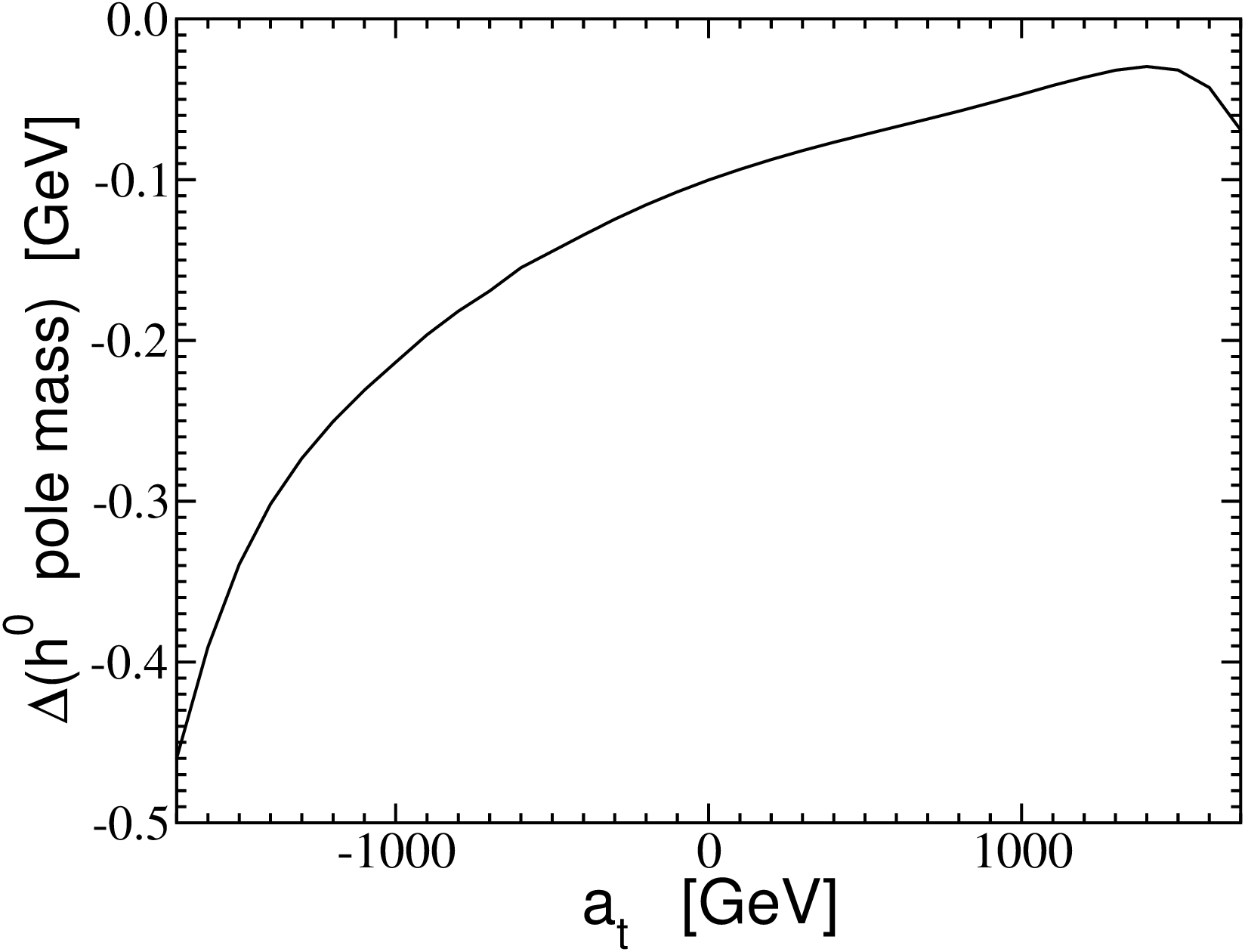}
\caption{\label{fig:atpole} The dependence of the computed $h^0$ pole mass
on the parameter $a_t$, for the model described in the text. The left
panel is the same approximation as the solid line of fig.~\ref{fig:hpole}. 
The right panel shows the change in the $h^0$ pole mass
induced by including the momentum-dependent self-energy, compared
to the full two-loop effective potential approximation.}
\end{figure*}
The imaginary part of the pole mass can in principle be used to obtain the 
physical decay width of 
$h^0$. The contribution from various decay channels can be identified by
isolating the imaginary parts due to each one-loop and two-loop 
contribution to the self-energy. In fig.~\ref{fig:gammahpole}, I show
the width corresponding to the decays $h^0 \rightarrow b \overline b$ and
$h^0 \rightarrow b \overline b g$. (Not included are spurious imaginary 
contributions of the self-energy coming from diagrams with Goldstone
bosons, which arise because we have not 
included all of the two-loop self-energy diagrams with non-zero $s$.) The 
dashed
line shows the result coming entirely from the imaginary parts of
one-loop bottom-quark diagrams, but using the (real) two-loop effective 
potential
approximation in order to get the kinematics correct by making a 
reasonable approximation for the real part of the pole mass $s = 
m^2_{h^0,{\rm pole}}$. The solid line incorporates the additional
parts from two-loop diagrams, which therefore includes the effects 
of gluon emission and one-loop corrections to the $h^0b\overline 
b$ vertex and the $b$-quark propagator. The complex pole mass is obtained 
by iteration of 
eq.~(\ref{eq:iteratepole}). In contrast, the dot-dashed line shows the 
same result,
but using the method of expanding the self-energies about the tree-level
mass, as in eqs.~(\ref{eq:expandpole})-(\ref{eq:taylorpole}). The 
latter method has a strong $Q$-dependence for the width
(although it only makes a difference
of at most a few tens of MeV in the real part of the pole mass). This is 
because 
the tree-level $h^0$ mass is only close to the two-loop mass for
renormalization scales near $Q=675$ GeV. Of course, the Higgs decay width 
is more accurately calculated using other methods 
(see e.g.~\cite{Djouadi:1997yw,Guasch:2003cv}
and references therein).

I have checked that comparable results obtain for a variety of other MSSM
model parameters, including some with large $\tan\beta$. As one 
illustration, consider the effect of the top squark mixing, which is
well-known to have a significant effect on the $h^0$ mass. 
Figure~\ref{fig:atpole} shows the dependence of the computed pole mass on 
the Lagrangian Higgs-$\tilde t_L$-$\tilde t_R$ coupling parameter
$a_t$, keeping all other parameters (except $\mu$ and $b$) fixed to the
values given above.  
Recall from the definition of ref.~\cite{Martin:1997ns}
or \cite{effpotMSSM} that the off-diagonal entries in the 
tree-level top-squark
squared-mass matrix are $v_u a_t - \mu y_t v_d$. Therefore, the top
squark mixing angle vanishes for $a_t = \mu y_t/\tan\beta$ (in this
model, about 45 GeV). Figure~\ref{fig:atpole} illustrates that the part of 
the 
$h^0$ pole
mass coming from momentum-dependent effects in the two-loop self-energy
is at most a few hundred MeV, and often much less.


In fig.~\ref{fig:atpole}, the maximum $h^0$ pole mass is obtained for
negative $a_t$, which at first sight might appear to differ from the
results obtained in refs.~\cite{Espinosa:1999zm,Espinosa:2000df,%
Heinemeyer:1998jw,Degrassi:2001yf}.  The reason is that different
quantities are being held constant while varying $a_t$. In those papers,
the on-shell masses are chosen to be held constant, while in this paper
the running parameters at the input renormalization scale are held
constant instead.  The fact that these two slices through parameter space
give opposite results for the condition that maximizes the $h^0$ pole mass
can be immediately seen by comparing eq.~(21) with $Q^2 = m_{\tilde t}^2$
and eq.~(27), both in ref.~[22].

\begin{figure}[tb]
\includegraphics[width=8.6cm]{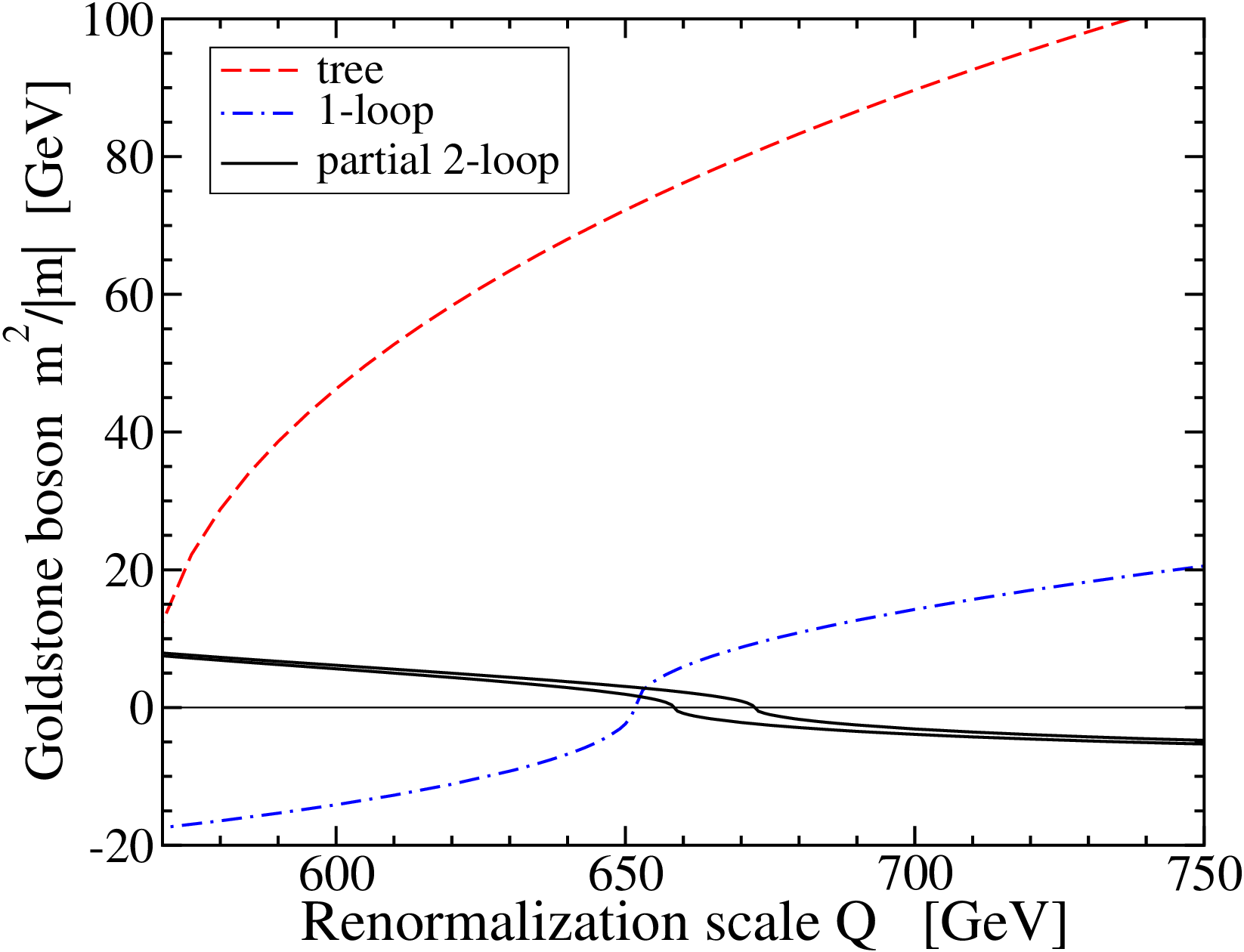}
\caption{\label{fig:gold} The Goldstone boson mass quantity 
$m^2_G/\sqrt{|m^2_G|}$ in GeV, in the tree-level, one-loop, and 
partial two-loop
approximations, for the model described in the text, as a function of the
choice of renormalization scale $Q$. 
The $G^0$ and $G^\pm$
lines are not visually distinguishable at tree-level (dashed) and 
one-loop 
(dot-dashed) 
order. The partial two-loop result for $G^0$ is the upper solid line and 
for $G^\pm$
is the lower solid line.
The full two-loop result for both $G^0$ and $G^\pm$ should be exactly 0 by
construction, since the fields are expanded around the minimum of the
Landau gauge two-loop effective potential.}
\end{figure}
As a numerical study of the effectiveness of the partial two-loop
self-energy corrections obtained in this paper, consider the masses of the
Goldstone bosons. Because the self-energies are obtained by expanding
the Higgs fields around VEVs that minimize the Landau gauge two-loop 
effective
potential, the Goldstone scalars $G^0$ and $G^\pm$ are exactly massless
at two loop order. This means that the matrices
\beq
m^2_{\phi^0_i} \delta_{ij} 
+ \frac{1}{16 \pi^2} \Pi^{(1)}_{\phi^0_i\phi^0_j} (0)
+ \frac{1}{(16 \pi^2)^2} \Pi^{(2)}_{\phi^0_i\phi^0_j} (0), &&
\label{eq:goldneut}
\\
m^2_{\phi^\pm_i} \delta_{ij} 
+ \frac{1}{16 \pi^2} \Pi^{(1)}_{\phi^+_i\phi^-_j} (0)
+ \frac{1}{(16 \pi^2)^2} \Pi^{(2)}_{\phi^+_i\phi^-_j} (0) &&
\label{eq:goldchar}
\eeq
each have one 0 eigenvalue. In figure \ref{fig:gold}, I show
the tree-level, one-loop and partial two-loop approximations to the
Goldstone boson mass quantity 
$m^2_G/\sqrt{|m^2_G|}$ as a function of the choice of renormalization 
scale $Q$. Here $m_G^2$ is defined to be the lowest 
eigenvalue of respectively
the first term, the first two terms, and all three terms with
$\Pi^{(2)}$ replaced by $\Pi^{(2)}_{\rm par}$, 
in eqs.~(\ref{eq:goldneut}) and (\ref{eq:goldchar}). 
Here $\Pi^{(2)}_{\rm par}$ is the partial two-loop approximation from 
sections 
\ref{sec:neutraltwoloop} and
\ref{sec:chargedtwoloop}. 
The effect of the approximation we have made for the two-loop 
self-energy is seen to be of 
order only tens of GeV$^2$ for the Goldstone boson squared masses
at $s=0$, and much smaller than for the one-loop and tree-level 
approximations. 

\begin{figure}[tb]
\includegraphics[width=8.6cm]{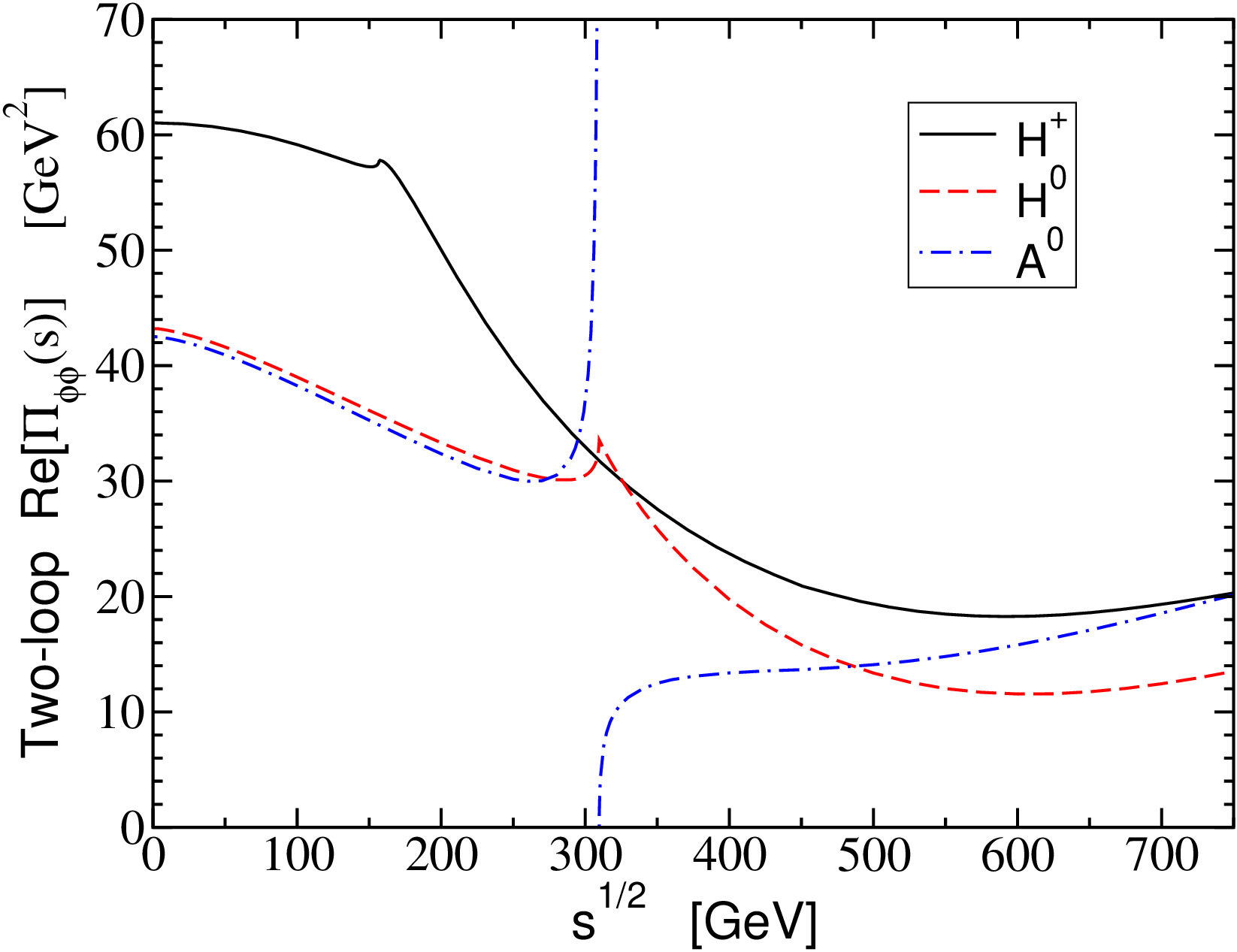}
\caption{\label{fig:Piphiphi} The two-loop contributions to the real parts
of the self-energy functions $\Pi_{H^0H^0}(s)$ and $\Pi_{A^0A^0}(s)$ 
(found in section \ref{sec:neutraltwoloop}) and
$\Pi_{H^+H^-}(s)$ (found in section \ref{sec:chargedtwoloop}), for the 
model described in the text, as a function of the momentum 
invariant $s$.}
\end{figure}
Let us now turn to the effects of the partial two-loop self energy 
corrections found in this paper on the 
heavier
Higgs scalar bosons $H^\pm$, $H^0$, and $A^0$. These corrections are
typically even smaller than for $h^0$, both in relative and absolute 
terms, in part because they have a weaker coupling to virtual top 
(s)quarks, but also because there are non-trivial cancellations.
Figure \ref{fig:Piphiphi} shows the dependence of the real parts of the
diagonal two-loop self-energies for $H^\pm$, $H^0$, and $A^0$, as a 
function of $s$. Since this model is not far from the decoupling limit,
these nearly form an isospin doublet, so the self-energy functions have
a similar behavior, especially at larger $s$. Note that the $A^0$ 
self-energy has a singular threshold at $\sqrt{s} = 2 m_t$, due to
the effects of massless gluon exchange. The diagrams of the type
$V_{FFFFV}$ and $M_{FFFFV}$ in Figure 1 cause threshold behavior 
proportional to
$(1 - s/4 m_t^2)^{-1/2}$ and ${\rm ln}(1 - s/4 m_t^2)$, respectively. 
If the pole mass were in the vicinity of this threshold, these 
singularities would have to be eliminated by re-summation, a topic 
beyond the scope 
of the
present paper. In contrast, the threshold behaviors of the
$H^0$ self-energy at $\sqrt{s}=2 m_t$ and of the $H^\pm$ self-energy at 
$\sqrt{s}=m_t + m_b$ are continuous (but not differentiable).
In all three cases, I have checked that there is a significant 
cancellation between the contributions of order $g_3^2 y_t^2$ and those of 
order $y_t^4$. The extent of this cancellation depends on the choice
of renormalization scale.

\begin{figure}[tb]
\includegraphics[width=8.6cm]{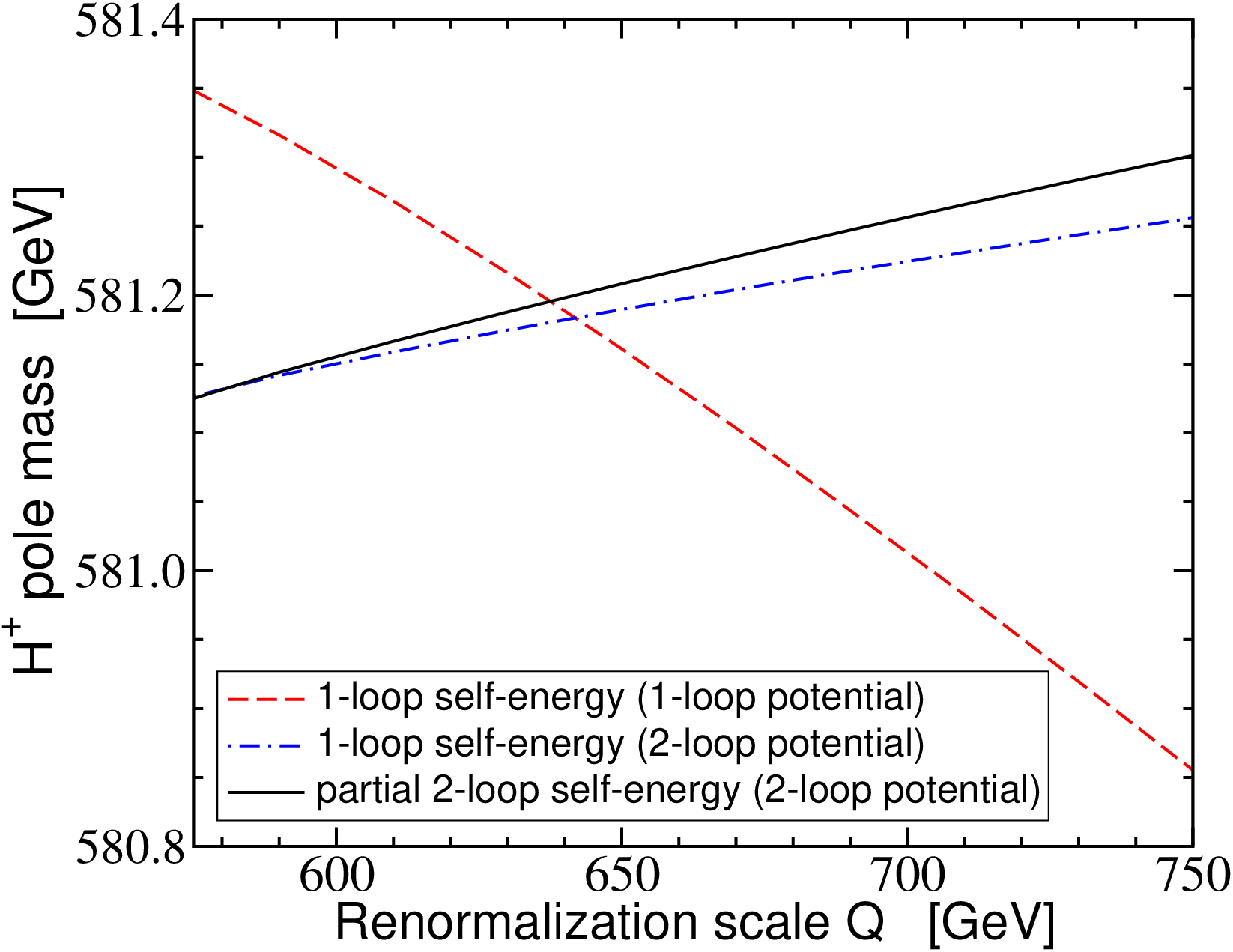}
\caption{\label{fig:Hppole} The computed $H^\pm$ pole mass for the model
described in the text,
in various approximations, as a function of the renormalization scale $Q$. 
The dashed line uses the one-loop effective
potential minimization conditions to determine parameters used in the 
one-loop self-energy. The
dot-dashed line uses the two-loop effective potential minimizations
condition, and the one-loop self-energy. The solid line uses the
two-loop effective potential minimization conditions, and the partial
two-loop self-energy as found in section \ref{sec:chargedtwoloop}.}
\end{figure}
\begin{figure}[t!h]
\includegraphics[width=8.6cm]{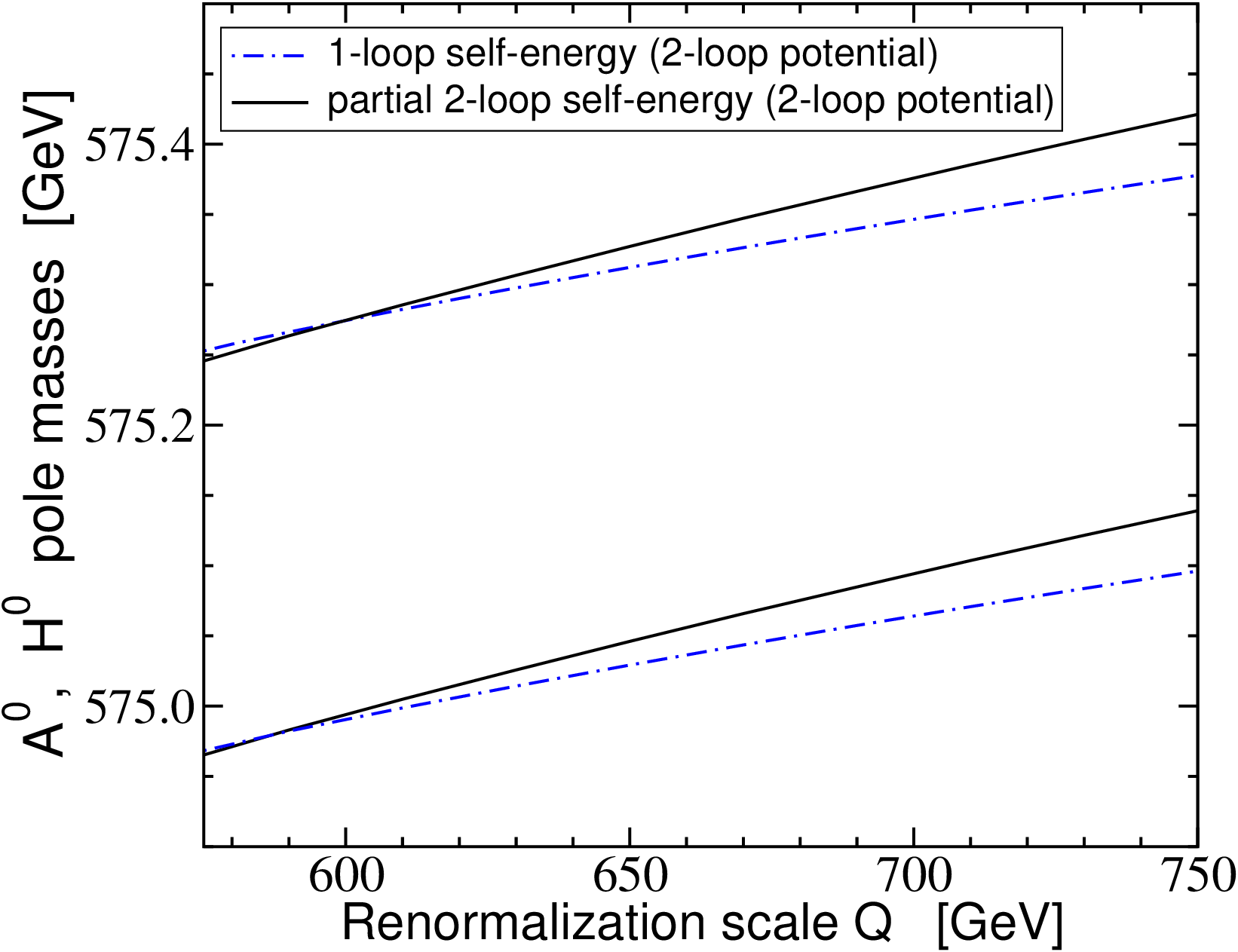}
\caption{\label{fig:AHpole} The computed pole masses of $A^0$ (lower pair
of lines) and $H^0$ (upper pair of lines), for the model described in the
text, as a function of the renormalization scale $Q$. The approximations
are as in figure \ref{fig:Hppole}.}
\end{figure}
The resulting effect of the partial two-loop self-energies on the $H^\pm$, 
$H^0$, 
and $A^0$
pole masses is rather small. Figure \ref{fig:Hppole} shows the 
renormalization scale
dependence of the calculated pole mass for the charged Higgs scalars.
Here, I do 
not use the trick of incorporating the effective potential results as
was done for $h^0$ in eq.~(\ref{eq:approxPI}), since the effective 
potential 
approximation to the self-energy is not close to valid for the heavier 
Higgs scalar 
bosons. 
Here the dashed line shows the result of a purely one-loop calculation,
meaning that the parameters $\mu$, $b$ are fixed from the VEVs by using
the one-loop effective potential, and the pole mass is computed using the
one-loop self-energy. The dot-dashed line uses the two-loop effective
potential to fix $\mu$, $b$, but then uses the one-loop self-energy 
function to get the pole mass. This is seen to remove much of the 
renormalization group scale dependence. Using the two-loop self-energy 
contributions as
found in this paper changes the pole mass by only a small amount,
(and actually makes the $Q$-dependence slightly worse). 
The 
change is much smaller than the dependence on $Q$. The 
remaining two-loop diagrams involving electroweak gauge couplings and
perhaps the three-loop contributions to electroweak symmetry breaking are 
therefore more important than the diagrams calculated here
for this case, and in particular should
remove most of the remaining $Q$ dependence in the calculated pole mass.
However, the remaining theoretical error is probably already much smaller
than future experimental uncertainties 
\cite{Battaglia:2001be,HiggsLCe}.

Very similar results follow for the $A^0$ and $H^0$ pole masses. They are
shown in Figure \ref{fig:AHpole}. The same remarks apply here as for
$H^\pm$.
 
\section{Outlook}\label{sec:outlook}
\setcounter{equation}{0}


In this paper, I have presented partial results for the two-loop
self-energy functions of the Higgs scalar bosons in minimal supersymmetry,
in the mass-independent and supersymmetric \DRbarprime~ renormalization
scheme. In the case of the lightest Higgs scalar, $h^0$, this allows an
improved calculation of the gauge-invariant pole mass, which should
correspond to the kinematic mass observed at colliders. The size of the
corrections was found in typical cases to be of order one to a few hundred
MeV. This is significant compared to the eventual experimental uncertainty
to be obtained at the LHC and especially at a LC.

To make further progress, it will be necessary to include the remaining
two-loop self-energy corrections involving electroweak couplings. This has
already been done in the effective potential approximation
\cite{effpotMSSM,Martin:2002wn}. However, it is precisely for these
contributions that the approximation $s=0$ is not always a very good one,
particularly for diagrams in which no momentum routing can avoid an
electroweak gauge boson. Therefore, it will certainly be necessary to
include these contributions in order to reduce the theoretical
uncertainties to acceptable levels. It also seems clear that the leading
(e.g.~$y_t^2 g_3^4$, $y_t^4 g_3^2$, and $y_t^6$) three-loop contributions
to the $h^0$ pole mass will be necessary, but can be included in the
effective potential approximation. These corrections can be estimated in a
leading-logarithm approach using the renormalization group,
as has recently been done in ref.~\cite{recentMSSMhiggs}. 
However, we have seen above that the non-logarithmic pieces are not always 
small compared to the logarithmic ones.

The size of the two-loop effects found above on the heavier Higgs boson
masses $H^\pm$, $H^0$, and $A^0$ do not seem to be significant compared to
the expected experimental uncertainties. However, I have not conducted an
exhaustive search of all of parameter space, and in any case the marginal
cost in human effort to include all of the Higgs scalar self-energies at
two-loop order is not great, once the two-loop self-energy for $h^0$ is
included.

Besides calculations in the Higgs sector, it will be necessary to
calculate two-loop corrections for the other superpartner masses in order 
to interpret the results above in realistic situations. This
issue is particularly acute in the mass-independent renormalization scheme
adopted here, since e.g.~the top-quark Yukawa coupling and the top-squark
tree-level masses are used as inputs, rather than the physical top-quark
and top-squark masses.  In order to make meaningful comparisons with
higher-order
calculations for the Higgs masses done in the on-shell schemes and to
future experimental constraints or (hopefully) data, the two-loop mass
corrections for the top and bottom quark, the squarks, and the gluino, at
least, will be needed. Fortunately, these results are definitely not out 
of reach.

\vspace{0.25in}

This work was supported by the National Science Foundation under Grant No.
0140129.

\end{document}